%% file: paper.tex
\documentclass{acmtrans2m}

\usepackage{xspace}
\usepackage{graphicx}
\usepackage{amsmath,amssymb}
\usepackage{boxedminipage}     
\usepackage{proof}             
\usepackage[usenames]{color}
\input{macros}

\newdef{definition}[theorem]{Definition}
\newdef{remark}[theorem]{Remark}
\newdef{example}[theorem]{Example}
           
\title{Nominal Logic Programming}
            
\author{JAMES CHENEY\\ University of Edinburgh
  \and CHRISTIAN URBAN\\ Technische Universit\"at, M\"unchen}

\begin{abstract} 
\input{abstract}
\end{abstract}
            

            
            
\begin{document}
\begin{bottomstuff} 
  The first author was supported by AFOSR grant F49620-01-1-0298,
ONR grant N00014-01-1-0968,
AFOSR grant F49620-03-1-0156,
  and EPSRC grant R37476 while performing this research.  The second
  author was supported by a fellowship from the Alexander-von-Humboldt
  foundation and an Emmy-Noether fellowship from the German Research Council.\medskip

  This paper expands and improves upon material presented in several
  earlier publications, primarily
  \cite{gabbay04lics,cheney04iclp,urban05tlca,cheney06iclp}.\medskip
\end{bottomstuff}

\maketitle

\hfill 
\begin{minipage}{0.9\textwidth}
Declarative Programming enables one to concentrate on the essentials 
of a problem, without getting bogged down in too much operational detail.\\
\mbox{}\hfill ---David Warren in \cite{SterlingShapiro94}\medskip 
\end{minipage}

\input{introduction}

\input{examples}


\input{background}


\input{semantics}


\input{applications}

\input{comparison}

\input{conclusions}

\bibliographystyle{acmtrans}
\bibliography{nominal,logicprog,logic,fp,lf,lambda,har}

\begin{received}
  Received Month Year;
  revised Month Year;
  accepted Month Year
\end{received}

\newpage
\appendix

\input{appendix}

\end{document}

%% file: macros.tex
\newcommand{\FOLDNabla}{\ensuremath{FO\lambda^{\Delta\nabla}}\xspace}

\newcommand{\supp}{\mathrm{supp}}
\newcommand{\alphacaml}{C$\alpha$ml}

\newcommand{\intTy}{\mathbf{int}}
\newcommand{\charTy}{\mathbf{char}}
\newcommand{\listTy}{\mathbf{list}}
\newcommand{\wf}[3]{#1\nd #2 : #3}

\newcommand{\elab}{\leadsto}
\newcommand{\reduce}[3]{#1\ab{#2}\Downarrow #3}
\newcommand{\eq}{\approx}
\newcommand{\ev}{\sim}

\newcommand{\labelEq}[1]{\label{eqn:#1}}
\newcommand{\refEq}[1]{(\ref{eqn:#1})}

\newcommand{\labelSec}[1]{\label{sec:#1}}
\newcommand{\refSec}[1]{Section~\ref{sec:#1}}
\newcommand{\labelFig}[1]{\label{fig:#1}}
\newcommand{\refFig}[1]{Figure~\ref{fig:#1}}

\newcommand{\labelRmk}[1]{\label{rmk:#1}}
\newcommand{\refRmk}[1]{Remark~\ref{rmk:#1}}
\newcommand{\labelThm}[1]{\label{thm:#1}}
\newcommand{\refThm}[1]{Theorem~\ref{thm:#1}}
\newenvironment{restateThm}[1]
  {\begin{@begintheorem}{Theorem}
  {\ref{thm:#1}}}{\end{@begintheorem}}
\newcommand{\labelLem}[1]{\label{lem:#1}}
\newcommand{\refLem}[1]{Lemma~\ref{lem:#1}}
\newenvironment{restateLem}[1]
  {\begin{@begintheorem}{Lemma}
  {\ref{lem:#1}}}{\end{@begintheorem}}
\newcommand{\labelProp}[1]{\label{prop:#1}}
\newcommand{\refProp}[1]{Proposition~\ref{prop:#1}}
\newenvironment{restateProp}[1]
  {\begin{@begintheorem}{Proposition}
  {\ref{prop:#1}}}{\end{@begintheorem}}
\newcommand{\labelCor}[1]{\label{cor:#1}}

\newcommand{\labelDef}[1]{\label{def:#1}}
\newcommand{\refDef}[1]{Definition~\ref{def:#1}}
\newcommand{\labelEx}[1]{\label{ex:#1}}
\newcommand{\refEx}[1]{Example~\ref{ex:#1}}
\newcommand{\labelApp}[1]{\label{app:#1}}
\newcommand{\refApp}[1]{Appendix~\ref{app:#1}}

\newcommand{\conc}{\mathop{@}}

\newcommand{\LL}{\mathcal{L}}

\newcommand{\D}{\mathcal{D}} 
\newcommand{\E}{\mathcal{E}}
\newcommand{\MM}{\mathcal{M}} 
\newcommand{\HH}{\mathcal{H}}
 
\newcommand{\VV}{\mathcal{V}}

\newcommand{\SB}[1]{[\![#1]\!]}
\newcommand{\lfp}{\mathrm{lfp}}
\renewcommand{\emptyset}{\varnothing}
\newcommand{\Union}{\bigcup}
\newcommand{\union}{\cup}
\newcommand{\Isect}{\bigcap}

\newcommand{\new}[1][]{\reflectbox{\sf{#1{}N}}}
\newcommand{\smallnew}{\new[\scriptsize]}
\newcommand{\BA}{\mathbb{A}}

\newcommand{\trueR}{{\top}R}

\newcommand{\impL}{{\impp}L}

\newcommand{\andL}{{\andd}L}
\newcommand{\andR}{{\andd}R}

\newcommand{\orR}{{\orr}R}
\newcommand{\allL}{{\forall}L}

\newcommand{\exR}{{\exists}R}
\newcommand{\newL}{{\new}L}
\newcommand{\newR}{{\new}R}

\newcommand{\hyp}[2][]{\infer[#1]{#2}{}} 

\newcommand{\app}[2]{#1\,#2}
\newcommand{\tran}[2]{(#1~#2)}
\newcommand{\act}{{\boldsymbol{\cdot}}}

\newtheorem{theorem}{Theorem}[section]

\newtheorem{corollary}[theorem]{Corollary}
\newtheorem{proposition}[theorem]{Proposition}
\newtheorem{lemma}[theorem]{Lemma}

\newcommand{\aprolog}{\ensuremath\alpha\text{Prolog}\xspace}
\newcommand{\lprolog}{\ensuremath\lambda\text{Prolog}\xspace}

\newcommand{\seq}{\Rightarrow}
\newcommand{\nd}{\vdash}
\renewcommand{\models}{\vDash}

\newcommand{\red}[1]{\xrightarrow{#1}}
\newcommand{\notsat}[3]{[#1]~ #2\not\models#3}
\newcommand{\sat}[3]{[#1]~ #2\models#3}
\newcommand{\satsn}[1]{\sat{\Sigma}{\nabla}{#1}}
\newcommand{\upf}[4]{[#1]~ #2;#3\Longrightarrow#4}
\newcommand{\upfsd}[2]{\upf{\Sigma}{\Delta}{#1}{#2}}
\newcommand{\upfsdn}[1]{\upf{\Sigma}{\Delta}{\nabla}{#1}}
\newcommand{\apf}[5]{[#1]~ #2;#3\red{#4}#5}
\newcommand{\apfsd}[3]{\apf{\Sigma}{\Delta}{#1}{#2}{#3}}
\newcommand{\apfsdn}[2]{\apf{\Sigma}{\Delta}{\nabla}{#1}{#2}}
\newcommand{\rupf}[4]{[#1]~ #2\Longrightarrow#3~\backslash~#4}
\newcommand{\rupfsd}[1]{\rupf{\Sigma}{\Delta}{#1}}
\newcommand{\rapf}[5]{[#1]~ #2\red{#3}#4~\backslash~#5}
\newcommand{\rapfsd}[2]{\rapf{\Sigma}{\Delta}{#1}{#2}}
\newcommand{\equpf}[4]{[#1]~ #2;#3\Longrightarrow_\eq#4}
\newcommand{\equpfsdn}[1]{\equpf{\Sigma}{\Delta}{\nabla}{#1}}
\newcommand{\eqapf}[5]{[#1]~ #2;#3\red{#4}_\eq#5}
\newcommand{\eqapfsdn}[2]{\eqapf{\Sigma}{\Delta}{\nabla}{#1}{#2}}

\newcommand{\eqrapf}[5]{[#1]~ #2\red{#3}_\eq#4~\backslash~#5}
\newcommand{\eqrapfsd}[2]{\eqrapf{\Sigma}{\Delta}{#1}{#2}}

\newcommand{\st}[3]{#1\ab{#2\mid#3}}

\newcommand{\trans}[1]{\ulcorner{#1}\urcorner}
\newcommand{\true}{\top}
\newcommand{\false}{\bot}

\newcommand{\fresh}{\mathrel\#}

\newcommand{\too}{\longrightarrow}
\newcommand{\To}{\Rightarrow}
\newcommand{\ent}{\mathrel{{:}-}}

\newcommand{\ab}[1]{\langle #1 \rangle}

\newcommand{\Aa}{\mathsf{a}}
\newcommand{\Ab}{\mathsf{b}}
\newcommand{\Ac}{\mathsf{c}}

\newcommand{\Aq}{\mathsf{q}}
\newcommand{\Ar}{\mathsf{r}}
\newcommand{\Av}{\mathsf{v}}
\newcommand{\Ax}{\mathsf{x}}
\newcommand{\Ay}{\mathsf{y}}
\newcommand{\Aw}{\mathsf{w}}
\newcommand{\Az}{\mathsf{z}}
\newcommand{\impp}{\Rightarrow}
\newcommand{\Andd}{\bigwedge}
\newcommand{\andd}{\wedge}
\newcommand{\orr}{\vee}
\newcommand{\Orr}{\bigvee}

\newcommand{\powerset}[1]{\mathcal{P}(#1)}
\newcommand{\PP}{\mathcal{P}}
\newcommand{\lto}[1]{\stackrel{#1}{\longrightarrow}}

\newcommand{\query}{\mathrel{?\mbox{--}}}

\newcommand{\type}{\mathbf{type}}
\newcommand{\nametype}{\mathbf{name\_type}}
\newcommand{\abs}[2]{\langle #1 \rangle #2}

\renewcommand{\bar}[1]{\overline{#1}}
\newcommand{\swap}[3]{(#1~#2)\cdot#3}

\newcommand{\newgoal}{$\new$-goal\xspace}

\newcommand{\NPTIME}{\mathbf{NP}}
\newcommand{\lam}{\lambda}


%% file: abstract.tex
{  Nominal logic is an extension of first-order logic which provides a
  simple foundation for formalizing and reasoning about abstract
  syntax modulo consistent renaming of bound names (that is,
  $\alpha$-equivalence).  This article investigates logic programming
  based on nominal logic.  
  We describe some typical nominal logic programs, and
  develop the model-theoretic, proof-theoretic, and operational
  semantics of such programs.  Besides being of interest
  for ensuring the correct behavior of implementations, these results
  provide a rigorous foundation for techniques for analysis and
  reasoning about nominal logic programs, as we illustrate via 
  examples.
}

%% file: introduction.tex
\section{Introduction}\labelSec{introduction}

As stated by Warren the ideal of logic programming is that all the
programmer needs to do is describe the problem suitably, and let the
computer deal with the search for solutions.  Thus, logic programming
languages such as Prolog are very well-suited to problem solving
situations in which a problem can be formulated as a set of inference
rules describing a solution.  All the programmer has to do is describe
the problem and ask the system to search for solutions.

Unfortunately, for some problems this ideal is not achievable in
Prolog, the most well-known logic programming language, even in areas
where this language is regarded as superior.  Consider for example the
usual three inference rules by which the type-system for lambda-terms
is specified:

\begin{center}
\begin{tabular}{c@{\hspace{7mm}}c@{\hspace{7mm}}c}
 \infer{\Gamma\nd x : \tau}{x : \tau \in\Gamma}
&\infer{\Gamma\nd e_1~e_2: \tau'}
       {\Gamma\nd e_1: \tau\to \tau' &
        \Gamma\nd e_2\!: \tau}
&\infer{\Gamma\nd \lambda x.e: \tau\to \tau'}
       {\{x:\tau\}\cup\Gamma\nd e: \tau'}
\end{tabular}
\label{typing-rules}
\end{center}

\noindent
In the third rule it is often implicitly assumed that $x$ is a
variable not already present in $\Gamma$.  Inferring a type for the
term $e$ in the context $\Gamma$ should fit Prolog's declarative
programming paradigm very well.  However, a direct, na\"ive
implementation of such typing rules, as for example given in
\cite[Page~489]{Mitchell03}:
\[
\begin{array}{lcl}
mem(X,[X|T]).\\
mem(X,[Y|T]) &\ent& mem(X,T).\\
\\
tc(G,var(X),T) &  \ent & mem((X,T),G).\\
tc(G,app(E_1,E_2),T') &  \ent &  tc(G,E_1,{arrTy}(T,T')), tc(G,E_2,T).\\
tc(G,lam(X,E),{arrTy}(T,T'))  & \ent &
    tc([(X,T)|G],E,T').
\end{array}
\]
\noindent
behaves incorrectly on terms in which a lambda-bound name
``shadows'' another binding occurrence of a name.  For example, 
typechecking the lambda-term $\lambda x.\lambda x. (x\;x)$ via the query
\[
\query {tc}([],lam(x,lam(x,app(var(x),var(x)))),U)
\]
  yields two answers: 
  \begin{eqnarray*}
    U &=& arrTy(T,arrTy(arrTy(T,T'),T'))\\
    U &=& arrTy(arrTy(T,T'),arrTy(T,T'))\;.
  \end{eqnarray*}
  The first answer corresponds to
  binding the first bound occurrence of $x$ to the inner binder and
  the second to the outer binder; the second corresponds to the
  reverse binding.  Neither is correct, since this term is not
  well-typed.  This assumes that the implementation performs occurs
  checks---if the checks are omitted, this query may diverge instead.

  This problem can be worked around in several ways, including
  judicious use of the ``cut'' pruning operator to ensure that only
  the most recent binding of a repeated variable can be used (e.g. in
  the first clause of $mem$), or by defining a \texttt{gensym}
  predicate, defining capture-avoiding substitution, and performing
  explicit $\alpha$-renaming~(see \cite{clocksin03programming}), but
  both solutions rely on nonlogical, nondeclarative features of
  Prolog, and the resulting programs generally only work properly in
  the ``forward'' direction (when used with ground $G$ and $E$).
  Thus, one loses declarativeness and becomes ``bogged down in
  operational detail'' almost immediately even for the simplest
  problems involving name-binding.

The problems with the {na{\"\i}ve} implementation stem from the lack of
support for names, name-binding and alpha-equivalence in Prolog.  A
number of techniques for incorporating such support into logic
programming languages have been investigated, including higher-order
logic programming~\cite{nadathur98higher},
Qu-Prolog~\cite{staples89meta}, and logic programming with binding
algebras~\cite{hamana01tacs}.  

Of these approaches, higher-order
logic programming may be the most convenient and compelling.  For example,
the typechecking relation can be implemented in \lprolog as follows:
\[\begin{array}{lcl}
tc~(app~E_1~E_2)~T'& \ent& tc~E_1~({arrTy}~T~T'), tc~E_2~T.\\
tc~(lam~(\lambda x.E~x))~({arrTy}~T~T') &\ent& \Pi x.~ tc~x~T \impp {tc~(E~x)~T'}.
\end{array}\]
Here, meta-language variables and $\lambda$-bindings are used to
represent object-language variables and bindings; object language
application and lambda-abstraction are represented using constants
$app: exp \to exp \to exp$ and $lam:(exp \to exp)\to exp$.  Moreover,
local parameters {(introduced using the universal quantifier
  $\Pi$)} and local assumptions {(introduced using the
  implication connective $\impp$)} are used to represent the scope
restrictions on the local variable and its type assumption.  Thus, the
meta-language's context is used to implement locally-scoped parameters
and hypotheses of the object language.

{
Higher-order abstract syntax is a very elegant technique for programming with and reasoning about
languages with binding syntax.}  Unfortunately, there are some situations in which
higher-order encodings are no simpler than first-order equivalents; sometimes,
the use of higher-order features even obstructs natural-seeming programming
techniques.
  As a case in point, consider the following informal definition
  of the alpha-inequivalence relation $\not\equiv_\alpha$:
\newcommand{\aneq}{\not\equiv_\alpha}
\[
\begin{array}{c}
\infer{x \aneq y}{x \neq y} \quad \infer{e_1~e_2 \aneq e_1'~e_2'}{e_1\aneq e_1'}\quad \infer{e_1~e_2 \aneq e_1'~e_2'}{e_2\aneq e_2'} \quad \infer{\lambda x.e \aneq \lambda x.e'}{e \aneq e'}\\
\smallskip\\
\infer{e_1~e_2 \aneq \lambda x.e}{} \quad\infer{\lambda x.e\aneq e_1~e_2 }{} \quad \infer{x \aneq e_1~e_2 }{} \quad \infer{e_1~e_2 \aneq x}{} \quad \infer{\lambda x.e \aneq y}{} \quad \infer{y \aneq \lambda x.e}{} 
\end{array}
\]
Most of the
clauses are easy to implement in, for example, \lprolog; in
particular, the implicit use of the Barendregt renaming convention in
the $\lambda$-$\lambda$-rule can be used to provide an elegant, direct
translation:
\[
\begin{array}{lcll}
aneq~(lam~(\lambda x.E~x))~(lam~(\lambda x.E'~x)) &\ent & \Pi x.  aneq~(E~x)~(E'~x)
\end{array}
\]
However, we appear ``stuck'' when we wish to encode the
$var$-$var$-rule, since there is no obvious way of translating the
informal side-condition $x \neq y$ to a predicate $neq : exp \to exp
\to o$ that succeeds only when its arguments are distinct
eigenvariables.  

It is, nevertheless, still possible to define the $\aneq$ relation
between closed terms in \lprolog, in terms of a auxiliary predicates
$aneq':list~exp\to exp \to exp \to o$, $freshFor : exp \to list~exp
\to o$, and $neq:exp \to exp \to o$:
\[
\begin{array}{lcll}
aneq~E~N &\ent& aneq'~[]~E~N\\
aneq'~L~X~Y &\ent& neq~X~Y\\
...\\
aneq'~L~(lam~E)~(lam~E') &\ent & \Pi x. freshFor~x~L \impp aneq'~(x::L)~(E~x)~(E'~x)
\end{array}
\]
where the auxiliary predicate $freshFor$ has no defining
clauses and $neq$ is defined as
\[
\begin{array}{lcll}
neq~X~Y &\ent& freshFor~X~L, mem~Y~L\\
neq~X~Y &\ent& freshFor~Y~L, mem~X~L
\end{array}
\]

  We believe that this example illustrates that, just as first-order
  syntax is often too low-level because of the absence of first-class
  support for names and binding, higher-order syntax is sometimes too
  high-level because it abstracts away from the ability to compare and
  generate names as first-class data.  Thus, there are cases where
  neither first-order nor higher-order logic programming enables us to
  simply ``concentrate on the essentials of a problem'' involving
  names and binding.

In this paper, we investigate a new approach in which both of the
above examples (and a wide variety of other programs) can be
implemented easily and (we argue) intuitively.  Our approach is based
on \emph{nominal logic}, an extension of first-order logic introduced
by \citeN{pitts03ic}, and based on the novel approach to abstract
syntax developed by \citeN{gabbay02fac}.  In essence, nominal logic
axiomatizes an inexhaustible collection of \emph{names} $\Ax,\Ay$ and
provides a first-order axiomatization of a name-binding operation
$\abs{\Ax}{t}$ (called \emph{abstraction}) in terms of two primitive
operations, \emph{swapping} ($\swap{a}{b}{t}$) and \emph{freshness}
($a \fresh t$).  In addition, nominal logic includes a novel quantified
formula $\new\Aa.\phi$ (``for fresh $\Aa$, $\phi$ holds'') which
quantifies over fresh names.

In nominal logic, names and binding are abstract data types admitting only
swapping, binding, and operations for equality and freshness testing.
Name-abstractions $\abs{\Ax}{t}$ are considered equal up to
$\alpha$-equivalence, defined in terms of swapping and freshness.  For
example, object variables $x$ and lambdas $\lambda x.t$ can be encoded as
nominal terms $var(\Ax)$ and abstractions $lam (\abs{\Ax}{t})$ where $var : id
\to exp$ and $lam : \ab{id} exp \to exp$.  We can obtain a correct
implementation of the {$tc$} relation above by replacing the {third clause of $tc$} with
\[
{tc}(G,lam(\abs{\Ax}{E}),arrTy(T,U)) \ent \Ax \fresh G, {tc}({[(\Ax,T)|G]},E,U).
\]
which we observe corresponds closely to the {third inference
  rule} (reading $lam(\abs{\Ax}{E})$ as $\lambda x.E$, {$\Ax
  \fresh G$} as $x \not\in FV(\Gamma)$, 
and {$[(\Ax,T)|G]$} as $\{x{:}\tau\}\cup\Gamma$ ).  
Similarly, the $var$-$var$ clause of $aneq$ can be implemented directly
as 
\[
aneq(var(X),var(Y)) \ent X \fresh Y
\]
where the inequality side-condition $x \neq y$ is captured by the
constraint $X \fresh Y$; all of the other clauses of $aneq$ are also
direct translations of their informal versions.  

We refer to this approach to programming with names and binding modulo
$\alpha$-equivalence as \emph{nominal abstract syntax}.  This approach
provides built-in $\alpha$-equiv\-a\-lence and fresh name generation,
while retaining a clear declarative interpretation.  Names are
sufficiently abstract that the low-level details of name generation
{and $\alpha$-conversion} can be hidden from the programmer,
yet still sufficiently concrete that there is no difficulty working
with open terms, freshness constraints, or inequalities among names
precisely as is done ``on paper''.  Nominal abstract syntax and
nominal logic make possible a distinctive new style of
meta-programming, which we call \emph{nominal logic programming}.

  It is important to emphasize that we are not attempting to make or
  defend a claim that nominal techniques are ``superior'' in some
  sense to other techniques in all cases.  Instead, we argue only that
  that nominal techniques provide an interesting and different
  approach which, in some cases (such as $aneq$ above), does seem more
  convenient than other extant techniques.  However, higher-order and
  some other techniques certainly have advantages that are not shared
  by our approach, such as the presence of built-in, efficient
  capture-avoiding substitution.  It seems an open question whether
  the advantages of nominal and higher-order abstract syntax can be
  combined within a single system.

In this paper, we describe a particular implementation of nominal
logic programming, called \aprolog.  We also investigate the semantics
of nominal logic programs and discuss applications of these results.
\begin{itemize}
\item We first (\refSec{examples}) illustrate nominal logic
  programming via several examples written in \aprolog, drawing on
  familiar examples based on the $\lambda$-calculus and
  $\pi$-calculus.  The aim of these examples is to show that, in
  contrast to all other known approaches, \aprolog programs can be
  used to encode calculi correctly yet without essential alterations
  to their paper representations.  Thus, \aprolog can be used as a
  lightweight prototyping tool by researchers developing new systems,
  or by students learning about existing systems.  
  This section provides only a high-level discussion of nominal logic;
  readers who wish to understand the formal details before seeing
  examples may prefer to read Sections~\ref{sec:background} and
  \ref{sec:semantics} first.

\item We next (\refSec{background}) provide a summary of nominal
  abstract syntax and nominal logic needed for the rest of the paper.
  We introduce the domain of nominal terms, which plays a similar role
  to ordinary first-order terms in Prolog or lambda-terms in \lprolog,
  then review the semantics of term models of nominal logic
  (previously developed in \cite{cheney06jsl}){, and finally
    define a core nominal logic programming language.}

\item \refSec{semantics} develops the semantics of nominal logic
  programs.  This is crucial for justifying our claim that the
  notation and concepts of nominal logic match our intuition, and that
  nominal logic programs capture the informal meaning we assign to
  them.  Using the foundations introduced in \refSec{background}, we
  provide a model-theoretic semantics of nominal logic programs
  following \citeN{lloyd87foundations}.  We also introduce a uniform
  proof-theoretic semantics~\cite{miller91apal} via a variation of the
  proof-theoretic semantics of CLP, investigated by
  \citeN{darlington94lpar} and \citeN{leach01tplp}.  Finally, we
  present an operational semantics that models the low-level proof
  search behavior of an interpreter more directly.  We prove
  appropriate soundness and completeness results relating these
  definitions along the way.

\item In \refSec{applications}, we consider some applications of the
  semantics to issues arising in an implementation such as \aprolog.
  {We discuss how to
    use the semantics to check the correctness (``adequacy'') of
    \aprolog programs,} and verify the correctness of a standard
  ``elaboration'' transformation and an optimization which permits us
  to avoid having to solve expensive, $\NPTIME$-complete nominal
  constraint solving problems during execution.  This result
  supersedes an earlier characterization of \citeN{urban05tlca}.

\item \refSec{comparison} presents a detailed comparison of our work
  with previous techniques for incorporating support for name-binding
  into programming languages and \refSec{concl} concludes.
\end{itemize}

In order to streamline the exposition, many routine cases in proofs
in the body of the paper have been omitted. Complete proofs are
available in appendices.



%% file: examples.tex
\section{Programming in \aprolog}\labelSec{examples}

\subsection{Syntax}

  Before presenting examples, we sketch the concrete syntax we shall
  employ in this section for \aprolog programs, shown
  in~\refFig{concrete-syntax}.  The concrete syntax includes
  facilities for declarations of constants, function symbols, types
  and type abbreviations, clause declarations, and queries in this
  paper.
To improve readability, the syntax employed in the paper differs
slightly from the ASCII syntax employed in the current implementation.
The nominal terms used in \aprolog include standard first-order
variables $X$, constants $c$, and function symbols $f$; also, we have
new syntax for names $\Aa$, name-abstractions $\abs{a}{t}$, and
swappings $\swap{a}{b}{t}$.

\begin{figure}[tb]
  \[\begin{array}{llcl}
    \text{Terms} & t,u &::=& X \mid c \mid f(\vec{t}) \mid\Aa \mid  \abs{a}{t} \mid {\swap{a}{b}{t}} \mid  i \mid \text{'c'} \mid [] \mid t::t' \mid [t_1,\ldots,t_n|t'] \mid (t,t')\\
    \text{Constructor types} & \tau,\nu & ::= & tid~\vec{\sigma} \mid \sigma \to \tau\\
    \text{Types} & \sigma &::=& \alpha \mid \tau \mid \abs{\nu}{\sigma} \mid\intTy \mid \charTy \mid \listTy~\sigma \mid \sigma\times \sigma' \mid \sigma \to \sigma \mid o\\
    \text{Basic Kinds} & \kappa_0 & ::= & \nametype \mid \type \\
    \text{Kinds} & \kappa & ::= & \kappa_0 \mid \kappa_0 \to \kappa\\
    \text{Atomic formulas} & A &::=& p(\vec{t}) \mid f(\vec{t}) = u\\
    \text{Goals} & G & ::=& A \mid a \fresh t \mid t \eq u \mid G,G' \mid G;G' \mid \exists X{:}\sigma.G \mid \new \Aa{:}\nu.G\\
    \text{Declarations} & D & ::= &  tid : \kappa  \mid defid :: \sigma  \mid \type~tid~\vec{\alpha} = \sigma \mid conid : \tau \mid A \ent G
\end{array}\]
\caption{{Concrete} syntax summary}\labelFig{concrete-syntax}
\end{figure}

  Names and name-abstractions are used to represent syntax with bound
  names in \aprolog.  The unification algorithm used by \aprolog
  solves equations modulo an equational theory that equates terms
  modulo $\alpha$-renaming of names bound using abstraction.
  Swappings are a technical device (similar to explicit
  substitutions~\cite{abadi91jfp}) which are needed in constraint
  solving; they do not often appear in programs, but may appear in
  answer substitutions.  
We will present the details of the equational theory in
\refSec{background}.

\aprolog also contains standard built-in types for pairing, lists,
integers, and characters.  Note that $[t_1,\ldots,t_n|t']$ is a
standard Prolog notation for matching against an initial segment of a
list; it is equivalent to $t_1::\cdots::t_n::t'$.

User-defined types, including name types, can be introduced using 
declarations such as
\[
tid : \type. \qquad 
ntid : \nametype.
\]
Also, using functional kinds, we can introduce new type constructors
used for user-defined parametrized types.  For example, $\listTy$
could be declared as
\[\listTy : \type \to \type.\]
Similarly, abstraction $\abs{\nu}{\sigma}$ could be declared as 
\[\abs{-}- : \nametype \to \type \to \type.\] 
Only first-order kinds are supported in the current implementation.

Type abbreviations (possibly with parameters) can be introduced using
the syntax
\[\type~tid~\alpha_1~\cdots~\alpha_n=~\sigma(\alpha_1,\ldots,\alpha_n).\]

Likewise, uninterpreted constants and function symbols (which we call
\emph{(term) constructors}) are declared using a similar notation:
\[
conid : \tau.
\]
here $\tau$ is a ``constructor type'', that is, either a
user-defined type constructor application $tid~\vec{\sigma}$ or a
function type returning a constructor type.  These restrictions ensure
that user-defined term constructors cannot be added to built-in types,
including name-types, lists and products.  Constants and function
symbols must return a user-defined data type; so, there can be no
constants, function symbols, or other user-defined terms in a name
type, only name-constants.

Interpreted function and predicate symbols can be defined using 
the syntax 
\[
defid :: \sigma
\]
for example, 
\[
p :: \sigma \times \sigma \to o \qquad f :: \sigma \to \sigma
\]
introduce constants for a binary relation $p$ on type $\sigma$ and a
unary function $f$ on type $\sigma$.  There is no restriction on the
return types of defined symbols.

As in Prolog, programs are defined using Horn clauses $A \ent G$ where
$A$ is an atomic formula and $G$ is a goal formula.  Atomic formulas
include user-defined predicates $p(\vec{t})$ as well as equations
$f(\vec{t}) = u$; in either case $p$ or $f$ must be a defined symbol
of appropriate type, not a constructor.

Goal formulas $G$ can be built up out of atomic formulas $A$,
freshness constraints $a \fresh t$, equations $t \eq u$, conjunctions
$G,G'$, disjunctions $G;G'$, existential quantification $\exists X.G$,
or $\new$-quantification $\new \Aa.G$.  The freshness constraint $a
\fresh t$ holds if the name $a$ does not appear free (that is, outside
an abstraction) in $t$; equality $t \eq u$ between nominal terms is
modulo $\alpha$-renaming of name-abstractions.  For example,
$\abs{\Aa}{(\Aa,\Ab)} \eq \abs{\Ac}{(\Ac,\Ab)} \not\eq
\abs{\Ab}{(\Ab,\Ab)}$.

  \textbf{Polymorphism}.  \aprolog permits type variables in
  declarations, which are treated polymorphically, following previous
  work on polymorphic typing in logic
  programming~\cite{mycroft84ai,hanus91tcs,DBLP:conf/lpar/NadathurQ05}.
  Polymorphic type checking is performed in the standard way by
  generating equational constraints and solving them using
  unification.  As observed by Hanus, handling general polymorphism in
  logic programming may require performing typechecking at run-time.
  To avoid this, the current implementation \aprolog rules out
  ``non-parametric'' polymorphic program clauses that specialize type
  variables, {and requires all datatype constructors to be ``type-preserving''}.  For example, the second clause in
\[\begin{array}{lcl}
head &::& \alpha\times \listTy~\alpha \to o.\\
head(X,X::L).\\
head(1,1::L).
\end{array}\]
works only for $\alpha = int$, not for arbitrary $\alpha$, so is ruled
out.  
{
Similarly, a ``heterogeneous list'' datatype such as
\[hnil :: hlist. \qquad hcons :: \alpha \times hlist \to hlist.\]
is not allowed.  
}

\textbf{Function definitions}.  As in other Prolog-like languages, it is often
convenient to have a notation for writing predicates which are easier
written as functions.  For example, the functional definition
\[\begin{array}{lcl}
append &::& \listTy~\alpha\times\listTy~\alpha \to \listTy~\alpha.\\
append([],M) &=& M.\\
append(X::L,M) &=&  X::append(L,M).
\end{array}\]
can be viewed as an abbreviation for the relational definition
\[\begin{array}{lcl}
appendp&::&\listTy~\alpha\times\listTy~\alpha\times\listTy~\alpha \to o.\\
appendp([],M,M).\\
appendp(X::L,M,X::N) &\ent& appendp(L,M,N).
\end{array}\]
Using {this} notation for functional definitions can
considerably simplify a program. 
{It is
  well-understood how to translate programs that use function notation
  to equivalent purely relational programs, via a translation called
  \emph{flattening}~\cite{hanus94jlp}}.  More sophisticated techniques
such as \emph{narrowing} that have been investigated in functional
logic programming could also be used; however, doing so will require
extending equational unification techniques to nominal logic.

In \aprolog, it turns out to
  be convenient to generalize this notation slightly to permit
  function definition clauses qualified by subgoals or constraints.
  An example is the $subst$ program (discussed in \refEx{subst}), in
  which the declaration
\[
subst(var(Y),E, X) = var (Y) \ent X \fresh Y.
\]
is flattened to the clause
\[
substp(var(Y),E, X,var (Y)) \ent X \fresh Y.
\]

\subsection{The $\lambda$-calculus and variants}\labelSec{lam}

\begin{figure}
\[
\begin{array}{lrcl}
\text{Terms} & e &::=& x \mid \lambda x.e \mid e~e'\\
\text{Types} & \tau &::=& b \mid \tau \to \tau'\\
\text{Contexts} & \Gamma &::=& \cdot \mid \Gamma,x{:}\tau
\end{array}
\quad 
\begin{array}{rcll}
x\{e/x\} &=& e\\
y\{e/x\} &=& y & (x \neq y)\\
(e_1~e_2)\{e/x\} &=& e_1\{e/x\}~e_2\{e/x\}\\
(\lambda y.e')\{e/x\} &=& \lambda y.e'\{e/x\} & (y \not\in FV(x,e))
\end{array}
\]
\[\infer{\wf{\Gamma}{x}{\tau}}{x{:}\tau \in \Gamma}\quad
\infer{\wf{\Gamma}{e~e'}{\tau'}}{\wf{\Gamma}{e}{\tau \to \tau'} & \wf{\Gamma}{e'}{\tau}}\quad
\infer{\wf{\Gamma}{\lambda x.t}{\tau\to\tau'}}{\wf{\Gamma,x{:}\tau}{e}{\tau'} & (x \not\in Dom(\Gamma))}\]
\caption{Lambda-calculus: syntax, substitution, and typing}\labelFig{lambda}
\end{figure}

The prototypical example of a language with variable binding is the
$\lambda$-calculus.  In \aprolog, the syntax of $\lambda$-terms may be
described with the following type and constructor declarations:
\begin{quote}
\begin{tabular}{lll}
$id  : \nametype.$ &$\qquad exp : \type.$\\
$var : id \to exp.$ &$\qquad app : exp \times exp \to exp.$ &$\qquad  lam : \ab{id} exp \to exp.$
\end{tabular}
\end{quote}
{Note that for
  this and other examples in this section, it is important to check
  the correctness of the representation of the object system (often
  called \emph{adequacy}~\cite{Pfenning:HandbookAR:framework:2001}).
  Establishing adequacy requires first understanding the semantics of
  nominal logic programs given in \refSec{semantics}. We will discuss
  adequacy further in \refSec{adequacy}.}

\begin{example}[Typechecking and inference]
  First, for comparison with higher-order encodings, we consider the
  problem of typechecking $\lambda$-terms.  The syntax of types can be
  encoded as follows:
\[\begin{array}{cccc}
tid : \nametype. \qquad ty : \type. \qquad varTy : tid \to ty. \qquad arrTy : ty\times ty \to ty.
\end{array}
\]
We define contexts $ctx$ as lists of pairs of identifiers and types,
and the 3-ary relation {$tc$} relating a context, term, and type:
\[
\begin{array}{@{}lcl@{}}
\type~ctx &=& \listTy~(id\times ty).\\
tc &::& ctx\times exp \times ty \to o. \\
tc({G},var(X), T) &\ent&mem((X,T),{G}).\\
tc({G},app(E_1,E_2),T') &\ent&tc({G},E_1,{arrTy}(T,T')), tc({G},E_2,T).\\
tc({G},lam(\abs{\Ax}{E}),{arrTy}(T,T')) &\ent&\Ax \fresh {G}, tc([(\Ax,T)|{G}],E,T').
\end{array}
\]
The predicate $mem::\alpha \times [\alpha] \to o$ is the usual
predicate for testing list membership ($x:\tau \in\Gamma$).
{ The freshness constraint $\Ax \fresh G$
  expresses the (often implicit) side-condition $x \not\in
  Dom(\Gamma)$.  Note that for simply-typed lambda terms, it is
  immediate that $x \not\in Dom(\Gamma)$ is equivalent to $\Ax \fresh
  G$ whenever $G$ encodes $\Gamma$.  }

Consider the query $\query tc([],lam(\abs{\Ax}{ lam(\abs{\Ay}{var(\Ax)})}),T)$.  We can reduce this goal by backchaining against the
suitably freshened rule
\[tc({G}_1,lam(\abs{\Ax_1}{E_1}), arr(T_1,U_1)) \ent 
\Ax_1 \fresh {G}_1, tc([(\Ax_1,T_1)|{G}_1],E_1,U_1)\]
which unifies with the goal with $[{G}_1 = [], E_1 = lam(\abs{\Ay}{var(\Ax_1)}), T = arr(T_1,U_1)]$.  This yields subgoal 
$\Ax_1 \fresh {G_1}, tc([(\Ax_1,T_1)|G_1],E_1,U_1)$.  The first conjunct is trivially
valid since {${G}_1 = []$} is a constant.  The second is solved by backchaining
against the third 
${tc}$-rule again, producing unifier $[{G}_2 =
[(\Ax_1,T_1)],E_2 = var(\Ax_1), U_1 = arr(T_2,U_2)]$ and subgoal
$\Ax_2 \fresh [(\Ax_1,T_1)] , tc([(\Ax_2,T_2),(\Ax_1,T_1)],
var(\Ax_1),U_2)$.  The freshness subgoal reduces to the constraint
$\Ax_2 \fresh T_1$, and the ${tc}$ subgoal can be solved by
backchaining against
\[tc({G}_3,var(X_3),T_3) \ent mem((X_3,T_3),{G}_3)\]
using unifier $[{G}_3=[(\Ax_2,T_2),(\Ax_1,T_1)], X_3=\Ax_1, T_3=U_2]$.
Finally, the remaining subgoal
$mem((\Ax_1,U_2),[(\Ax_2,T_2),(\Ax_1,T_1)])$ clearly has most general
solution $[U_2 = T_1]$.  Solving for $T$, we have $T = arr(T_1,U_1) =
arr(T_1,arr(T_2,U_2)) = arr(T_1,arr(T_2,T_1))$.  This solution
corresponds to the principal type of $\lambda x. \lambda y. x$.

There are no other possible solutions.
\end{example}

  \begin{example}
    Returning to the example discussed in the introduction, the query
    \[
    \query tc([],lam(\abs{\Ax}{lam(\abs{\Ax}{app(var(\Ax),var(\Ax))})}),T)
    \]
    fails with no solutions in \aprolog.  The following derivation
    steps show why this is the case:
    \[\begin{array}{l}
      \Longrightarrow T \eq arrTy(T_1,T_2), \Ax_1 \fresh [], tc([(\Ax_1,T_1)],lam(\abs{\Ax}{app(var(\Ax),var(\Ax))}),T_2)\\
      \Longrightarrow \cdots T_1 \eq arrTy(T'_1,T'_2),  \Ax_1 \fresh [(\Ax_1,T_1)], tc([(\Ax_2,T'_1),(\Ax_1,T_1)],app(var(\Ax),var(\Ax)),T'_2)\\
      \Longrightarrow \cdots tc([(\Ax_2,T'_1),(\Ax_1,T_1)],var(\Ax_2),arrTy(T_3,T'_2)), \\
\qquad\qquad tc([(\Ax_2,T'_1),(\Ax_1,T_1)],var(\Ax_2),T_3)\\
      \Longrightarrow \cdots T'_1 \eq arrTy(T_3,T'_2)), T'_1 \eq T_3
      \end{array}
    \]
    The final two equations are unsatisfiable (since the occurs check
    will fail), and no other derivation steps are possible.
  \end{example}

\begin{example}[Capture-avoiding substitution]\labelEx{subst}
  Although capture-avoiding substitution is not a built-in operator in
  \aprolog, it is easy to define via the clauses:
\[
\begin{array}{lclcl}
subst &::& exp\times exp \times id \to exp.\\
subst(var(X),E, X)    &=& E.\\
subst(var(Y),E, X)    &=& var (Y) \\&\ent& X \fresh Y.\\
subst(app(E_1,E_2),E,X) &= &app(subst(E_1,E,X),subst(E_2,E,X)).\\
subst(lam(\abs{\Ay}{ E'}),E,X)  &=& lam(\abs{\Ay}{ subst(E',E,X)}) \\&\ent& \Ay \fresh (X,E).
\end{array}
\]
\noindent
Note the two freshness side-conditions: the constraint $X\fresh Y$
prevents the first and second clauses from overlapping; the constraint
$\Ay \fresh (X,E)$ ensures capture-avoidance, by restricting the
application of the fourth clause to when $\Ay$ is fresh for $X$ and
$E$.  Despite these side-conditions, this definition is total and
deterministic.  Determinism is immediate: no two clauses overlap.
Totality follows because, by nominal logic's \emph{freshness
  principle}, the bound name $\Ay$ in $lam (\abs{\Ay}{ E'})$ can always be
renamed to a fresh $\Az$ chosen so that $\Az \fresh (X,E)$.  

Consider the goal $\query X = subst(lam(\abs{\Ax}{ var(\Ay)}), var(\Ax), \Ay)$.
The substitution on the right-hand side is in danger of capturing the
free variable $var(\Ax)$.  How is capture avoided in \aprolog? First,
recall that function definitions are translated to a flattened clausal
form in \aprolog, so we must solve the equivalent goal
\[substp(lam(\abs{\Ax}{ var(\Ay)}), var(\Ax), \Ay,X)\]
subject to an appropriately translated definition of $substp$.  The
freshened, flattened clause
\[substp(lam(\abs{\Ay_1}{E'_1}),E_1,X_1,{lam(\abs{\Ay_1}{E''_1})}) \ent
\Ay_1 \fresh E_1,substp(E'_1,E_1,X_1,E''_1) \] 
unifies with substitution 
\[[E'_1 = var(\Ay), X_1=\Ay,E_1=var(\Ax),X=lam(\abs{\Ay_1}{E''_1})].\]
The freshness constraint $\Ay_1 \fresh var(\Ax)$ guarantees
that $var(\Ax)$ cannot be captured.  It is easily verified, so the
goal reduces to $substp(var(\Ay),var(\Ax),\Ay,E''_1)$.  Using the
freshened rule $substp(var(X_2),E_2, X_2,E_2)$ with unifying
substitution $[X_2 = \Ay, E_2 = var(\Ax),E''_1 = var(\Ax)]$, we obtain
the solution $X= lam(\abs{\Ay_1}{var(\Ax)})$.  

  We can also easily implement \emph{simultaneous} substitution, 
$ssubst$, as follows:
\[\begin{array}{lclcl}
ssubst &::& exp\times \listTy~(exp \times id) \to exp.\\
ssubst(var(X),[])    &=& var (X)\\
ssubst(var(X),[(E,Y)|S])    &=& ssubst(var(X),S) &\ent& X \fresh Y\\
ssubst(var(X),[(E,X)|S])    &=& E\\
ssubst(app(E_1,E_2),S) &= &app(ssubst(E_1,S),ssubst(E_2,S)).\\
ssubst(lam(\abs{\Ax}{ E}),S)  &=& lam(\abs{\Ax}{ subst(E,S)}) &\ent& \Ax \fresh S.
\end{array}
\]
\end{example}

\subsubsection{References}\labelSec{references}

All imperative languages, and some functional languages such as ML,
provide support for ``pointers'' or ``references''.  The semantics of
references typically involves threading some state (a \emph{heap}
$\mu$ mapping memory locations to values) through the evaluation.
When a new reference cell is allocated, a \emph{fresh} location must
be obtained.  Also, when reference is assigned a new value, the heap
must be updated.  Thus, a typical small-step semantics for
references~\cite[Ch. 13]{pierce02types} includes rules for allocating and updating
references, such as
\[\small\begin{array}{c}
  \infer[\mathit{ref}]{\mathsf{ref}~v \mid \mu \too l \mid (\mu,l \mapsto v)}{l \not\in dom(\mu)}
\quad
\infer[\mathit{assn}]{l := v \mid \mu \too () \mid \mu[l:=v]}{}
\quad
\infer[\mathit{deref}]{!l \mid \mu \too v \mid \mu}{\mu(l) = v}
\end{array}
\]
where implicitly $v$ is a value and $l$ is a memory location.

In \aprolog, we can use a name-type $loc$ for memory locations and
implement these rules easily as follows, using auxiliary predicate
$value: exp \to o$ and function $update :: [(loc,exp)] \times loc
\times exp \to [(loc,exp)]$:
\[
\begin{array}{lcl}
\multicolumn{3}{l}{step :: (exp\times [(loc,exp)])\times (exp\times [(loc,exp)]) \to o.}\\
step((ref(V),M),(loc(L),[(L,V)|M])) &\ent& value(V), L \fresh M.\\
step((assn(loc(L),V),M),(unit,update(M,L,V))) &\ent& value(V).\\
step((deref(loc(L),M),(V,M)) &\ent& value(V), mem((L,V),M).
\end{array}
\]

  \subsubsection{Dependent types}\labelSec{dependent-types}

  In the previous section, we considered a simply-typed language, in
  which term variables cannot occur in types.  We can also handle
  dependent types in \aprolog.  The
  dependent function type constructor $\Pi x{:}\tau.\tau'$ typically
  has well-formedness and introduction rules:
  \[ \infer[\text{$\Pi$-formation}]{\Gamma\vdash \Pi
    x{:}\tau.\tau'~\mathsf{type}}{\Gamma \vdash \tau ~\mathsf{type}& \Gamma,x{:}\tau \vdash \tau'~\mathsf{type}}
  \quad \infer[\text{$\Pi$-introduction}]{\Gamma \vdash \lambda x.e :
    \Pi x{:}\tau.\tau'}{\Gamma,x{:}\tau \vdash e:\tau'}
  \]
  As with simple types, the $\Pi$-formation rule carries an implicit
  caveat that $x$ does not already appear in the domain of
  $\Gamma$.  The freshness constraint $\Ax \fresh G$ in the
  following rule is exactly what is needed again here because in a
  well-formed context $\Gamma$, we have $Dom(\Gamma) \supseteq
  \bigcup_{x \in Dom(\Gamma)}FV(\Gamma(x))$.  Hence the following rule
  suffices:
  \[
  wfty(G,piTy(T,\abs{\Ax}{T'})) \ent wfty(G,T), \Ax \fresh G,
  wfty([(\Ax,T)|G],T')
  \]

  Similarly, the $\Pi$-introduction rule has an implicit constraint
  that $x \not\in Dom(\Gamma)$.  If $\Gamma$ is well-formed, then this
  is equivalent to $\Ax \fresh G$; moreover, if $\Gamma \vdash \Pi
  x{:}\tau.\tau'$, then $\Ax \fresh G$ implies that $\Ax \fresh T$
  as well (although $\Ax$ may still occur in $T'$).  So the following
  rule suffices:
  \[
  tc(G,lam(\abs{\Ax}{E}),piTy(T,\abs{\Ax}{T'})) \ent x \fresh G,
  tc([(\Ax,T)|G],E,T')).
  \]

  \subsubsection{Substructural type systems}\labelSec{substructural}
  Substructural type systems (or associated logics) such as linear
  logic~\cite{girard88} or bunched implications~\cite{ohearn99bsl} can
  also be implemented directly in \aprolog.  For example,
  ``multiplicative'' linear logic rules such as
\[
\infer[\text{$\otimes$-introduction}]{\Gamma_1,\Gamma_2 \nd (e_1,e_2): \tau_1 \otimes \tau_2}{\Gamma_1 \nd e_1: \tau_1 & \Gamma_2 \nd e_2: \tau_2}
\]
can be translated to program clauses such as
\[
tc(merge(G_1,G_2),lpair(E_1,E_2),tensorTy(T_1,T_2)) \ent tc(G_1,E_1,T_1), tc(G_2,E_2,T_2).
\]
where we define $merge$ as follows:
\[\begin{array}{lclcl}
merge([],G) &=& G.\\
merge([(X,T)|G],G') &=& [(X,T)|merge(G,G')] &\ent& X \fresh G'.
\end{array}
\]
Note the use of freshness to enforce that the domains of the two
contexts do not overlap; again, the constraint $X \fresh G'$ is
equivalent to $x \not\in Dom(\Gamma')$ for the well-formed contexts in
which we are interested.

Bunched type systems can also be implemented in \aprolog, but we
cannot use lists to represent bunched contexts; instead we have to
define the bunches as a new data type, and define appropriate
operations for splitting and merging contexts.

\subsection{The $\pi$-calculus}\label{sec:pi}

The $\pi$-calculus is a calculus of concurrent, mobile processes.  Its
syntax (following \citeN{milner92ic}) is described by the grammar
rules shown in Figure~\ref{fig:pi}.  The symbols $x, y, \ldots$ are
\emph{channel names}.  The inactive process $0$ is inert.  The
$\tau.p$ process performs a \emph{silent action} $\tau$ and then does
$p$.  Parallel composition is denoted $p|q$ and nondeterministic
choice by $p+q$.  The process $x(y).p$ inputs a channel name from $x$,
binds it to $y$, and then does $p$.  The process $\bar{x}y.p$ outputs
$y$ to $x$ and then does $p$.  The match operator $[x=y]p$ is $p$
provided $x = y$, but is inactive if $x \neq y$.  {The mismatch
  operator $[x \neq y]p$, in contrast, is $p$ provided $x$ and $y$
  differ, and inactive otherwise.} The restriction operator $(y)p$
restricts $y$ to $p$.  Parenthesized names (e.g. $y$ in $x(y).p$ and
$(y)p$) are binding, and $fn(p)$, $bn(p)$ and $n(p)$ denote the sets
of free, bound, and all names occurring in $p$.  Capture-avoiding
renaming is written $t\{x/y\}$.

\begin{figure}[tb]
\[
\begin{array}{lrcl}
\text{Process terms}&p &::= & 
0 \mid\tau.p \mid p|q \mid p+q  \mid x(y).p \mid \bar{x}y.p  
\mid [x = y]p\mid [x \neq y]p\mid (x)p\\
\text{Actions}&
a &::=& \tau \mid x(y) \mid \bar{x}y \mid \bar{x}(y)
\end{array}\]

\[\begin{array}{lcl}
chan      &:& \nametype.\\
proc &:& \type.           \\ 
ina       &:& proc.                  \\    
tau  &:& proc \to proc.\\
par   &:& proc\times proc \to proc.   \\   
sum   &:& proc\times proc \to proc.   \\   
in   &:& chan\times \ab{chan} proc \to proc.  \\
out &:& chan\times chan\times proc \to proc.\\  
match &:& chan\times chan\times proc \to proc.\\  
mismatch &:& chan\times chan\times proc \to proc.\\  
res  &:& \ab{chan}proc \to proc.\\
\end{array}\quad
\begin{array}{lcl}
act      &:& \type.\\           
tau\_a &:& act.     \\
in\_a&:& chan\times chan \to act. \\
fout\_a&:& chan\times chan \to act. \\
bout\_a  &:& chan\times chan \to act.
\end{array}\]
\caption{The $\pi$-calculus: syntax and \aprolog declarations}\labelFig{pi}
\end{figure}

\begin{figure}[tb]
\[\begin{array}{c}
\tau.p\lto{\tau} p 
\quad
\infer{p|q \lto{a} p'|q}
      {p \lto{a} p' & bn(a) \cap fn(q) = \emptyset}
\quad
\infer{p|q \lto{\tau} p'|q'\{y/z\}}
      {p \lto{\bar{x}y} p' & q \lto{x(z)} q'}
\smallskip\\
\infer{p+q\lto{a} p'}
      {p \lto{a} p'}
\quad
\bar{x}y.p \lto{\bar{x}y} p
\quad\infer{x(z).p \lto{x(w)} p\{w/z\}}
{w \notin fn((z)p)}
\quad
\infer{[x = x]p \lto{a} p'}{p \lto{a}p'}
\quad
\infer{[x \neq y]p \lto{a} p'}{(x \neq y) & p \lto{a}p'}
\smallskip\\
\infer{p|q \lto{\tau} (w)(p'|q')}
      {p \lto{\bar{x}(w)} p' & q \lto{x(w)} q'}
\quad
\infer{(y)p\lto{a} (y)p'}
      {p \lto{a} p' & y \notin n(a)}
\quad\infer{(y)p\lto{\bar{x}(w)} p'\{w/y\}}
      {p \lto{\bar{x}y} p' & y \neq x & w \notin fn((y)p)}

\end{array}
\]\caption{{$\pi$-calculus transitions}}\labelFig{pitrans}
\end{figure}
\begin{figure}
\[
\begin{array}{l}
\begin{array}{lcl}
ren\_p &::& proc\times chan\times chan \to proc.~\text{(* definition omitted *)}\\
safe &::& act\times pr \to o.~\text{(* tests $bn(A) \cap fn(P) = \emptyset$ *)}\\
safe(tau\_a,P).\\
safe(fout\_a(X,Y),P).\\
safe(bout\_a(X,Y),P) &\ent& Y \fresh P.\\
safe(in\_a(X,Y),P)   &\ent& Y \fresh P.
\end{array}\\
\begin{array}{lcl}
step &::& pr\times act \times pr \to o.~\text{(* encodes $p \lto{a} p'$ *)}\\
step(tau(P),tau\_a,P).\\
step(par(P,Q),A,par(P',Q))                 &\ent& step(P,A,P'), safe(A,Q).\\
step(par(P,Q),tau\_a,par(P',ren\_p(Q',Y,Z))) &\ent& step(P,fout\_a(X,Y),P'),\\
                                              &&step(Q,in\_a(X,Z),Q').\\
step(sum(P,Q),A,P')                        &\ent& step(P,A,P').\\
step(out(X,Y,P),fout\_a(X,Y), P).\\
step(in(X,\abs{\Az}{P}), in\_a(X,W),ren\_p(P,W,\Az))    &\ent& W \fresh \abs{\Az}{P}. \\
step(match(X,X,P),A,P')                    &\ent& step(P,A,P').\\
step({mis}match(X,Y,P),A,P')                    &\ent& X \fresh Y, step(P,A,P').\\
step(par(P,Q),tau\_a,res(\abs{\Az}{par(P',Q')}))     &\ent& step(P,bout\_a(X,\Az),P'), \\
                                              &&step(Q,in\_a(X,\Az),Q').\\
step(res(\abs{\Ay}{P}),A,res(\abs{\Ay}{P'}))                 &\ent& \Ay \fresh A, step(P,A,P').\\
step(res(\abs{\Ay}{P}),bout\_a(X,W),ren\_p(P',W,\Ay))   &\ent& step(P,fout\_a(X,\Ay),P'),\Ay \fresh X,\\
                                              && W \fresh \abs{\Ay}{ P}.
\end{array}
\end{array}
\]
\caption{{\aprolog implementation of the $\pi$-calculus}}\label{fig:picode}
\end{figure}

\citeN{milner92ic}'s original operational semantics (shown in
{\refFig{pitrans}}, symmetric cases omitted) is a labeled transition
system with relation $p \lto{a} q$ indicating ``$p$ steps to $q$ by
performing action $a$''.  Actions $\tau$, $\bar{x}y$, $x(y)$,
$\bar{x}(y)$ are referred to as \emph{silent}, \emph{free output},
\emph{input}, and \emph{bound output} actions respectively; the first
two are called \emph{free} and the second two are called \emph{bound}
actions.  For an action $a$, $n(a)$ is the set of all names appearing
in $a$, and $bn(a)$ is empty if $a$ is a free action and is $\{y\}$ if
$a$ is a bound action $x(y)$ or $\bar{x}(y)$.  Processes and actions
can be encoded using the declarations shown in \refFig{pi}.

Much of the complexity of the rules is due to the need to handle
\emph{scope extrusion}, which occurs when restricted names ``escape''
their scope because of communication.  In $((x) \bar{a}x.p) |
(a(z).z(x).0) \lto{\tau} (x') (p | x'(x).0))$, for example, it is
necessary to ``freshen'' $x$ to $x'$ in order to avoid capturing the
free $x$ in $a(z).z(x).0$.  Bound output actions are used to lift the
scope of an escaping name out to the point where it is received.  The
rules can be translated directly into \aprolog (see
Figure~\ref{fig:picode}).  The function $ren\_p(P,Y,X)$ performing
capture-avoiding renaming is not shown, but easy to define.

We can check that this implementation of the operational semantics
produces correct answers for the following queries:
\[\begin{array}{l}
\query step(res(\abs{\Ax}{ par(res(\abs{\Ay}{ out(\Ax,\Ay,ina)}),in(\Ax,\abs{\Az}{out(\Az,\Ax,ina)}))}),A,P).\\
 A = tau\_a,  P = res (\abs{\Ay_{58}}{res(\abs {\Az_{643}}{par(ina,out(\Az_{643},\Ay_{58},ina))})}) \\
 \query step(res(\abs{\Ax}{ out(\Ax,\Ay,ina)}),A,P).\\
 No.
\end{array}\]
This \aprolog session shows that $(x)((y)\bar{x}y.0\mid
x(y).\bar{y}x.0) \lto{\tau} (x)(y)(0\mid\overline{y}x.0)$, but
$(x)(x(y).0)$ cannot make any transition.  Moreover, the answer to the
first query is unique (up to renaming).

\citeN{rockl01merlin} and \citeN{gabbay03automath} have
also considered encodings of the $\pi$-calculus using nominal abstract
syntax.  R\"ockl considered only modeling the syntax of terms up to
$\alpha$-equivalence using swapping, whereas Gabbay went further,
encoding transitions and the bisimulation relation and proving basic
properties thereof.  By \cite[Thm 4.5]{gabbay03automath}, Gabbay's
version of the $\pi$-calculus is equivalent to our conventional
representation.  In fact, Gabbay's presentation is a bit simpler to
express in \aprolog, but we have chosen \citeN{milner92ic}'s original
presentation to emphasize that informal ``paper'' presentations (even
for fairly complicated calculi) can be translated directly to \aprolog
programs.

\subsubsection{Dyadic $\pi$-calculus}

The polyadic $\pi$-calculus adds to the $\pi$-calculus the ability to
send and receive $n$-tuples of names, not just single names.  It is a
useful intermediate stage for translations form other languages (such
as the $\lambda$-calculus, object calculi, or the ambient calculus) to
the pure $\pi$-calculus.  We can easily define a special case of dyadic
$\pi$-terms (that can send and receive pairs of names) in \aprolog:
\[\begin{array}{lcl}
in2   &:& chan\times \abs{chan}{\abs{chan}{proc}} \to proc.  \\
out2 &:& chan\times chan \times chan \times proc \to proc.\\  
\\
unpoly &::& proc \to proc.\\
unpoly(out2(C,X,Y,P)) &=& res(\abs{\Az}{out(C,\Az,out(\Az,X,out(\Az,Y,unpoly(P))))}) \\
&\ent& \Az \fresh (C,X,Y,P).
\\
unpoly(in2(C,\abs{\Ax}{\abs{\Ay}{P}})) &=& in(C,\abs{\Az}{in(\Az,\abs{\Ax}{in(\Az,\abs{\Ay}{unpoly(P)})})})\\
&\ent& \Az \fresh (C,P).
\end{array}\]

\subsubsection{Translation from $\lambda$-calculus to $\pi$-calculus}

Both call-by-value and call-by-name translations from the
$\lambda$-calculus to (dyadic) $\pi$-calculus can be developed.  We
assume that the $\lambda$-calculus variables and $\pi$-calculus names
coincide.
\[\begin{array}{lclll}
cbv&::&\multicolumn{3}{l}{exp\times chan \to proc.}\\
cbv(var(X),P) &=& out(P,X).\\
cbv(app(M,N),P) &=& res(\abs{\Aq}{par(&cbv(M,\Aq),\\
&&&in(\Aq,\abs{\Av}{res(\abs{\Ar}{par(&cbv(N,\Ar), \\
&&&&in(\Ar,\abs{\Aw}{out2(\Av,\Aw,P,ina)}))})}))})\\
cbv(lam(\abs{\Ax}{M}),P) &=& \multicolumn{3}{l}{res(\abs{\Ay}{out(P,\Ay,rep(in2(\Ay,\abs{\Ax}{\abs{\Aq}{cbv(M,\Aq)}})))}).}
\end{array}\]

This can be seen to be equivalent to an informal definition (paraphrasing
\cite[Table 15.2]{sangiorgi01pi}):
\[
\begin{array}{rcl}
\VV\SB{x}p &=& \overline{p}x\\
\VV\SB{M~N}p &=& (q) \left(\VV\SB{M}q \mid q(v).(r)(\VV\SB{N}r \mid r(w).\overline{v}\ab{w,p})\right)\\
\VV\SB{\lambda x.M}p &=& \overline{p}(y).!y(x,q).\VV\SB{M}q
\end{array}
\]

\subsection{Discussion}\labelSec{discussion}

We conclude our high-level exposition by discussing \aprolog in the
context of other logic programming systems.  As reflected by our
choice of examples, at present we view \aprolog as rather narrowly
focused on the domain of prototyping and experimenting with logics,
operational semantics for programming and concurrency calculi, and
type systems and other program analyses.  We believe that this is a
rich domain containing certain classes of problems for which
\aprolog's uniform and declarative treatment of names, binding and
generativity is an especially good fit (although not providing support
for substitution and contexts comparable to that offered by
higher-order abstract syntax).  At present, our prototype interpreter
aims to support rapid prototyping and experimentation with such
systems, not general-purpose programming, just as several constraint
logic programming languages are oriented towards particular domains.

Nevertheless, it is an interesting question whether nominal logic
programming features are advantageous in general-purpose logic
programming.  We believe the case for this is presently ambiguous, at
best.  Even for highly symbolic programs such as compilers and theorem
provers, programmers typically rely on direct access to variable names
for some operations (e.g., printing out informative error messages);
moreover, names sometimes have additional structure (as in module
systems).  Thus, significant changes to such programs may be
necessary to accommodate nominal logic's abstract treatment of names.
In particular, nominal logic's \emph{equivariance}
principle~\cite{pitts03ic} guarantees that \emph{there is no linear
  ordering on names} (from the point of view of nominal logic).  This
means that efficient data structures indexed by names (such as symbol
tables) can hardly be implemented directly as nominal logic programs
and would instead have to be provided as built-in operations.

On the other hand, other systems such as
\lprolog~\cite{DBLP:conf/cade/NadathurM99},
Qu-Prolog~\cite{staples89meta} and FreshML~\cite{shinwell03icfp} have
demonstrated that support for name-binding is useful as a
general-purpose programming feature even if access to names is limited
(as in \lprolog or FreshML).  Besides obvious applications to symbolic
programming, names have been used in Qu-Prolog in multithreading and
message passing.  Moreover, \citeN{pitts07popl} have shown that
certain functions and relations on names (including linear orderings)
can be added to FreshML without damaging its semantics.  Such results
have yet to be extended or specialized to nominal logic proper,
however, and this is an area for future work.

The only way to find out how well nominal techniques work in general
logic programming is to try to use them to develop significant
programs.  This appears to first require developing a
production-quality compiler or interpreter for \aprolog, together with
libraries and other programming support.  As our semantics
(\refSec{semantics}) demonstrates, nominal logic programming can be
viewed as ``constraint logic programming over the domain of nominal
terms''; thus, it may be possible to add support for some features of
nominal logic programming to an existing mature CLP system simply by
implementing it as an additional constraint domain.  However, we leave
this and other implementation concerns for future work.



%% file: background.tex
\section{Nominal Logic, Herbrand Models, and Logic Programs}\labelSec{background}

\subsection{Syntax}

\begin{figure}[tb]
\begin{boxedminipage}{\textwidth}
\[\begin{array}{lrcl}
\text{(Types)}&  \sigma &::=& \nu \mid \delta \mid \abs{\nu}{\sigma}\\
\text{(Contexts)}&  \Sigma &::=& \cdot \mid \Sigma,X{:}\sigma \mid \Sigma\#\Aa{:}\nu\\
\text{(Terms)}&  t &::=& \Aa \mid c \mid f(\vec{t}) \mid {X} \mid \swap{a}{b}{t} \mid \abs{a}{t}\\
\text{(Formulas)}&  \phi &::=& \true \mid \false \mid p(\vec{t}) \mid t \eq u \mid a \fresh t \mid t \ev u\\
&&\mid& \phi \impp \psi \mid  \phi \andd \psi  \mid \phi \orr \psi  \\
&&\mid& \forall X{:}\sigma.\phi \mid \exists X{:}\sigma.\phi \mid \new \Aa{:}\nu.\phi
\end{array}\]
\end{boxedminipage}
\caption{Syntax of nominal logic}\labelFig{nl}
\end{figure}

\begin{figure}
\begin{boxedminipage}{\textwidth}
\[\begin{array}{c}
\infer{\Sigma \nd \Aa:\nu}{\Aa:\nu \in\Sigma}\quad
\infer{\Sigma \nd x:\sigma}{x:\sigma\in\Sigma}\quad
\infer{\Sigma \nd c:\delta}{c:\delta\in\LL}\quad
\infer{\Sigma \nd f(\vec{t}):\delta}{f:\vec{\sigma}\to\delta\in\LL & \Sigma \nd \vec{t}:\vec{\sigma}}
\quad
\infer{\Sigma \nd \abs{a}{t}:\abs{\nu}{\sigma}}{\Sigma\nd a : \nu & \Sigma \nd t : \sigma}
\smallskip\\
\infer{\Sigma \nd \swap{a}{b}{t}:{\sigma}}{\Sigma\nd a : \nu &\Sigma\nd b : \nu & \Sigma \nd t : \sigma}\quad
\infer{\Sigma \nd t \eq u, t \ev u: o}{\Sigma \nd t, u:\sigma}
\quad
\infer{\Sigma \nd a \fresh t: o}{\Sigma \nd a:\nu & \Sigma \nd t:\sigma}\quad
\smallskip\\
\infer{\Sigma \nd \true, \false: o}{}
\quad
\infer{\Sigma \nd \phi \andd \psi, \phi \orr \psi, \phi \impp \psi:o}{\Sigma \nd \phi,\psi :o}
\quad
\infer{\Sigma \nd \forall X{:}\sigma.\phi,\exists X{:}\sigma.\phi:o}{\Sigma,X{:}\sigma \nd \phi :o}
\quad
\infer{\Sigma \nd \new\Aa{:}\nu.\phi:o}{\Sigma\#\Aa{:}\nu \nd \phi :o}
\end{array}\]
\end{boxedminipage}\caption{Well-formedness for nominal terms and formulas}\labelFig{termwf}
\end{figure}

The syntax of nominal logic is shown in \refFig{nl}.  We assume fixed
{countably infinite} sets of \emph{variables} $\VV$ and \emph{names}
$\BA$.  A \emph{language} $\LL$ consists of a set of \emph{data types}
$\delta$, \emph{name types} $\nu$, \emph{constants} $c:\delta$,
\emph{function symbols} $f:\vec{\sigma} \to \delta$, and
\emph{relation symbols} $p:\vec{\sigma}\to o$, where we write $o$ for
the type of propositions.  Types $\sigma$ also include abstraction
types $\abs{\nu}{\sigma}$; additional type constructors such as
pairing are omitted to simplify the presentation.  First-class
function types are not included, although the declarations of function
and relation symbols in $\LL$ employ suggestive notation.
The novel term constructors include names $\Aa
\in \BA$, name-abstractions $\abs{a}{t}$ denoting $\alpha$-equivalence
classes, and name-swapping applications $\swap{a}{b}{t}$.  
  The formulas of nominal logic include all connectives and
  quantifiers of (sorted) first-order logic with equality; additional
  formulas include freshness ($a \fresh t$) and the $\new$-quantified
  formulas ($\new \Aa{:}\nu.\phi$).  Quantification over types
  mentioning $o$ is not allowed.  Well-formedness is defined for terms
  and formulas in \refFig{termwf}.  {Most cases are standard;
    note that $\new$-quantified names are added to the context using
    the $\Sigma\#\Aa{:}\nu$ context form.}

  Context bindings include ordinary variable bindings
  $\Sigma,X{:}\sigma$ and name bindings $\Sigma\#\Aa{:}\nu$.  As
  usual, we adopt the convention that names and variables are not
  repeated in a context, so that it is impossible to write
  $X{:}\sigma,X{:}\sigma'$ or $\Aa{:}\nu\#\Aa{:}\nu'$.  This
  convention implicitly constrains many inference rules with the side
  condition that $X$ or $\Aa$ does not appear in some context
  $\Sigma$.  We sometimes use the notations $\forall \Sigma[\phi]$, 
  $\exists \Sigma[\phi]$, and $\Sigma,\Sigma'$, defined as follows:
\[\begin{array}{rcl}
\forall \cdot[\phi] &=& \phi\\ 
\forall \Sigma,X{:}\sigma[\phi] &=& \forall \Sigma[\forall X{:}\sigma.\phi]\\
\forall \Sigma,\Aa{:}\nu[\phi] &=& \forall \Sigma[\new {:}\nu.\phi]
\end{array}
\begin{array}{rcl}
\exists \cdot[\phi] &=& \phi\\ 
\exists \Sigma,X{:}\sigma[\phi] &=& \exists \Sigma[\exists X{:}\sigma.\phi]\\
\exists \Sigma,\Aa{:}\nu[\phi] &=& \exists \Sigma[\new {:}\nu.\phi]
\end{array}
\begin{array}{rcl}
\Sigma,\cdot &=& \Sigma\\
\Sigma,(\Sigma',X{:}\sigma) &=& (\Sigma,\Sigma'),X{:}\sigma\\
\Sigma,(\Sigma',\Aa{:}\nu) &=& (\Sigma,\Sigma'),\Aa{:}\nu
\end{array}\]

  Contexts play two roles in our presentation of nominal logic
  (following~\citeN{cheney05fossacs}).  First, as usual they track the
  types of scoped variables as well as those of names introduced by
  $\new$-quantifiers.  Abusing notation, we sometimes identify a
  context with the corresponding set of bindings, and write $\Aa{:}\nu
  \in \Sigma$ or $X{:}\sigma \in \Sigma$ to indicate that a name $\Aa$
  has type $\nu$ or variable $X$ has type $\sigma$ in $\Sigma$.
  Second, contexts track freshness information.  In nominal logic, a
  name introduced by the $\new$-quantifier can always be assumed fresh
  for all other values in scope; thus, contexts need to track the
  order in which names and variables were introduced.  This is the
  reason why we write name-bindings as $\Sigma\#\Aa{:}\nu$.

\subsection{Semantics}

\begin{figure}[tb]
\begin{boxedminipage}{\textwidth}
\[
\begin{array}{rcl}
  \swap{\Aa}{\Ab}{\Aa} &=& \Ab\\
  \swap{\Aa}{\Ab}{\Ab} &=& \Aa\\
  \swap{\Aa}{\Ab}{\Aa'} &=& \Aa' \quad(\Aa \neq \Aa' \neq \Ab)
\end{array}
\qquad
\begin{array}{rcl}
  \swap{\Aa}{\Ab}{c} &=& c\\
  \swap{\Aa}{\Ab}{f(\vec{t})} &=& f(\swap{\Aa}{\Ab}{\vec{t}})\\
  \swap{\Aa}{\Ab}{\abs{\Aa'}{t}} &=& \abs{\swap{\Aa}{\Ab}{\Aa'}}{\swap{\Aa}{\Ab}{t}}
\end{array}
\]
\[\begin{array}{c}
\infer{\models \Aa \fresh \Ab}{(\Aa \neq \Ab)}\quad
\infer{\models\Aa \fresh c}{}\quad
\infer{\models\Aa \fresh f(t_1^n)}{\Andd _{i=1}^n\models\Aa \fresh t_i }\quad
\infer{\models\Aa \fresh \abs{\Ab}{t}}{\models\Aa \fresh \Ab & \models\Aa \fresh t}\quad
\infer{\models\Aa \fresh \abs{\Aa}{t}}{}
\end{array}\]
\[\begin{array}{c}
\hyp{\models\Aa \eq \Aa}\quad
\hyp{\models c \eq  c}\quad
\infer{\models f(t_1^n) \eq f(u_1^n)}{\Andd _{i=1}^n\models t_i \eq u_i}\quad
\infer{\models\abs{\Aa}{t} \eq \abs{\Aa}{u}}{\models t \eq u }\quad
\infer{\models\abs{\Aa}{t} \eq \abs{\Ab}{u}}{\models\Aa \fresh u & \models t \eq \swap{\Aa}{\Ab}{u}}
\end{array}\]
\[\begin{array}{c}
\infer{t \ev u}{t \eq u}
\qquad 
\infer{t \ev u}{\swap{\Aa}{\Ab}{t} \ev u}
\end{array}\]
\end{boxedminipage}
\caption{Swapping, freshness, equality, and equivariance for ground nominal terms}\labelFig{termops}
\end{figure}

\refFig{termops} defines the {meaning of
  the} swapping, freshness, equality, {and equivariance}
operations on ground terms.  Swapping exchanges two syntactic
occurrences of a name in a term (including occurrences such as $\Aa$
in $\abs{\Aa}{t}$.)  The freshness relation defines what it means for
a name to be ``not free in'' (or \emph{fresh} for) a term.
Intuitively, a name $\Aa$ is fresh for a term $t$ (that is, $\Aa
\fresh t$) if $t$ possesses no occurrences of $\Aa$ unenclosed by an
abstraction of $\Aa$.  {The equality relation on nominal terms
  is defined using freshness and swapping.  }  The
only interesting cases are for abstractions; the second rule for
abstractions is equivalent to more standard forms of
$\alpha$-renaming, as has been shown
elsewhere~\cite{gabbay02fac,pitts03ic}.   {Finally, the equivariance
  relation $t \ev u$ indicates that two terms are equal up to a
  permutation of names; it is needed for nominal resolution.}

We sometimes refer to the set of ``free'' names of a term $supp(t)
=\BA - \{\Aa \mid \Aa \fresh t\}$ as its \emph{support}.  Also,
swapping and support are extended to formulas by setting
$\swap{\Aa}{\Ab}{Q X.\phi[X]} = QX.\swap{\Aa}{\Ab}{\phi}[X]$ for $Q
\in \{\forall,\exists\}$ and $\swap{\Aa}{\Ab}{\new \Aa'.\phi} = \new
\Aa'.\swap{\Aa}{\Ab}{\phi}$, provided $\Aa' \not\in \{\Aa,\Ab\}$;
thus, using $\alpha$-renaming, we have {$\swap{\Aa}{\Ab}{\forall X.\new
  \Aa.p(\Aa,\Ab,X)} = \forall X.  \new \Aa'.p(\Aa',\Aa,X)$}.  Likewise,
swapping can be extended to sets of terms or formulas by setting
$\swap{\Aa}{\Ab}{S} = \{\swap{\Aa}{\Ab}{t}\mid t \in S\}$.

\begin{figure}[tb]
\begin{boxedminipage}{\textwidth}
\[\begin{array}{lcl}
\HH \models \true\\
\HH \not\models \false\\
\HH \models A&\iff&A \in \HH\\
\HH \models t \eq u&\iff&\models t \eq u\\
\HH \models a \fresh u&\iff&\models a \fresh u\\
\HH \models \phi \andd \psi &\iff& \text{$\HH \models \phi$
  and $\HH \models \psi$}
\end{array}
\begin{array}{lcl}
  \HH \models \phi \orr \psi&\iff&\text{$\HH \models \phi$
    or $\HH \models \psi$}\\
  \HH \models\phi \impp \psi&\iff&\text{$\HH \models \phi$
    implies $\HH \models \psi$}\\
  \HH \models \forall X{:}\sigma.\phi &\iff& \text{for all $t : \sigma$, $\HH \models \phi[t/X]$}\\
  \HH \models \exists X{:}\sigma.\phi &\iff& \text{for some $t : \sigma$, $\HH \models \phi[t/X]$}\\
  \HH \models \new \Aa{:}\nu.\phi&\iff& \text{for all $\Ab:\nu \not\in \supp(\new \Aa{:}\nu.\phi)$,}\\
  &&\HH
  \models \swap{\Ab}{\Aa}\phi.
\end{array}
\]
\end{boxedminipage}
\caption{Term model semantics of nominal logic}\labelFig{sem}
\end{figure}

For the purposes of this paper, it suffices to restrict attention to
\emph{term models} of nominal logic in which the domain elements are
nominal terms with equality and freshness defined as in
\refFig{termops}.  We write $B_\LL$ for the \emph{Herbrand base}, that
is, the set of all ground instances of user-defined predicates $p$.

We view {a Herbrand} model $\HH$ as a subset
of $B_\LL$ that is \emph{equivariant}, or closed under swapping (that
is, $ \HH\subseteq \swap{\Aa}{\Ab}{\HH}$ for any $\Aa,\Ab$.)  The
semantics of nominal logic formulas over term models is defined as
shown in \refFig{sem}.  The only nonstandard case is that for $\new$.
{The meaning of the $\new$-quantifier can be defined in several
  equivalent ways:}

\begin{lemma}\labelLem{some-any-new}
  The following are equivalent:
\begin{enumerate}
\item $\HH \models \new \Aa{:}\nu.\phi$, that is, $\HH \models
  \swap{\Aa}{\Ab}{\phi}$ for every $\Ab \not\in supp(\new \Aa{:}\nu.\phi)$
\item $\HH \models \phi$ 
\item The set $\{\Ab \mid \HH \models \swap{\Aa}{\Ab}\phi\}$ is
  cofinite
\item $\HH \models \swap{\Aa}{\Ab}{\phi}$ for some $\Ab \not\in supp(\new \Aa{:}\nu.\phi)$
\end{enumerate}
\end{lemma}
\begin{proof}
  It is immediate that (1) implies (2,3) and that (2) implies (4).
  Case (3) implies (4) because the sets $\{\Ab \mid \HH \models
  \swap{\Aa}{\Ab}\phi\}$ $\{\Ab \mid \Ab\not\in supp(\new \Aa.\phi)\}$
  are both cofinite so have nonempty intersection.  Case (4) is
  equivalent to (1) because
\[
\exists a. a \fresh \vec{x} \andd \phi(a,\vec{x}) \iff \forall a. a \fresh \vec{x} \impp \phi(a,\vec{x})
\]
is a theorem of nominal logic for any $\phi$ such that $FV(\phi) \subseteq \{a,\vec{x}\}$~\cite[Prop. 4]{pitts03ic}.
\end{proof}
\begin{remark}
  In light of \refLem{some-any-new}, we could instead have defined
  $\HH \models \new \Aa.\phi$ in several alternative ways, such as
  (2).  However, definition (1) is preferable for the subsequent
  developments because it corresponds closely to a natural ``one-step
  deduction operator'' on Herbrand models for $\new$-quantified
  formulas; see \refDef{one-step-deduction}.
\end{remark}

We define ground substitutions $\theta$ as functions from $\VV$ to
ground terms.  Given a context $\Sigma$, we say that a ground
substitution $\theta$ \emph{satisfies $\Sigma$} (written $\theta :
\Sigma$) when
\[\infer{\cdot : \cdot}{} \quad \infer{\theta,X \mapsto v:\Sigma,X{:}\sigma}{\theta(X) : \sigma & \theta : \Sigma} \quad \infer{\theta :\Sigma\#\Aa{:}\nu}{\Aa \fresh \theta & \theta : \Sigma}\]
where $\Aa \fresh \theta$ abbreviates $\Aa \fresh \theta(X)$ for each
$X \in Dom(\theta)$.  For example, $[X \mapsto \Aa]$ satisfies $\Sigma
= X{:}\nu$ and $\Sigma' = \Aa{:}\nu,X{:}\nu$, but not $\Sigma'' =
X{:}\nu\#\Aa{:}\nu$.  Since contexts grow to the right, we should read
a (sub-)context $\Sigma\#\Aa{:}\nu$ as saying that $\Aa$ is fresh for
all the names in $\Sigma$ and for (the values of) all variables in
$\Sigma$; but, $\Aa$ may appear in variables occurring to the right of
$\Aa$ (that is, values introduced after $\Aa$ was introduced).

We generalize the satisfiability judgments as follows.  Given sets of
formulas $\Gamma,\Delta$, we write
\begin{itemize}
\item $\HH \models \Gamma$ (for $\Gamma$ closed) to
indicate that $\HH \models \phi$ for each $\phi \in \Gamma$
\item $\Gamma \models \Delta$ (for $\Gamma,\Delta$ closed) to indicate
  that for every $\HH$, $\HH \models \Gamma$ implies $\HH \models \Delta$
\item $\sat{\Sigma}{\theta}{\Delta}$ to indicate
  that   $\theta:\Sigma$  and $\models \theta(\Delta)$ 
\item $\sat{\Sigma}{\Gamma,\theta}{ \Delta}$ to indicate that
  $\theta:\Sigma$ and $\theta(\Gamma) \models \theta(\Delta)$
\item $\sat{\Sigma}{\Gamma}{\Delta}$ to indicate that
  $\sat{\Sigma}{\Gamma,\theta}{\Delta}$ for every $\theta:\Sigma$
\item $\forall \Sigma[\phi]$ (or $\exists \Sigma[\phi]$) for the
  formula obtained by $\forall$-quantifying (or $\exists$-quantifying)
  all variables and $\new$-quantifying all names in $\Sigma$, in
  order.
\end{itemize}
Note that, for example, $X \#\Aa:\cdot \models \Aa \fresh X$ but
$\Aa,X:\cdot \not\models \Aa \fresh X$.

We enumerate a number of basic properties of satisfiability, most of
which are standard.
\begin{lemma}\labelLem{deduction}
  If $\sat{\Sigma}{\Gamma}{\phi}$ and
  $\sat{\Sigma}{\Gamma,\phi}{\psi}$ then $\sat{\Sigma}{\Gamma}{\psi}$.
\end{lemma}
\begin{lemma}\labelLem{exists-sat}
  If $\sat{\Sigma}{\Gamma}{\exists X{:}\sigma.\psi}$ and
  $\sat{\Sigma,X{:}\sigma}{\Gamma,\psi}{\phi}$ hold then $\sat{\Sigma}{\Gamma}{\exists X{:}\sigma.\phi}$ holds.
\end{lemma}
\begin{lemma}\labelLem{new-sat}
  If $\sat{\Sigma}{\Gamma}{\new \Aa{:}\nu.\psi}$ and
  $\sat{\Sigma\#\Aa{:}\nu}{\Gamma,\psi}{\phi}$ hold then $\sat{\Sigma}{\Gamma}{\new \Aa{:}\nu.\phi}$ holds.
\end{lemma}
\begin{lemma}\labelLem{and-left}
If $\sat{\Sigma}{\Gamma,\psi_i}{\phi}$ then $\sat{\Sigma}{\Gamma,\psi_1 \andd \psi_2}{\phi}$.
\end{lemma}
\begin{lemma}\labelLem{imp-left}
If $\sat{\Sigma}{\Gamma}{\psi_1}$ and $\sat{\Sigma}{\Gamma,\psi_2}{\phi}$ then $\sat{\Sigma}{\Gamma,\psi_1 \impp \psi_2}{\phi}$.
\end{lemma}
\begin{lemma}\labelLem{all-left}
  If $\sat{\Sigma,X{:}\sigma}{\Gamma,\psi,\theta[X \mapsto t]}{\phi}$ where
  $\wf{\Sigma}{t}{\sigma}$ and $X \not\in FV(\Gamma,\phi)$
  then $\sat{\Sigma}{\Gamma,\forall X{:}\sigma.\psi,\theta}{\phi}$.
\end{lemma}
\begin{lemma}\labelLem{new-left}
  If $\sat{\Sigma\#\Aa{:}\nu}{\Gamma,\psi}{\phi}$ for some $\Aa \not\in \supp(\Gamma,\phi)$ then $\sat{\Sigma}{\Gamma,\new \Aa{:}\nu.\psi}{\phi}$.
\end{lemma}

\subsection{Nominal logic programs}\labelSec{nomlogic-programs}

  In \refSec{examples}, we employed a concrete syntax for \aprolog
  programs that is more convenient for writing programs, but less
  convenient for defining the semantics and reasoning about programs.
  We take the view that \aprolog programs are interpreted as theories
  in nominal logic, just as pure Prolog programs can be viewed as
  theories of first-order logic.  Consequently, we will now adopt an
  abstract syntax for \aprolog programs that is based on the syntax of
  nominal logic.

\begin{figure}[tb]
\begin{boxedminipage}{\textwidth}
\[\begin{array}{lrcl}
\text{(Constraints)}&  C &::=&  \true \mid t \eq u \mid  a \fresh t  \mid t \ev u \mid C \andd C' \mid \exists X{:}\sigma.C \mid \new \Aa{:}\sigma.C\\
\text{(Goals)}&G &::=& \true \mid A \mid C  \mid G \andd G' \mid G \orr G' \mid \exists X{:}\sigma.G \mid \new \Aa{:}\sigma.G \\
\text{(Program clauses)}&D &::=& \true \mid A \mid D \andd D' \mid G \impp D \mid \forall X{:}\sigma.D \mid \new \Aa{:}\sigma.D 
\end{array}\]
\end{boxedminipage}\caption{Constraints, goals, and program clauses}\labelFig{constr-goal-program}
\end{figure}

  \refFig{constr-goal-program} displays three special classes of
  formulas used frequently in the rest of the article.
  \emph{Constraints} $C$ consist of formulas built using only atomic
  constraints, conjunction, and existential and $\new$-quantification.
  We consider atomic constraints including equality, freshness, and
  \emph{equivariance}, that is, equality modulo a permutation of
  names.  \emph{Nominal Horn goal formulas} $G$ include atomic
  formulas, constraints, conjunctions, disjunctions, and existential
  and $\new$-quantification; \emph{program clauses} $D$ include atomic
  formulas, conjunctions, subgoal implications, and universal and
  $\new$-quantification.  A nominal logic program is a set $\PP$ of
  closed program clauses $D$.

  As usual in logic programming, we interpret a program clause $A \ent
  G$ with free variables $\vec{X}$ as an implicitly quantified, closed
  formula $\forall \vec{X}.G \impp A$.  Moreover, if the program
  clause contains free names $\vec{\Aa}$, they are interpreted as
  implicitly $\new$-quantified outside of the scope of the
  universally-quantified variables; thus, a clause $A \ent G$ with
  free variables $\vec{X}$ and free names $\vec{\Aa}$ is considered
  equivalent to the nominal logic formula $\new \vec{\Aa}.\forall
  \vec{X}. G \impp A$.



%% file: semantics.tex
\section{Semantics}\labelSec{semantics}

So far, we have motivated \aprolog purely in intuitive terms, arguing
that \aprolog concepts such as freshness and name-abstraction behave
as they do ``on paper''.  However, in order to prove the correctness
of the example programs we have considered, it is important to provide
a semantic foundation for reasoning about such programs.  We shall
investigate model-theoretic, proof-theoretic, and operational semantics
for nominal logic programs.

Classical model-theoretic semantics for logic
programming~\cite{vanemden76jacm,lloyd87foundations} defines the
meaning of a program as a Herbrand model constructed as the least
fixed point of a continuous operator.  We take for granted the theory
of Herbrand models for nominal logic introduced in the previous
section (full details are presented in \cite{cheney06jsl}).  We then
define an appropriate least fixed point semantics for nominal logic
programs and prove that the least fixed point model and the least
Herbrand model coincide.

While model-theoretic semantics is convenient for relating formal and
informal systems, it is not as useful for implementation purposes.
Instead, syntactic techniques based on proof theory are more
appropriate because they provide a declarative reading of connectives
as proof search operations in constructive logic.
\citeN{miller91apal} introduced the concept of \emph{uniform proof}; a
collection of program clauses and goal formulas is considered an
\emph{abstract logic programming language} if goal-directed proof
search is complete with respect to the underlying logic.  Accordingly,
we introduce a proof theory for a fragment of intuitionistic nominal
logic which performs goal-directed proof search (decomposes complex
goals to simple atomic formulas) and focused resolution (searches
systematically for proofs of atomic formulas based on the syntax of
program clauses).  We prove the soundness and completeness of this
system with respect to the model-theoretic semantics.

Finally, we consider the operational semantics of nominal logic
programs at an abstract level.  The proof theoretic semantics contains
a number of ``don't-know'' nondeterministic choices.  We provide an
operational semantics (following the semantics of constraint logic
programming~\cite{jaffar98jlp,darlington94lpar,leach01tplp}) which
delays these choices as long as possible, and { models the
behavior of an abstract interpreter.}  

Along the way we prove appropriate soundness and completeness results
relating the model-theoretic, proof-theoretic, and operational
semantics.  These results ensure the correctness of a low-level
interpreter based on the operational semantics relative to the
high-level approaches, and provide a rich array of tools for analyzing
the behavior of nominal logic programs.  The model-theoretic semantics
is especially useful for relating informal systems with nominal logic
programs, while the proof-theoretic semantics is convenient for
proving {properties of program transformations.}  We shall
consider such applications in \refSec{applications}.

\subsection{Model-theoretic semantics}\labelSec{model-theoretic}

In this section we define the model-theoretic semantics of nominal logic
programs.  We show that least Herbrand models exist for nominal Horn
clause programs and that the least Herbrand model is the least fixed
point of an appropriate continuous one-step deduction operator,
following \citeN{lloyd87foundations}.  This section also relies
on standard definitions and concepts from lattice
theory~\cite{davey02introduction}.  

Although the overall structure of our proof follows Lloyd, it differs
in some important technical details.  Most importantly, we do not
assume that clauses have been normalized to the form $A \ent G$.
Instead, all definitions and proofs are by induction over the
structure of goals and program clauses.  This is advantageous because
it permits a much cleaner treatment of each logical connective
independently of the others; this is especially helpful when
considering the new cases arising for the $\new$-quantifier, and when
relating the model-theoretic semantics to the proof-theoretic and
operational semantics.  Most proofs are in \refApp{model-theoretic-proofs}

\subsubsection{Least Herbrand Models}

It is a well-known fact that least Herbrand models exist for Horn
clause theories in first-order logic.  This is also true for nominal
Horn clause theories.  We rely on a previous development of Herbrand
model theory for nominal logic~\cite{cheney06jsl}, culminating in the
completeness of Herbrand models for Horn clause theories:

\begin{theorem}[Completeness of nominal Herbrand models]\labelThm{herbrand}
  A collection of program clauses is satisfiable in
 nominal logic if and only if it has {a Herbrand} model.
\end{theorem}

\begin{lemma}\labelLem{herbrand-intersection}
  Let $\Delta$ be a program and $\MM$ a nonempty set of
  Herbrand models of $\Delta$.  Then $\HH = \Isect\MM$ is also {a Herbrand} model of $\Delta$.
\end{lemma}

An immediate consequence is that a least Herbrand model
$\HH_\Delta = \Isect\{\HH \mid \HH \models \Delta\}$
exists for any nominal Horn theory $\Delta$.  Moreover, $\HH_\Delta$
consists of all ground atoms entailed by $\Delta$, as we now show.
\begin{theorem}\labelThm{herbrand-models-atoms}
  Let $\Delta$ be a program.  Then $\HH_\Delta = \{A \in
  B_\LL \mid \Delta \models A\}$.
\end{theorem}

\subsubsection{Fixed Point Semantics}

Classical fixed point theorems assert the existence of a fixed point.
However, to ensure that the fixed point of an operator on nominal Herbrand
models is still {a Herbrand} model we need an additional constraint: we
require that the operator is also equivariant, in the following sense.
\begin{definition}
  A set operator $T: \powerset{B_\LL} \to \powerset{B_\LL}$ is called
  \emph{equivariant} if
$\swap{\Aa}{\Ab}{T(S)} = T(\swap{\Aa}{\Ab}{S})$.
\end{definition}
\begin{theorem}\labelThm{nominal-fix}
  Suppose $T:\powerset{B_\LL} \to \powerset{B_\LL}$ is equivariant and monotone.  Then 
  $\lfp(T) = \bigcap\{S \in \powerset{B_\LL}\mid T(S) \subseteq S\}$
  is the least fixed point of $T$ and is equivariant.  If, in
  addition, $T$ is continuous, then
  $\lfp(T) = T^\omega = \bigcup^\omega_{i = 0}T^i(\emptyset)$.
\end{theorem}

\begin{definition}\labelDef{one-step-deduction}
  Let $S$ be {a Herbrand} interpretation and $D$ a closed program
  clause.  The \emph{one-step deduction operator}
  $T_D:\powerset{B_\LL} \to \powerset{B_\LL}$ is defined as
  follows:
  \[\begin{array}{rcl}
    T_\true(S) &=& S\\
    T_A(S) &=& S \cup \{A\}\\
    T_{D_1 \andd D_2}(S) &=& T_{D_1}(S) \cup T_{D_2}(S)\\
    T_{G \impp D}(S) &=& \left\{\begin{array}{ll}
      T_{D}(S) & \text{if }S \models G\\
      S & \text{otherwise}
    \end{array}\right.\\
    T_{\forall X{:}\sigma.D}(S) &=& \bigcup_{t:\sigma} T_{D[t/X]}(S)\\
    T_{\smallnew \Aa{:}\nu.D}(S) &=& \bigcup_{\Ab:\nu \not\in \supp(\smallnew \Aa.D)} T_{\swap{\Aa}{\Ab}{D}}(S)\\
  \end{array}
\]
We define $T_\Delta$ as $T_{D_1 \andd \cdots \andd D_n}$ provided
$\Delta = \{D_1,\ldots,D_n\}$ and each $D_i$ is closed.
\end{definition}

\begin{remark}\labelRmk{problem-with-previous-pt}
  Many prior expositions of the model-theoretic semantics of logic
  programs treat ``open'' Horn clauses $A \ent B_1,\ldots,B_n$ as the
  basic units of computation.  For example, the one-step deduction
  operator is usually formulated as
\[
T(S) = \{\theta(A) \mid \exists (A \ent B_1,\ldots,B_n \in P), \theta. S \models \theta(B_1),\ldots,\theta(B_n)\}
\]
This definition is not straightforward to extend to nominal logic
programming because of the presence of the $\new$-quantifier.
Although it can be done~\cite[Chapter 6]{cheney04phd}, the resulting
model-theoretic semantics is difficult to relate to the proof-theoretic
and operational semantics.  Instead, we prefer to define $T$
by induction on the structure of program clauses.  This necessitates
reorganizing our proofs, but the resulting argument is more modular
with respect to extensions based on connectives.
\end{remark}

\begin{lemma}\labelLem{one-step-mono-cont}
  For any program $\Delta$, $T_\Delta$ is monotone and continuous.
\end{lemma}

\begin{lemma}\labelLem{one-step-equivariant}
  For any $\Aa,\Ab\in \BA$, $\swap{\Aa}{\Ab}T_D(S) =
  T_{\swap{\Aa}{\Ab} D}(\tran{\Aa}{\Ab} \act S)$.  In particular, if
  $\Delta$ is a closed program with $FV(\Delta) = \supp(\Delta) =
  \emptyset$, then $T_\Delta$ is equivariant.
\end{lemma}

\begin{lemma}\labelLem{fixed-points-are-models}
  If  $\MM$ is a fixed point of $T_\Delta$, then $\MM
  \models \Delta$.
\end{lemma}

\begin{lemma}\labelLem{models-are-fixed-points}
  If $\MM \models \Delta$ then $\MM$ is a fixed point of $T_\Delta$.
\end{lemma}

\begin{theorem}\labelThm{lfp-equals-least-Herbrand}
$\HH_\Delta = \lfp(T_\Delta) = T_\Delta^\omega$.
\end{theorem}
\begin{proof}
  Clearly $T_\Delta^\omega=\lfp(T_\Delta)$ by~\refThm{nominal-fix}.
  Moreover, by \refLem{models-are-fixed-points} and
  \refLem{fixed-points-are-models}, the set of models of $\Delta$
  equals the set of fixed points of $T_\Delta$, so we must have
  $H_\Delta = \lfp(T_\Delta)$, since $\HH_\Delta$ is the least model
  of $\Delta$ and $\lfp(T_\Delta)$ is the least fixed point of
  $T_\Delta$.
\end{proof}

\subsection{Proof-theoretic semantics}\labelSec{proof-theoretic}

  In proof-theoretic semantics, an approach due to
  \citeN{miller91apal}, well-behaved logic programming languages are
  characterized as those for which \emph{uniform} (or
  \emph{goal-directed}) proof search is complete.  Uniform proofs were
  defined by \citeN{miller91apal} as sequent calculus proofs in which
  right-introduction rules are always used to decompose non-atomic
  goal formulas before any other proof rules are considered.

\begin{remark}
  Uniform proofs have been investigated previously for nominal logic
  programming \cite{gabbay04lics,cheney05fossacs,cheney06iclp}.  Our
  presentation is based on that of \citeN{cheney06iclp}; this approach
  resolves various problems in earlier work, principally the problem
  of making proof search goal-directed for (valid) nominal logic
  formulas such as $\new \Aa.\exists X. A \fresh X$.

  For example, the system $NL^\seq$ of \cite{cheney05fossacs} contains
  a \emph{freshness rule} $(F)$ that asserts that a fresh name can be
  introduced at any point in an argument:
  \[\infer[F]{[\Sigma]~\Gamma \seq \phi}{[\Sigma\#\Aa]~\Gamma \seq \phi}\quad (\Aa \not\in \Sigma)\]
  Here, the judgment $[\Sigma]~\Gamma \seq \phi$ can be read as ``For any
  valuation satisfying $\Sigma$, if all the formulas of $\Gamma$ hold
  then $\phi$ holds.''  As the following partial derivation suggests,
  the goal formula $\new \Aa.  \exists X. \Aa \fresh X$ cannot be
  derived in $NL^\seq$ without using $(F)$ before
  $\exR$, because otherwise there is no way to obtain a ground name
  $\Ab$ distinct from $\Aa$ with which to instantiate $X$:
  \[\infer[\newR]{\cdot:\cdot \seq \new \Aa.  \exists X. \Aa \fresh
    X}{\infer[F]{\Aa:\cdot \seq \exists X. \Aa \fresh
      X}{\infer[\exR]{\Aa\#\Ab:\cdot \seq \exists X. \Aa \fresh
        X}{\deduce{\Aa\#\Ab:\cdot \seq \Aa \fresh \Ab}{\vdots}}}}\]
\end{remark}

We adopt a variation of the $NL^\seq$ proof theory
of~\cite{cheney05fossacs} that solves this problem: specifically, we
define an ``amalgamated'' proof system $NL^\seq_\models$ that
separates the term-level constraint-based reasoning from logical
reasoning and proof search.  This technique was employed by
\citeN{darlington94lpar} and further developed by \citeN{leach01tplp}
in studying the semantics of constraint logic programs.

In this section we introduce the amalgamated proof system
$NL^\seq_\models$ and relate it to the model-theoretic semantics in
the previous section.  We also introduce a second \emph{residuated}
proof system that eliminates the nondeterminism involved in the
constraint-based rules; this system forms an important link between
the proof theory and the operational semantics in the next section.

\begin{figure}[tb]
\begin{boxedminipage}{\textwidth}
\[
\begin{array}{c}
    \infer[\trueR]{\upfsdn{\true}}{}
  \quad
  \infer[con]{\upfsdn{ C}}
  {\satsn{C}}
  \quad
  \infer[\andR]{\upfsdn{G_1 \andd G_2}}
  {\upfsdn{G_1} & \upfsdn{G_2}}
\medskip\\
  \infer[\orR_i]{\upfsdn{G_1 \orr G_2}}
  {\upfsdn{G_i}}
\quad
  \infer[\exR]{\upfsdn{\exists X{:}\sigma.G}}
  { \sat{\Sigma}{\nabla}{\exists X{:}\sigma.C} & \upf{\Sigma,X{:}\sigma}{\Delta}{\nabla,C}{G} }
  \medskip\\
  \infer[\newR]{\upfsdn{\new\Aa{:}\nu.G}}
  {\sat{\Sigma}{\nabla}{\new \Aa{:}\nu.C} & \upf{\Sigma\#\Aa{:}\nu}{\Delta}{\nabla,C}{G} }
  \quad
  \infer[sel]{\upfsdn{A}}{\apfsdn{D}{A} & (D \in \Delta)}
  \medskip\\
\hline
  \medskip\\
  \infer[hyp]{\apfsdn{A'}{A}}{\satsn{A' \ev A}}
  ~~ 
  \infer[\andL_i]{\apfsdn{D_1 \andd D_2}{A}}{\apfsdn{D_i}{A}}
  ~~ 
  \infer[\impL]{\apfsdn{G \impp D}{ A}}
  {\apfsdn{D}{A} & \upfsdn{G}}
  \medskip\\
  \infer[\allL]{\apfsdn{\forall X{:}\sigma. D}{A}}
  {\sat{\Sigma}{\nabla}{\exists X{:}\sigma.C} & \apf{\Sigma,X{:}\sigma}{\Delta}{\nabla,C}{D}{A} }
  \quad
  \infer[\newL]{\apfsdn{\smallnew \Aa{:}\nu. D}{A}}
  {\sat{\Sigma}{\nabla}{\new \Aa{:}\nu.C} & \apf{\Sigma\#\Aa{:}\nu}{\Delta}{\nabla,C}{D}{A} }
\end{array}
\]
\end{boxedminipage}
\caption{Uniform/focused proof search for intuitionistic nominal
  logic}\labelFig{uapf-aprolog}
\end{figure}

\subsubsection{The amalgamated system $NL^\seq_\models$}

The proof rules in \refFig{uapf-aprolog} describe a proof system that
first proceeds by decomposing the goal to an atomic formula, which is
then solved by refining a program clause.  The uniform derivability
judgment $\upfsdn{G}$ indicates that $G$ is derivable from $\Delta$
and $\nabla$ in context $\Sigma$, while the focused proof judgment
$\apfsdn{D}{A}$ indicates that atomic goal $A$ is derivable from
$\Delta$ and $\nabla$ by refining the program clause $D$ (using
$\Delta$ to help solve any residual goals).  The judgment $\satsn{C}$
is the ordinary constraint entailment relation defined in
\refSec{background}.

These rules are unusual in several important respects.  
  First, the $hyp$ rule requires solving an equivariance constraint of
  the form $A \ev B$; thus, from $p(\Aa,\Ab)$ we can conclude
  $p(\Ab,\Aa)$ since $\tran{\Aa}{\Ab}$ maps $p(\Aa,\Ab)$ to
  $p(\Ab,\Aa)$.
In contrast usually the hypothesis rule requires only that $A \eq A'$.
Our rule accounts for the fact that equivalent atomic formulas may not
be syntactically equal as nominal terms, but only equal modulo a
permutation, {due to nominal logic's \emph{equivariance}
  principle~\cite{pitts03ic}.}  Second, the proof system treats
constraints specially, separating them into a context $\nabla$.  This
is necessary because the role of constraints is quite different from
that of program clauses: the former are used exclusively for
constraint solving whereas the latter are used in backchaining.
  Third, the $\newL,\newR,\exR$ and $\allL$ rules are permitted to
  introduce a constraint on the quantified name $\Aa$ or variable $X$
  rather than providing a witness term.  Although these rules resemble
  the ``cut'' sequent calculus rule, which is typically excluded from
  uniform proof systems, these rules do not implicitly build ``cut''
  into the system; rather, they merely generalize the ability to
  instantiate a variable in a $\allL$ or $\exR$ rule to a constraint
  setting in an appropriate way.

  This treatment compartmentalizes all reasoning about the constraint
  domain in the judgment $\sat{\Sigma}{\nabla}{C}$, and makes it
  possible to retain ``uniform'' proofs in the presence of constraints
  for which a term instantiation of a quantified variable in rules
  $\exR, \allL$ may not be available.  For example, in constraint
  logic programming over the real numbers, the goal $\exists x. x^2 =
  2$ has no witnessing term.

  Furthermore, this approach solves the problem discussed in
  \refRmk{problem-with-previous-pt}, because the goal $\new
  \Aa.\exists X. \Aa \fresh X$ now has the following uniform
  derivation:
  \[\small
  \infer[\newR]{\upfsdn{\new \Aa.\exists X. \Aa \fresh X}}{\sat{\Sigma\#\Aa}{\nabla}{ \new \Aa.\true} & \infer[\exR]{\upf{\Sigma\#\Aa}{\Delta}{\nabla,\true}{\exists X. \Aa \fresh X}}{\sat{\Sigma\#\Aa}{\nabla,\true}{ \exists X. \Aa \fresh X} & \upf{\Sigma\#\Aa,X}{\Delta}{\nabla,\true,\Aa \fresh X}{\Aa \fresh X}}}
  \]
since $\sat{\Sigma\#\Aa}{\nabla}{ \exists X. \Aa \fresh X}$ is clearly
valid for any $\nabla$ (take $X$ to be any ground name besides $\Aa$).
The price we pay is the introduction of nondeterministic choices of
constraints in the quantifier rules.  We will show how to eliminate
this source of nondeterminism using a residuated proof theory in
\refSec{residuated-semantics}.

  We state without proof the following basic ``weakening'' properties.
  For brevity, here and elsewhere, we frequently say ``if
  $J_1,\ldots,J_n$ then $J'_1,\ldots,J'_m$'' rather than ``if $J_1,\ldots,J_n$ have
  derivations then $J'_1,\ldots,J'_m$ has a derivation'', for
  judgments $J_1,\ldots,J_n$, $J'_1,\ldots,J'_m$.
\begin{lemma}
~
\begin{enumerate}
\item If $\upfsdn{G}$ (or $\apfsdn{D}{A}$) and $\Sigma,\Sigma'$ is a
  well-formed context then $\upf{\Sigma,\Sigma'}{\Delta}{\nabla}{G}$
  (or $\apf{\Sigma,\Sigma'}{\Delta}{\nabla}{D}{A}$).
\item 
If $\upfsdn{G}$ (or $\apfsdn{D}{A}$) and $\Delta \subseteq \Delta'$
  then $\upf{\Sigma}{{\Delta'}}{\nabla}{G}$
  (or $\apf{\Sigma}{{\Delta'}}{\nabla}{D}{A}$).
\item If $\upfsdn{G}$ (or $\apfsdn{D}{A}$) and $\sat{\Sigma}{\nabla'}{\nabla}$
  then $\upf{\Sigma}{\Delta}{\nabla'}{G}$
  (or $\apf{\Sigma}{\Delta}{\nabla'}{D}{A}$).
\end{enumerate}
\end{lemma}

We first show that the restricted system
is sound with respect to the model-theoretic semantics.
\begin{theorem}[(Soundness)]\labelThm{soundness}
~
\begin{enumerate}
\item
  If $\upfsdn{G}$ is derivable then $\sat{\Sigma}{\Delta,\nabla}{G}$.  
\item
If $\apfsdn{D}{G}$ is derivable then 
  $\sat{\Sigma}{\Delta,D,\nabla}{G}$.
\end{enumerate}
\end{theorem}

\begin{proof}
\begin{enumerate}
\item For part (1), proof is by induction on derivations; the only
  novel cases involve $\new$.
  \begin{itemize}
\item  Suppose we have derivation
  \[\infer[\newR]{\upfsdn{\new \Aa.G}}{\satsn{\new \Aa.C} & \upf{\Sigma\#\Aa}{\Delta}{\nabla,C}{G}}\]
  By induction we have that $\sat{\Sigma\#\Aa}{\Delta,\nabla,C}{ G}$.
  Appealing to \refLem{new-sat}, we conclude
  $\sat{\Sigma}{\Delta,\nabla}{ \new \Aa.G}$.
  
\item   Suppose we have derivation
  \[  \infer[sel]{\upfsdn{A}}{\apfsdn{D}{A} & (D \in \Delta)}\]
  Then by induction hypothesis (2), we have that
  $\sat{\Sigma}{\Delta,D,\nabla}{ A}$.  Since $D \in \Delta$, clearly
  $\sat{\Sigma}{\Delta}{ D}$ so we can deduce
  $\sat{\Sigma}{\Delta,\nabla}{ A}$.
\end{itemize}
\item
  For the second part, proof is by induction on the derivation of $\apfsdn{D}{G}$.  The interesting cases are $hyp$ and $\newL$.
\begin{itemize}
\item Suppose we have derivation 
\[
  \infer[hyp]{\apfsdn{A'}{A}}{\satsn{A' \ev A}}
\]
We need to show $\sat{\Sigma}{\Delta,A',\nabla}{ A}$.  To see this,
suppose $\theta$ satisfies $\nabla$ and $\HH$ is {a Herbrand} model of
$\Delta,\theta(A')$.  Since $\satsn{A' \ev A}$, there must be a
permutation $\pi$ such that $\pi \act \theta(A') = \theta(A)$.
Moreover, since $\HH \models \theta(A')$, by the equivariance of $\HH$
we also have $\HH \models \pi \act \theta(A') $ so $\HH \models
\theta(A)$.  Since $\theta$ and $\HH$ were arbitrary, we conclude that
$\sat{\Sigma}{\Delta,A',\nabla}{A}$.

\item Suppose we have derivation 
  \[\infer[\newL]{\apfsdn{\smallnew \Aa.D}{A}}{\satsn{\new \Aa.C} & \apf{\Sigma\#\Aa}{\Delta}{\nabla,C}{D}{A}}\]
  By induction, we know that $\sat{\Sigma\#\Aa}{\Delta,D,\nabla,C}{
  A}$.  Since $\satsn{\new \Aa.C}$ it follows that
  $\sat{\Sigma\#\Aa}{\Delta,D,\nabla,C}{ A}$, so by \refLem{deduction}
  we have $\sat{\Sigma\#\Aa}{\Delta,D,\nabla}{ A}$.  Moreover, by
  \refLem{new-left}, we can conclude $\sat{\Sigma}{\Delta,\new \Aa.D,
  \nabla}{ A}$.
\end{itemize}
\end{enumerate}
This completes the proof.
\end{proof}

We next show a restricted form of completeness relative to the
model-theoretic semantics.  Since the model-theoretic semantics is
classical while the proof theory is constructive, it is too much to
expect that classical completeness holds.  For example,
$\sat{X{:}\nu,Y{:}\nu}{\cdot}{ X \eq Y \orr X \fresh Y}$ is
valid, but $\upf{A,B}{\cdot}{\cdot}{A \eq B \orr A\fresh B}$ is not
derivable (and indeed not intuitionistically valid).  Instead,
however, we can prove that any valuation $\theta$ that satisfies a
goal $G$ also satisfies a constraint which entails $G$.
\begin{proposition}\labelProp{completeness-step}
  For any $\Sigma,\Delta,G,D,i \geq 0$:
\begin{enumerate}
\item If $\sat{\Sigma}{T_\Delta^i,\theta}{G}$ then there exists
  $\nabla$ such that $\sat{\Sigma}{\theta}{\nabla}$ and $\upfsdn{G}$
  is derivable.
\item If $\sat{\Sigma}{T_{\theta(D)}(T_\Delta^i),\theta}{ A}$ but
  $\notsat{\Sigma}{T_\Delta^i,\theta}{ A}$ then there exists $\nabla$
  such that $\sat{\Sigma}{\theta}{ \nabla}$ and $\apfsdn{D}{A}$.
\end{enumerate}
\end{proposition}
\begin{proof}
  For the first part, proof is by induction on $i$ and $G$; most cases
  are straightforward.  We give two illustrative cases.
\begin{itemize}
\item If $G = A$ and $i > 0$, then there are two further cases.  If
  $\sat{\Sigma}{T_\Delta^{i-1},\theta}{ A}$ then we use part (1) of the
  induction hypothesis.  Otherwise $\notsat{\Sigma}{T_\Delta^{i-1},\theta}{
   A}$.  This implies that $\theta(A) \in
  T_\Delta(T_\Delta^{i-1}) = \bigcup_{D \in
    \Delta}T_D(T_\Delta^{i-1})$, so we must have $\theta(A) \in
  T_D(T_\Delta^{i-1})$ for some $D \in\Delta$.  Observe that since $D$
  is closed, $\theta(D) = D$.  Consequently
  $\sat{\Sigma}{T_{\theta(D)}(T_\Delta^{i-1}),\theta}{  A}$ but
  $\sat{\Sigma}{T_\Delta^{i-1}}{ A}$, so induction hypothesis (2)
  applies and we can obtain a derivation of $\apfsdn{D}{A}$.  The
  following derivation completes this case:
\[\infer[sel]{\upfsdn{A}}{\apfsdn{D}{A} & (D \in \Delta)}\;.\]

\item If $G = \new \Aa{:}\nu.G'$, assume without loss of generality
  $\Aa \not\in \Sigma$.  Then $\sat{\Sigma}{T_\Delta^i,\theta}{ \new
  \Aa.G'}$ implies $\sat{\Sigma\#\Aa}{T_\Delta^i,\theta}{
  G'}$.  By induction, there exists $\nabla$ such that
  $\upf{\Sigma\#\Aa}{\Delta}{\nabla}{G'}$ is derivable.  We can
  therefore derive
  \[
  \infer[\newR]{\upf{\Sigma}{\Delta}{\new \Aa.\nabla}{\new \Aa.G'}}{\sat{\Sigma}{\new \Aa.\nabla}{\new \Aa.\nabla} & 
    \upf{\Sigma{\#\Aa{:}\nu}}{\Delta}{\new \Aa.\nabla,\nabla}{G'}}
  \]
  using weakening to obtain the second subderivation.
\end{itemize}

Similarly, the second part follows by induction on $D$, unwinding the
definition of $T_D$ in each case.  We show the case for $\newL$.
\begin{itemize}

\item If $D = \new \Aa{:}\nu.D'$, assume without loss of generality
  that $\Aa \not\in \Sigma,\theta,A$.  Then $\theta(D)=\new
  \Aa.\theta(D')$ and since $T_{\smallnew \Aa.\theta(D')}(S) =
  \bigcup_{\Ab \not\in \supp(\smallnew
    \Aa.\theta(D'))}T_{\swap{\Aa}{\Ab}{\theta(D')}}(S)$, so we must
  have $\sat{\Sigma}{\bigcup_{\Ab \not\in \supp(\smallnew
    \Aa.\theta(D'))}T_{\swap{\Aa}{\Ab}\theta(D')}(T_\Delta^i),\theta}{
   A}$.  By definition, this means that $\theta(A) \in
  \bigcup_{\Ab \not\in \supp(\smallnew
    \Aa.\theta(D'))}T_{\swap{\Aa}{\Ab}\theta(D')}(T_\Delta^i)$.  Since
  by assumption $\Aa \not\in \Sigma,\theta,A$ and $\Aa \not\in
  \supp(\new \Aa.D')$, we must have $\theta(A) \in
  T_{\swap{\Aa}{\Aa}\theta(D')}(T_\Delta^i)$. Note that
  $\swap{\Aa}{\Aa}{\theta(D')} = \theta(D')$, and $\theta :
  \Sigma\#\Aa$, hence $\sat{\Sigma\#\Aa}{
  T_{\theta(D')}(T_\Delta^i),\theta}{ A}$.  Consequently, by
  induction, there exists a $\nabla$ such that
  $\sat{\Sigma}{\theta}{\nabla}$ and
  $\apf{\Sigma\#\Aa}{\Delta}{\nabla}{D'}{A}$.  Therefore, we have
  \[\infer[\newL]{\apf{\Sigma}{\Delta}{\new \Aa.\nabla}{\smallnew \Aa.D'}{A}}
  {\sat{\Sigma}{\new \Aa.\nabla}{\new \Aa.\nabla} & 
    \apf{\Sigma\#\Aa}{\Delta}{\new \Aa.\nabla,\nabla}{D'}{A}}\]
  Moreover, clearly $\sat{\Sigma\#\Aa}{\theta}{\nabla}$ implies
  $\sat{\Sigma}{\theta}{\new \Aa.\nabla}$.
\end{itemize}
The complete proof can be found in \refApp{proof-theoretic-proofs}.
\end{proof}

\begin{theorem}[(Algebraic Completeness)]\labelThm{algebraic-completeness}
  If $\sat{\Sigma}{\Delta,\theta}{ G}$ then there exists a constraint
  $\nabla$ such that $\sat{\Sigma}{\Delta,\theta}{ \nabla}$ and $\upfsdn{G}$
  is derivable.  
\end{theorem}
\begin{proof}
  If $\sat{\Sigma}{\Delta,\theta}{ G}$, then there is some $n$ such
  that $\sat{\Sigma}{T_\Delta^n,\theta}{ G}$, so
  \refProp{completeness-step} applies.
\end{proof}
We can also extend this to a ``logical'' completeness result
(following~\cite{jaffar98jlp}), namely that if an answer $C$
classically implies $G$, then there is a finite set of constraints
which prove $G$ and whose disjunction covers $C$.  We first establish
that a goal formula is classically equivalent to the disjunction
(possibly infinite) of all the constraints that entail it.
\begin{lemma}\labelLem{goal-equiv-constraints}
Let $\Sigma$ be a context, $\Delta$ a program, $G$ a goal, and $\Gamma_G = 
\{C \mid \upfsd{C}{G}\}$.  Then $\sat{\Sigma}{\Delta}{ G} \iff \Orr \Gamma_G$.
\end{lemma}
\begin{proof}
  For the forward direction, if $\sat{\Sigma}{\Delta,\theta}{ G}$ then
  by \refThm{algebraic-completeness} there exists a constraint
  $\nabla$ such that $\sat{\Sigma}{\theta }{ \nabla}$ and $\upfsdn{G}$.
  Hence, $\Andd \nabla \in \Gamma_G$, so $\sat{\Sigma}{\Delta,\theta }{
  \Orr \Gamma_G}$.  

  Conversely, if $\sat{\Sigma}{\Delta,\theta}{ \Orr
  \Gamma_G}$, then for some constraint $C \in \Gamma_G$,
  $\sat{\Sigma}{\Delta,\theta}{ C}$.  Consequently $\upfsd{C}{G}$
  holds, so by \refThm{soundness}, we have $\sat{\Sigma}{\Delta,C}{
  G}$.  Since $\sat{\Sigma}{\Delta,\theta}{ C}$, we conclude that
  $\sat{\Sigma}{\Delta,\theta}{ G}$.
\end{proof}
\begin{theorem}[(Logical Completeness)]\labelThm{logical-completeness}
  If $\sat{\Sigma}{\Delta,C}{ G}$ then there exists a finite set of
  constraints $\Gamma_0$ such that $\sat{\Sigma}{C}{ \Orr \Gamma_0}$ and for each $C' \in \Gamma_0$, 
  $\upf{\Sigma}{\Delta}{C'}{G}$.
\end{theorem}
\begin{proof}
  Again set $\Gamma_G = \{C' \mid \upf{\Sigma}{\Delta}{C'}{G}\}$.  By
  \refLem{goal-equiv-constraints}, $\sat{\Sigma}{\Delta,G}{ \Orr
  \Gamma_G}$.  Hence, $\sat{\Sigma}{\Delta,C }{ \Orr \Gamma_G}$.  By the
  Compactness Theorem for nominal logic~\cite[Cor. 4.8]{cheney06jsl},
  it follows that there is a finite subset $\Gamma_0 \subseteq
  \Gamma_G$ such that $\sat{\Sigma}{\Delta,C }{ \Orr \Gamma_0}$.  By
  definition, every $C' \in \Gamma_0 \subseteq \Gamma_G$ satisfies
  $\upf{\Sigma}{\Delta}{C'}{G}$.
\end{proof}

\begin{figure}[tb]
\begin{boxedminipage}{\textwidth}
\[
\begin{array}{c}
  \infer[con]{\rupfsd{C}{C}}
  {}
  \quad  
  \infer[\andR]{\rupfsd{G_1 \andd G_2}{C_1 \andd C_2}}
  {\rupfsd{G_1}{C_1} & \rupfsd{G_2}{C_2}}
  \medskip\\
  \infer[\orR_i]{\rupfsd{G_1 \orr G_2}{C}}
  {\rupfsd{G_i}{C}}
\quad 
    \infer[\trueR]{\rupfsd{\true}{\true}}{}
\quad
  \infer[\exR]{\rupfsd{\exists X{:}\sigma.G}{\exists X{:}\sigma.C}}
  {  \rupf{\Sigma,X{:}\sigma}{\Delta}{G}{C} }
  \medskip\\
  \infer[\newR]{\rupfsd{\new\Aa{:}\nu.G}{\new \Aa{:}\nu.C}}
  { \rupf{\Sigma\#\Aa{:}\nu}{\Delta}{G}{C} }
  \quad
  \infer[back]{\rupfsd{A}{C}}
  {\rapfsd{D}{A}{G} & \rupfsd{G}{C}& (D \in \Delta) }
\medskip\\
\hline
\medskip\\
\infer[hyp]{\rapfsd{A'}{A}{ A \ev A'}}{}
  \quad
  \infer[\andL_i]{\rapfsd{D_1 \andd D_2}{A}{G}}{\rapfsd{D_i}{A}{G}}
  \quad
  \infer[\impL]{\rapfsd{G \impp D}{ A}{G \andd G'}}
  {\rapfsd{D}{A}{G'}}
  \medskip\\
  \infer[\allL]{\rapfsd{\forall X{:}\sigma. D}{A}{\exists X{:}\sigma.G }}
  {\rapf{\Sigma,X{:}\sigma}{\Delta}{D}{A}{G }}
  \quad
  \infer[\newL]{\rapfsd{\smallnew \Aa{:}\nu. D}{A}{\new \Aa{:}\nu.G}}
  { \rapf{\Sigma\#\Aa{:}\nu}{\Delta}{D}{A}{G} }
\end{array}
\]
\end{boxedminipage}
\caption{Residuated uniform/focused proof search}\labelFig{ruapf-aprolog}
\end{figure}

\subsubsection{The residuated system
  $RNL^\seq_\models$}\labelSec{residuated-semantics}

  The rules in \refFig{uapf-aprolog} have the potential disadvantage
  that an arbitrary constraint $C$ is allowed in the rules $\exR$,
  $\allL$, $\newL$, $\newR$.  Such arbitrary constraints arguably
  correspond to a building a limited form of ``cut'' rule into proof
  search.
\refFig{ruapf-aprolog} shows a
\emph{residuated} proof system that eliminates this nondeterminism.  
(A similar idea is employed by \citeN{cervesato98icslp}).
Specifically, the judgment $\rupf{\Sigma}{\Delta}{G}{C}$ means that
given context $\Sigma$ and program $\Delta$, goal $G$ reduces to
constraint $C$; similarly, $\rapf{\Sigma}{\Delta}{D}{A}{G}$ means that
goal formula $G$ suffices to prove $A$ from $D$.

To see why this residuated system reduces nondeterminism, recall the goal $\new \Aa.\exists X. \Aa \fresh X$ from the previous section.  Using the residuated system, we can derive:
\[\small
\infer[\newR]{\rupfsd{\new \Aa.\exists X. \Aa \fresh X}{\new \Aa.\exists X.\Aa \fresh X}}{ \infer[\exR]{\rupf{\Sigma\#\Aa}{\Delta}{\exists X. \Aa \fresh X}{\exists X.\Aa \fresh X}}{ \infer[con]{\rupf{\Sigma\#\Aa,X}{\Delta}{\Aa \fresh X}{\Aa \fresh X}}{}}}
\]
Note that this simply says that in order to solve the \emph{goal}
$\new \Aa.\exists X. \Aa \fresh X$, it suffices to solve the
\emph{constraint} $\new \Aa.\exists X. \Aa \fresh X$, which is valid
so equivalent to $\true$.

\begin{theorem}[(Residuated Soundness)]\labelThm{residuated-soundness}
~
\begin{enumerate}
  \item If $\rupf{\Sigma}{\Delta}{G}{C}$ then
    $\upf{\Sigma}{\Delta}{C}{G}$.  
  \item If $\upf{\Sigma}{\Delta}{\nabla}{G}$
    and $\rapf{\Sigma}{\Delta}{D}{A}{G}$ then
    $\apf{\Sigma}{\Delta}{\nabla}{D}{A}$.
\end{enumerate}
\end{theorem}

\begin{theorem}[(Residuated Completeness)]\labelThm{residuated-completeness}
~
\begin{enumerate}
  \item   If $\upf{\Sigma}{\Delta}{\nabla}{G}$ then there exists a constraint $C$ such that
    $\rupf{\Sigma}{\Delta}{G}{C}$ and $\satsn{C}$.
  \item If $\apf{\Sigma}{\Delta}{\nabla}{D}{A}$ then there exists goal $G$
    and constraint $C$ such that $\rapf{\Sigma}{\Delta}{D}{A}{G}$ and
    $\rupf{\Sigma}{\Delta}{G}{C}$ and $\satsn{C}$.
\end{enumerate}
\end{theorem}

Both proofs are straightforward structural inductions (see
\refApp{proof-theoretic-proofs}).

\subsection{Operational Semantics}\labelSec{operational}

We now give a CLP-style operational semantics for nominal logic
programs. The rules of the operational semantics are shown in
\refFig{opsem}.  A program state is a triple of the form
$\st{\Sigma}{\Gamma}{\nabla}$.  Note that the backchaining step is
defined in terms of residuated focused proof,
$\rapf{\Sigma}{\Delta}{D}{A}{G}$.

\begin{figure*}[tb]
\begin{boxedminipage}{\textwidth}
\[\begin{array}{lclcll}
  &(B)& \st{\Sigma}{A,\Gamma}{ \nabla}&\too&\st{\Sigma}{G, \Gamma }{ \nabla}
  & (\text{if }\exists D \in \Delta. \rapf{\Sigma}{\Delta}{D}{A}{G})\smallskip\\
  &(C)&  \st{\Sigma}{C, \Gamma}{ \nabla}& \too& \st{\Sigma}{\Gamma }{ \nabla, C}  & \text{($\nabla,C$ consistent)}\smallskip\\
  &(\true)&  \st{\Sigma}{ \true,\Gamma}{ \nabla}& \too& \st{\Sigma}{\Gamma }{ \nabla}   \smallskip\\
  &(\andd)& \st{\Sigma}{ G_1\andd G_2,\Gamma}{ \nabla} &\too& \st{\Sigma}{G_1,G_2,\Gamma }{ \nabla}   \smallskip\\
  &(\orr_i)& \st{\Sigma}{G_1\orr G_2,\Gamma}{ \nabla} &\too& \st{\Sigma}{G_i,\Gamma }{ \nabla}  \smallskip\\
  &(\exists)& \st{\Sigma}{ \exists X{:}\sigma. G,\Gamma}{ \nabla} &\too& \st{\Sigma,X{:}\sigma}{G,\Gamma }{ \nabla}   \smallskip\\
  &(\new)& \st{\Sigma}{ \new \Aa{:}\nu. G,\Gamma}{ \nabla} &\too& \st{\Sigma\#\Aa{:}\nu}{G,\Gamma }{ \nabla} 
 \end{array}\]
\end{boxedminipage}
\caption{Operational semantics transitions for nominal logic programs}\labelFig{opsem}
\end{figure*}

{The operational semantics is quite close to the residuated
  proof system.  We now state the operational soundness and
  completeness properties.  The proofs are straightforward by cases or
  induction; some details are presented in an appendix.}  To simplify
notation, we write $\rupf{\Sigma}{\Delta}{\vec{G}}{\vec{C}}$ where
$\Gamma=G_1,\ldots,G_n$ and $\vec{C} = C_1,\ldots,C_n$ to abbreviate
$\rupf{\Sigma}{\Delta}{G_1}{C_1},\ldots,\rupf{\Sigma}{\Delta}{G_n}{C_n}$.
In addition, we will need to reason by well-founded induction on such
ensembles of derivations.  We define the subderivation relation $\D <
\E$ to indicate that $\D$ is a strict subderivation of $\E$, and write
$\vec{\D} <^* \vec{\E}$ for the multiset ordering generated by $<$.

\refProp{trans-sound} amounts to showing that each operational transition
corresponds to a valid manipulation on (multisets of) residuated
proofs. 
\begin{proposition}[(Transition Soundness)]\labelProp{trans-sound}
  If $\st{\Sigma}{\vec{G}}{\nabla} \too \st{\Sigma'}{\vec{G'}}{\nabla'}$
  and $\rupf{\Sigma'}{\Delta}{\vec{G'}}{\vec{C'}}$ then there exist
  $\vec{C}$ such that 
\begin{enumerate}
\item $\rupf{\Sigma}{\Delta}{\vec{G}}{\vec{C}}$ and
\item  $\sat{\Sigma'}{\nabla',\vec{C'}}{ \nabla,\vec{C}}$.
\end{enumerate}
\end{proposition}

\begin{theorem}[(Operational Soundness)]\labelThm{op-sound}
  if  $\st{\Sigma}{\vec{G}}{\nabla} \too^*
  \st{\Sigma'}{\emptyset}{\nabla'}$ then there exists $\vec{C}$
  such that $\sat{\Sigma'}{\nabla' }{ \nabla,\vec{C}}$ and
  $\rupf{\Sigma}{\Delta}{\vec{G}}{\vec{C}}$.  
\end{theorem}

The transition completeness property (\refProp{trans-complete}) states
that for any configuration $\st{\Sigma}{\Gamma}{\nabla}$ such that the
goals $\Gamma$ have appropriate derivations in the residuated proof
system, there is an operational transition step to a new state with
appropriately modified derivations.  This is essentially the
(complicated) induction hypothesis for proving completeness of the
operational semantics with respect to the other systems
(\refThm{op-complete}).

\begin{proposition}[(Transition Completeness)]\labelProp{trans-complete}
  For any nonempty $\vec{G}$ and satisfiable $\nabla$, $\vec{C}$, if we
  have derivations $\vec{\D} $ of $
  \rupf{\Sigma}{\Delta}{\vec{G}}{\vec{C}}$ then for some $\Sigma'$,
  $\nabla'$, and $\vec{C'}$ we have
\begin{enumerate}
\item $\st{\Sigma}{\vec{G}}{\nabla} \too \st{\Sigma'}{\vec{G'}}{\nabla'}$,
\item There exist derivations $\vec{\D'}$ of
  $\rupf{\Sigma'}{\Delta}{\vec{G'}}{\vec{C'}}$, where $\vec{\D'} <^*
  \vec{\D}$
\item $\exists \Sigma[\nabla] \models \exists \Sigma'[\nabla']$
\end{enumerate}
\end{proposition}

\begin{theorem}[(Operational Completeness)]\labelThm{op-complete}
  If $\rupf{\Sigma}{\Delta}{\vec{G}}{\vec{C}}$ and $\nabla,\vec{C}$ is
  satisfiable then for some $\Sigma'$ and $\nabla'$, we have
  $\st{\Sigma}{\vec{G}}{\nabla} \too^*
  \st{\Sigma'}{\emptyset}{\nabla'}$ and $\exists \Sigma[\nabla,\vec{C}] \models\exists\Sigma'[\nabla']$.
\end{theorem}

\subsection{Summary}

The goal of this section has been to present and show the equivalence
of model-theoretic, proof-theoretic, and operational presentations of
the semantics of nominal logic programs.  
We abbreviate
$\st{\Sigma}{G}{\emptyset} \too^* \st{{\Sigma,\Sigma'}}{\emptyset}{C}$
as $\reduce{\Sigma}{G}{\exists\Sigma'[C]}$.  The soundness and
completeness theorems we have established can be chained together as
follows to summarize these results:

\begin{corollary}\labelCor{cor-sound}
  If $\reduce{\Sigma}{G}{\nabla}$ then:
  \begin{enumerate}
  \item there exists C such that $\sat{\Sigma}{\nabla}{ C}$ and $\rupf{\Sigma}{\Delta}{G}{C}$;
  \item $\upf{\Sigma}{\Delta}{\nabla}{G}$; and
  \item $\sat{\Sigma}{\Delta,\nabla}{G}$
  \end{enumerate}
\end{corollary}
\begin{proof}
Immediate using \refThm{op-sound}, \refThm{residuated-soundness}, and \refThm{soundness}.
\end{proof}

\begin{corollary}
~
\begin{enumerate}
\item  If $\rupf{\Sigma}{\Delta}{G}{C}$ and $C$ is satisfiable then for some 
  $\nabla$, we have $\reduce{\Sigma}{G}{\nabla}$
  and $\sat{\Sigma}{C}{ \nabla}$.
\item If $\upf{\Sigma}{\Delta}{\nabla}{G}$ and $\nabla$ is satisfiable then for some 
  $\nabla'$, we have $\reduce{\Sigma}{G}{\nabla'}$ and $\sat{\Sigma}{\nabla}{
   \nabla'}$.
\item If $\sat{\Sigma}{\Delta,\theta}{ G}$ then for some $\nabla$, we
  have $\reduce{\Sigma}{G}{\nabla}$ and $\sat{\Sigma}{\theta}{
 \nabla}$.
\item If $\sat{\Sigma}{\Delta,C}{ G}$ then there exists a finite
  collection of constraints $\vec{\nabla}$ such that
  $\reduce{\Sigma}{G}{\nabla_i}$ for each $\nabla_i \in \vec{\nabla}$
  and $\sat{\Sigma}{C}{ \nabla_1 \orr \cdots \orr \nabla_n}$.
\end{enumerate}
\end{corollary}\labelCor{cor-complete}
\begin{proof}
  Immediate using \refThm{op-complete},
  \refThm{residuated-completeness}, \refThm{algebraic-completeness},
  \refThm{logical-completeness}.
\end{proof}

These results ensure that the operational semantics computes all (and
only) correct solutions with respect to nominal logic, so the
proof-theoretic and model-theoretic semantics can be used to reason
about the behavior of programs; this is often much easier than
reasoning about the operational semantics, as we shall now
demonstrate.



%% file: applications.tex
\section{Applications}\labelSec{applications}

\subsection{Adequacy}\labelSec{adequacy}

As discussed in \refSec{examples}, when we use \aprolog programs to
implement a formal system, it is important to ensure that the
relationship between the formal and informal system is correct.  To
some extent this property, often called
\emph{adequacy}~\cite{Pfenning:HandbookAR:framework:2001}, is in the
eye of the beholder, because the fact that the ``real'' system lacks a
precise formal characterization is often the problem we are trying to
solve by formalizing it.  Nevertheless, for nominal logic programs, we
can emulate typical adequacy arguments by checking that the
expressions, relations, and functions of the informal language being
formalized correspond to their representations in \aprolog.

For example, recall the encodings of informal $\lambda$-terms, types,
and contexts as \aprolog expressions, introduced in \refSec{lam}.
\noindent
We make the simplifying assumption that the variables of object
$\lambda$-terms and types are names of type $id$ and $tid$
respectively.  Then we can translate $\lambda$-terms, types, and
contexts as follows:
\[\begin{array}{rcl}
\trans{\Ax} &=&var(\Ax) \\
\trans{e_1~e_2} &=& app(\trans{e_1},\trans{e_2}) \\
\trans{\lambda  \Ax.e} &=& lam (\abs{\Ax}{\trans{e}})
\end{array}
\begin{array}{rcl}
\trans{\Aa} &=& tvar(\Aa) \\
\trans{\tau \to \tau'} &=& arrTy(\trans{\tau},\trans{\tau'})
\end{array}
\begin{array}{rcl}
\trans{\cdot} &=& [] \\ \trans{\Gamma,\Ax{:}\tau} &=& [(\Ax,\trans{\tau})|\trans{\Gamma}]
\end{array}
\]

We also introduce an auxiliary predicate $valid\_ctx : [(id,ty)] \to
o$, needed to characterize the ``well-formed'' contexts $\Gamma$ (in
which no variable is bound more than once).  It is defined by the
rules:
  \[
  valid\_ctx([]).\qquad valid\_ctx([(X,T)|G]) \ent X \fresh G, valid\_ctx(G).
  \]
Using the model-theoretic semantics introduced in
\refSec{model-theoretic}, it is straightforward (if tedious) to show
that:
\begin{proposition}
  Let $\mathrm{Exp},\mathrm{Ty},\mathrm{Ctx}$ be the sets of syntactic
  expressions, types, and contexts of the $\lambda$-calculus.  Let
  $FV(-)$ be the free-variables function, and ${-}~{\equiv_\alpha}~{-}$
  and ${-}[{-}:={-}] \equiv -$ the $\alpha$-equivalence and
  substitution relations respectively, defined in \citeN[Ch.
  2]{barendregt84}.

\begin{enumerate}
  \item   The following functions are bijective:
    \begin{eqnarray*}
      \trans{-} &: &\mathrm{Exp}/_{\equiv_\alpha} \to \{e \mid{} \vdash e : exp\}\\
      \trans{-} &:& \mathrm{Ty} \to \{t \mid {}\vdash t : exp\}\\
      \trans{-} &:& \mathrm{Ctx} \to \{g \mid {}\vdash g : [(id,ty)]\}
    \end{eqnarray*}
  \item $e \equiv_\alpha e'$ if and only if $\trans{e} \eq \trans{e'}$.
  \item $\trans{e[\Ax:=\Ay,\Ay:=\Ax]} = \swap{\Ax}{\Ay}{\trans{e}}$.
  \item $\Ax \not\in FV(e)$ if and only if $\Ax \fresh \trans{e}$.
  \item $e[\Ax:=e'] \equiv e''$ if and only if $subst(\trans{e},\trans{e'},\Ax) \eq \trans{e''}$ 
  \item $\Gamma$ is well-formed if and only if $valid\_ctx(\trans{\Gamma})$.
  \item For well-formed $\Gamma$, $\Ax \not\in Dom(\Gamma)$ if and
    only if $\Ax \fresh \trans{\Gamma}$.
  \item $\Gamma \vdash e : \tau$ if and only if
    $tc(\trans{\Gamma},\trans{e},\trans{\tau})$.
  \end{enumerate}
\end{proposition}

\subsection{Correctness of elaboration}\labelSec{applications-elaboration}
In an implementation, program clauses are often \emph{elaborated} into
a normal form $\forall \Sigma[G \impp A]$ which is easier to
manipulate and optimize.  We define the elaboration of a program
clause or program as the result of normalizing it with respect to the
following rewrite system:
\[
\begin{array}{rcl}
G \impp \true &\elab& \true\\
D \andd \true &\elab & D \\
\true \andd D &\elab& D\\
\forall X.\true &\elab& \true\\
\new \Aa.\true &\elab& \true\\
\Delta,D \andd D' &\elab& \Delta,D,D'\\
\Delta,\true &\elab& \Delta
\end{array}
\begin{array}{rcll}
G \impp G' \impp D &\elab& G \andd G' \impp D \\
G \impp D \andd D' &\elab& (G \impp D) \andd (G \impp D') \\
G \impp \forall X.D &\elab& \forall X.(G \impp D) &(X \not\in FV(G))\\
G \impp \new \Aa.D &\elab& \new \Aa.(G \impp D) &(\Aa \not\in supp(G))\\
\forall X. (D \andd D') &\elab&\forall X.D\andd \forall X.D'\\
\new \Aa. (D \andd D') &\elab&\new \Aa.D\andd \new \Aa.D'\\
{\forall X.\new \Aa.D} &{\elab}& {\new \Aa.\forall X. \Aa \fresh X \impp D}
\end{array}
\]
It is straightforward to show that this system is terminating and
confluent (up to $\alpha$- and multiset-equality) and that elaborated
programs consist only of closed formulas of the form $\forall \Sigma[G
\impp A]$ { where $\forall\Sigma$ is of the form
  $\new\vec{\Aa}\forall \vec{X}$.}
Moreover, this translation clearly preserves the meaning of the program since all of the rewrite rules correspond to valid equivalences in nominal logic.

\subsection{Avoiding expensive nominal constraint
  problems}\labelSec{applications-constraints}

  We have focused on reducing proof search for nominal logic programs
  to constraint solving over the theory of nominal terms.  The latter
  problem, while beyond the scope of this paper, is naturally central
  to an implementation.  Unfortunately, like many constraint domains
  encountered in constraint logic programming, full nominal constraint
  solving is $\NPTIME$-hard~\cite{cheney04icalp} and algorithmically
  involved~\cite{cheney05rta}.  In this section, we discuss the state
  of the art of nominal constraint solving and identify an
  optimization which can be used to avoid the need to handle
  $\NPTIME$-complete constraint problems in order to execute many
  typical programs (including all of the examples in this paper)
  efficiently in practice.

  At present, full equivariant resolution ($\ev$-resolution) is not
  implemented in \aprolog.  Instead, it uses \citeN{urban04tcs}'s
  \emph{nominal unification} algorithm, to which we refer in this
  paper as \emph{restricted nominal unification}.  This algorithm
  solves a tractable special case, specifically, it works for
  constraints involving only $\eq$ or $\fresh$
  {that satisfy} the following \emph{ground name
    restriction}:

\begin{definition}
  We say that a term, formula, or constraint is \emph{(ground)
    name-restricted} if, for every subformula or subterm of one of the
  forms
  \[a \fresh t \quad \swap{a}{b}{t} \quad \abs{a}{t}\] 
  the subterms $a,b$ are ground names.
\end{definition}

In the proof theoretic
semantics, we can model the use of equational unification for
resolution by replacing the $hyp$ rule with $hyp_\eq$
\[
\infer[hyp_\eq]{\apfsdn{A'}{A}}{\satsn{A \eq A'}}
\]
in which the stronger condition $\satsn{A \eq A'}$ is required to
conclude $\apfsdn{A'}{A}$.  We write $\equpfsdn{G}$ and
$\eqapfsdn{D}{A}$ for uniform or focused proofs in which $hyp_\eq$ is
used instead of $hyp$, and refer to such proofs as
\emph{$\eq$-resolution} proofs, to contrast with the $\ev$-resolution
proofs using the original $hyp$ rule.  It is easy to verify that
$\eq$-resolution proofs are sound with respect to ordinary derivations
and that all constraints arising in such proofs for name-restricted programs
and goals are name-restricted.

Unfortunately, $\eq$-resolution is incomplete relative to the full
system, because unlike in first-order logic, two ground atomic
formulas can be logically equivalent, but not equal as nominal terms.
Instead, because of the \emph{equivariance principle}, two ground
atomic formulas are equivalent if they are equal ``up to a
permutation'' (that is, related by $\ev$). Equational  resolution fails to find
solutions that depend on equivariance.  

\begin{example}\labelEx{bad-eg0}
  The simplest example is the single program clause $\new \Aa.p(\Aa)$.
  If we try to solve the goal $\exists X.p(X)$ against this program,
  then we get a satisfiable answer constraint $\new \Aa.  \exists X.
  p(\Aa) \eq p(X)$. However, if we pose the (logically equivalent)
  query $\new \Ab.p(\Ab)$ then proof search fails with the
  unsatisfiable $\new \Aa,\Ab. p(\Aa) \eq p(\Ab)$.
\end{example}

This example shows that equational resolution is incomplete for
name-restricted programs.  Moreover, $\ev$-resolution over
name-restricted terms remains $\NPTIME$-complete via an easy reduction
from the $\NPTIME$-completeness of equivariant
unification~\cite{cheney04icalp}.  Perhaps counterintuitively,
however, this does not appear to be a problem for most programs
encountered in practice.  In particular, all of the programs presented
in \refSec{examples}, even those including clauses such as
\begin{equation}
tc(G,lam( \abs{\Ax}{E}),arrTy(T,T')) \ent 
  \Ax \fresh G, tc([(\Ax,T)|G],E,T').\labelEq{lam-tc}
\end{equation}
 seem to work correctly using only
$\eq$-resolution, despite its incompleteness.

In previous work~\cite{urban05tlca}, the authors investigated this
situation and developed a (rather complicated) test for identifying
clauses for which $\eq$-resolution proof search is complete.
Informally, this test checks whether any names mentioned by a clause
are ``essentially'' free in its head.  However, this intuition is 
difficult to capture syntactically, as the following examples demonstrate:

\begin{example}\labelEx{bad-eg1}
  Suppose we require that names only appear in abstractions in the
  head of the clause.  This rules out the problematic clause $\new
  \Aa.p(\Aa)$.  However, $\eq$-resolution is still incomplete for such
  clauses.  For example, consider
  \begin{equation}\labelEq{bad-eg1}
    \new \Aa.\forall X.q(\abs{\Aa}{X},X).
  \end{equation}
  This clause can prove goal $q(\abs{\Aa}{\Aa},\Aa)$ for any name
  $\Aa$.  Since $\abs{\Aa}{\Aa} \eq \abs{\Ab}{\Ab}$ for any names
  $\Aa,\Ab$, the clause also proves $q(\abs{\Ab}{\Ab},\Aa)$ for any
  names $\Aa,\Ab$.  Yet $\eq$-resolution proof search for the goal
  $\new \Aa,\Ab.q(\abs{\Aa}{\Aa},\Ab)$ fails:
\[
\infer{\rupf{\Sigma}{\Delta}{\new \Aa,\Ab.q(\abs{\Aa}{\Aa},\Ab)}{\new \Aa,\Ab,\Aa'.\exists X. q(\abs{\Aa'}{X'},{X'})
\eq q(\abs{\Aa}{\Aa},\Ab)}}
{\infer{\rupf{\Sigma\#\Aa\#\Ab}{\Delta}{q(\abs{\Aa}{\Aa},\Ab)}{\new \Aa'.\exists X. q(\abs{\Aa'}{X'},X')
\eq q(\abs{\Aa}{\Aa},\Ab)}}
  {\infer{\rapf{\Sigma\#\Aa\#\Ab}{\Delta}{\smallnew \Aa.\forall
        X.q(\abs{\Aa}{X},X)}{q(\abs{\Aa}{\Aa},\Ab)}{q(\abs{\Aa'},{X'}{X'})
\eq q(\abs{\Aa}{\Aa},\Ab)}}
    {\infer{\rapf{\Sigma\#\Aa\#\Ab,\Aa',X'}{\Delta}{q(\abs{\Aa'}{X'},X')}{q(\abs{\Aa}{\Aa},\Ab)}{q(\abs{\Aa'}{X'},{X'})
\eq q(\abs{\Aa}{\Aa},\Ab)}}{}
      } 
    }
  }
\]
since the constraint $\new \Aa,\Ab,\Aa'.\exists X.
q(\abs{\Aa'}{X'},{X'}) \eq q(\abs{\Aa}{\Aa},\Ab)$ is unsatisfiable.
In contrast, the equivariance constraint $\new \Aa,\Ab,\Aa'.\exists X.
q(\abs{\Aa'}{X'},{X'}) \ev q(\abs{\Aa}{\Aa},\Ab)$ \emph{is}
satisfiable, since we may set $X = \Aa'$ to obtain problem
$q(\abs{\Aa'}{\Aa'},\Aa') \ev q(\abs{\Aa}{\Aa},\Ab)$ and then swap
$\Aa'$ and $\Ab$ to make the two terms equal.
\end{example}
\begin{example}\labelEx{bad-eg2}
  Suppose we forbid names \emph{anywhere} in the head of the clause,
  ruling out $\new \Aa.\forall X.q(\abs{\Aa}{X},X)$.  Incompleteness
  can still arise, as the following program illustrates:
  \begin{equation}
    \new \Aa. \forall X. r(X) \ent X \eq \Aa.\labelEq{bad-eg2}
  \end{equation}
  because this program logically implies goal $\new \Aa.r(\Aa)$ but
  $\eq$-resolution produces the unsatisfiable constraint $\new
  \Aa,\Aa'. r(\Aa) \eq r(X), X \eq \Aa'$.
\end{example}
\begin{example}\labelEx{bad-eg3}
  Suppose we forbid names \emph{anywhere} in a clause.  This means
  that only ``first-order'' Horn clauses not mentioning names,
  abstraction, freshness, or swapping can be used as program clauses.
  While this does mean that ordinary first-order logic programs can be
  executed efficiently over nominal terms, it rules out all
  interesting nominal logic programs.
\end{example}

In the rest of this section, we provide a new characterization of the
program clauses for which $\eq$-resolution is complete that is much
easier to prove correct and to check than the original formulation
in~\cite{urban05tlca}.  In this approach, we identify a class of
program clauses called \emph{\newgoal clauses}, show how to translate
arbitrary program clauses $D$ to \newgoal clauses $\|D\|$, show that
the behavior of an arbitrary program clause under $\eq$-resolution is
equivalent to that of its \newgoal translation, and finally show that
$\eq$-resolution is complete for \newgoal clauses.  Hence, if a clause
is equivalent to its \newgoal translation, then $\eq$-resolution proof
search is complete for the clause.

These results can be applied in two different ways.  First, in an
implementation that does not provide full $\ev$-resolution (as is the
case in the current implementation), they show that proof search is
complete for many typical programs anyway, and provide a systematic
way for the implementation to warn the programmer of a potential
source of incompleteness.  Second, in an implementation that does
provide full $\ev$-resolution, they can be used to recognize clauses
for which more efficient nominal unification can be used instead of
equivariant unification.

\subsubsection{\newgoal clauses}

We say that a program clause is \newgoal if it has no subformula of
the form $\new \Aa.D$.  However, $\new$-quantified goals $\new \Aa.G$
are allowed.  Such goals and program clauses are generated by the BNF
grammar:
\begin{eqnarray*}
G &::=& \true \mid A \mid C  \mid G \andd G' \mid G \orr G' \mid \exists X.G \mid \new \Aa.G \\
D &::=& \true \mid A \mid D \andd D' \mid G \impp D \mid \forall X.D 
\end{eqnarray*}

Arbitrary (normalized) program clauses of the form $\new
\vec{\Aa}\forall \vec{X}[ G \impp p(\vec{t})]$ can be translated to 
$\new$-goal clauses in the following way:
\[\|\new \vec{\Aa}\forall \vec{X}[ G \impp p(\vec{t})]\| 
= 
\forall
\vec{Z}[ (\new \vec{\Aa}.\exists \vec{X}. \vec{t} \eq
\vec{Z} \andd G )\impp p(\vec{Z})]\]
Note, however, that the \newgoal translation of a clause is \emph{not}
equivalent to the original clause, in general:
\begin{example}
  Recalling \refEx{bad-eg0}, consider the translation $\|\new
  \Aa.p(\Aa)\| = \forall Z[(\new \Aa.\Aa \eq Z) \impp p(Z)]$; the
  subgoal in the latter clause can never be satisfied since $\Aa$ must
  be fresh for $Z$.
\end{example}
\begin{example}
  Consider the translation of \refEq{bad-eg1}:
\[\|\new \Aa\forall X[ p(\abs{\Aa}{X},X)]\|
  = \forall Z_1,Z_2[(\new \Aa.\exists X.\abs{\Aa}{X} \eq Z_1 \andd X \eq
  Z_2) \impp p(Z_1,Z_2)]\;.
\] 
The latter clause cannot derive $p(\abs{\Aa}{\Aa},\Ab)$, so
  differs in meaning from the former.  In fact, the \newgoal clause is
  logically equivalent to $p(\abs{\Aa}{X},X) \ent \Aa \fresh X$.
\end{example}
\begin{example}
  Consider the translation  of \refEq{bad-eg2}:
\[\|\new \Aa\forall X[  \Aa \eq
  X \impp r(X)]\| = \forall Z[ (\new \Aa.\exists X. X \eq Z \andd  \Aa \eq X) \impp  r(Z)].\]
  The goal $\new \Aa.\exists X. X \eq Z \andd \Aa \eq X$ is never
  satisfiable since $\Aa$ will always be fresh for $X \eq Z$.
\end{example}
\begin{example}
  Any \newgoal program clause $\forall \vec{X}[ G \impp p(\vec{t})]$
  (including any purely first-order clause) is equivalent to its
  \newgoal translation $\forall \vec{Z}[(\exists \vec{X}. \vec{t} \eq
  \vec{Z} \andd G) \impp p(\vec{Z})]$.
\end{example}
\begin{example}
  Consider the  \newgoal translation of \refEq{lam-tc}:
  \[\begin{array}{lcl}
    \forall Z_1,Z_2,Z_3[\new \Ax.\exists G,E,T,T'.\\
    \quad 
    Z_1 \eq G \andd Z_2 \eq lam(\abs{\Ax}{E})\andd  Z_3 \eq arrTy(T,T')\andd  tc([(\Ax,T)|G],E,T') \\
    \qquad\impp tc(Z_1,Z_2,Z_3)].
  \end{array}\]
  Technically, the above clause is not literally equivalent to the
  original third clause; instead, it is equivalent to
  \[
  tc(G,lam( \abs{\Ax}{E}),arrTy(T,T')) \ent 
  \Ax \fresh(G,T,T'), tc([(\Ax,T)|G],E,T').
  \]
  which imposes the additional restriction that $\Ax \fresh T,T'$.
  These additional constraints clearly do not affect the meaning of the
  program in a simply typed setting where types cannot contain variable
  names; moreover, as discussed in \refSec{dependent-types}, if types
  can depend on term variables, these constraints follow from $\Ax
  \fresh G$ provided $G$ is well-formed and $T,T'$ are well-formed with
  respect to $G$.
\end{example}

As the above examples suggest, the \newgoal translation of a clause is
equivalent to the original clause precisely when the clause is
well-behaved with respect to $\eq$-resolution.  We now formalize this
observation, by showing that a clause has the same behavior as its
\newgoal translation under the $\eq$-resolution semantics, and then
showing that $\eq$-resolution proof search is complete for \newgoal
clauses.

\begin{proposition}\labelThm{eq-res-equals-newgoal}
  Let context $\Sigma$, $\Delta$, $D$, and $A$ be given, with $D$
  normalized to the form $\new \vec{\Aa}\forall \vec{X}[G \impp p(t)]$.
  Then there exist $G_1$ and $G_2$ such that $\eqrapfsd{D}{A}{G_1}$ and
  $\eqrapfsd{\|D\|}{A}{G_2}$ are derivable and such that
  $\sat{\Sigma}{}{G_1 \iff G_2}$.
\end{proposition}
\begin{proof}
  A derivation of a normalized $D = \new \vec{\Aa}\forall \vec{X}
  [G \impp p(\vec{t})]$ must be of the form
\[
\infer{\eqrapfsd{\smallnew \vec{\Aa}.\forall \vec{X}. G \impp p(\vec{t}) }{A}{\new \vec{\Aa}.\exists \vec{X}.p(\vec{t}) \eq A \andd G}}{
  \infer{\eqrapf{\Sigma\#\vec{\Aa}}{\Delta}{\forall \vec{X}. G \impp p(\vec{t}) }{A}{\exists \vec{X}.p(\vec{t}) \eq A \andd G}}
  {\infer{\eqrapf{\Sigma\#\vec{\Aa},\vec{X}}{\Delta}{ G \impp p(\vec{t}) }{A}{p(\vec{t}) \eq A \andd G}}
  {\hyp{\eqrapf{\Sigma\#\vec{\Aa},\vec{X}}{\Delta}{ p(\vec{t})}{A}{p(\vec{t}) \eq A}}
  }}}
\]
Similarly, $D$'s \newgoal translation $\|D\| = \forall
\vec{Z}.[(\new \vec{\Aa}.\exists \vec{X}.
\vec{t}\eq\vec{Z}  \andd G) \impp p(\vec{Z})]$, can only have a derivation of the form
\[
\infer{\eqrapf{\Sigma}{\Delta}{\forall
\vec{Z}[( \smallnew \vec{\Aa}.\exists \vec{X}.\vec{t}\eq\vec{Z}\andd G) \impp p(\vec{Z})]}{A}{\exists \vec{Z}.p(\vec{Z}) \eq A \andd \new \vec{\Aa}.\exists \vec{X}.\vec{t}\eq\vec{Z} \andd G}}
{\infer{\eqrapf{\Sigma,\vec{Z}}{\Delta}{(\smallnew \vec{\Aa}.\exists \vec{X}.\vec{t}\eq\vec{Z} \andd G) \impp p(\vec{Z})}{A}{p(\vec{Z}) \eq A \andd \new \vec{\Aa}.\exists \vec{X}.\vec{Z} \eq
\vec{t} \andd G}}
{\hyp{\eqrapf{\Sigma,\vec{Z}}{\Delta}{p(\vec{Z}) }{A}{p(\vec{Z}) \eq A }
}}}
\]
Now observe that
\begin{eqnarray*}
\exists \vec{Z}.p(\vec{Z}) \eq A \andd \new \vec{\Aa}.\exists \vec{X}. 
\vec{t}\eq \vec{Z}\andd G
&\iff&
\exists \vec{Z}.\new \vec{\Aa}.\exists \vec{X}.p(\vec{Z}) \eq A \andd
\vec{t} \eq \vec{Z} \andd G\\
&\iff&
\exists \vec{Z}.\new \vec{\Aa}.\exists \vec{X}.p(\vec{t}) \eq A 
 \andd G\\
&\iff&
\new \vec{\Aa}.\exists \vec{X}.p(\vec{t}) \eq A 
 \andd G
\end{eqnarray*}
This concludes the proof.
\end{proof}
Using the above lemma, it is straightforward to show that:
\begin{theorem}
  Let $\Sigma,\Delta,\nabla,G$ be given and suppose $\Delta'$ is the
  result of replacing some program clauses $D \in \Delta$ with
  $\|D\|$.  Then $\equpfsdn{G}$ is derivable if and only if
  $\equpf{\Sigma}{\Delta'}{\nabla}{G}$ is derivable.
\end{theorem}

\subsubsection{Completeness of $\eq$-resolution}\labelSec{applications-eq-resolution}

We now prove the completeness of $\eq$-res\-olu\-tion for \newgoal
programs. We first need a lemma showing that $\eq$-backchaining
derivations from \newgoal clauses and programs are stable under
application of permutations.  The full proofs can be found in
\refApp{eq-resolution-proofs}.
\begin{lemma}\labelLem{eq-swapping}
  Let $\Delta$ be a \newgoal program and $\pi$ be a type-preserving
  permutation of names in $\Sigma$.
\begin{enumerate}  
\item If $\equpfsdn{G}$ then $\equpfsdn{\pi \act G}$.  
\item If $\eqapfsdn{D}{A}$ then $\eqapfsdn{\pi \act D}{\pi \act A}$.
\end{enumerate}  
\end{lemma}

\begin{theorem}\labelThm{nice-eq-equivariance}
  If $\Delta$ is \newgoal then
\begin{enumerate}
\item If $\upfsdn{G}$ is derivable, then $\equpfsdn{G}$ is derivable.
\item If $\apfsdn{D}{A}$ is derivable, there exists a $\pi$ such that
  $\eqapfsdn{\pi \act D}{A}$ is derivable.
\end{enumerate}
\end{theorem}

Note that \refThm{nice-eq-equivariance} fails if $\newL$ is allowed:
for example, faced with a derivation
\[
\infer[\newL]{\apfsdn{\smallnew
  \Aa.D}{A}}{\apf{\Sigma\#\Aa}{\Delta}{\nabla}{D}{A}}
\]
we can obtain $\eqapf{\Sigma\#\Aa}{\Delta}{\nabla}{\pi \act D}{A}$ by
induction, but since $\pi$ may mention $\Aa$, it is not possible in
general to conclude $\eqapf{\Sigma}{\Delta}{\nabla}{\pi' \act
  \smallnew \Aa.D}{\pi' \act A}$ for some $\pi'$.  (This can be seen
for $D = p(\Aa), A = p(\Ab),\pi = \tran{\Aa}{\Ab}$ in
\refEx{bad-eg0}.)

\subsubsection{Discussion}

We introduced \newgoal programs above as a way of justifying using
\citeN{urban04tcs}'s name-restricted nominal unification.  However,
the completeness of $\eq$-resolution still holds if we consider full
nominal unification, in which variables may appear in place of names
anywhere in a term.  The current implementation also solves
constraints $a \fresh t$ where $a$ may also be a variable; constraints
such as $X \fresh Y$ used in $aneq$ and $subst$ are of this form.
Conjunctions of constraints of the form $X \fresh \pi \act X$ and $X
\fresh \pi \act Y$ can encode finite-domain set constraint problems,
so their satisfiability is $\NPTIME$-hard, but such constraints are
delayed until the end of proof search and then tested for
satisfiability by exhaustive search.

Although we have argued that many typical programs work fine using
$\eq$-resolution, it still seems worthwhile to investigate full
$\ev$-resolution.  We conclude this section with a discussion of
examples where full equivariant unification seems helpful.
\begin{example}
  The following program clauses 
\[\begin{array}{lclcl}
aneq(var(X),var(Y)) &\ent& X \fresh Y.\\
subst(var(X),N,var(Y)) &=& var(X) &\ent& X \fresh Y.\\
step(mismatch(X,Y,P),A,P')                    &\ent& X \fresh Y, step(P,A,P').
\end{array}\]
are equivalent to the clauses 
\[\begin{array}{lclcl}
aneq(var(\Ax),var(\Ay)).\\
subst(var(\Ax),N,var(\Ay)) &=& var(\Ax).\\
step(mismatch(\Ax,\Ay,P),A,P')      &\ent& step(P,A,P').
\end{array}\]
which require equivariant unification to execute correctly.  Thus,
equivariant unification allows us to write clauses using a convention
that syntactically distinct names are semantically distinct, instead
of explicitly needing to specify this using freshness constraints.
\end{example}
\begin{example}
  In a type inference algorithm such as Algorithm
  $\mathcal{W}$~\cite{DBLP:journals/jcss/Milner78}, consider the
  predicate $spec$ that relates a polymorphic type $\sigma$ to a list
  of distinct variables $\vec{\alpha}$ and monomorphic type $\tau$
  such that $\sigma = \forall\vec{\alpha}.\tau$.  This predicate is
  useful both for quantifying a monomorphic type by its unconstrained
  type variables and for instantiating a polymorphic type to some
  fresh type variables.  It can be implemented using the following
  \aprolog program clauses:
  \[
  \begin{array}{lcl}
    spec(monoTy(T),[],T).\\
    spec(polyTy(\abs{\Aa}{P}),[\Aa|L],T) & \ent & \Aa \fresh L, spec(P,L,T).
  \end{array}
  \]
  However, the second clause is not \newgoal, nor equivalent to its
  \newgoal form, because $\Aa$ can (and often will) occur free in $T$.
  Thus, it is not handled correctly in the current implementation.
  Correct handling of the above definition requires equivariant
  unification.
\end{example}



%% file: comparison.tex
\section{Comparison with previous work}\labelSec{comparison}

Several techniques for providing better handling of syntax with bound
names in logic programming settings have been considered:
\begin{itemize}
\item 
Higher-order logic programming and higher-order abstract
  syntax~\cite{DBLP:conf/slp/MillerN87,nadathur98higher,DBLP:conf/cade/NadathurM99,pfenning89pldi,pfenning91logic,pfenning99system}
\item Lambda-term abstract syntax, a variation on higher-order
  abstract syntax based on Miller's \emph{higher-order
    patterns}~\cite{miller91jlc}

\item Qu-Prolog, a first-order logic programming language with binding
  and substitution
  constraints~\cite{staples89meta,cheng91higher,nickolas96tcs,clark01multithreading}

\item Logic programming based on \emph{binding
    algebras}, an approach to the semantics of
  bound names based on functor
  categories~\cite{hamana01tacs,fiore99lics,hofmann99lics}.
\end{itemize}
We also relate our approach with functional programming languages that
provide built-in features for name-binding, such as
$ML_\lambda$~\cite{miller90lf},
FreshML~\cite{pitts00mpc,shinwell03icfp,shinwell05tcs,pitts07popl,pottier07lics}, and
Delphin~\cite{schuermann05tlca}, as well as recent efforts to provide
nominal abstract syntax as a lightweight language
extension~\cite{pottier05ml,cheney05icfp}.

\subsection{Logic programming with names and binding}

\subsubsection{Higher-order logic programming}

Higher-order abstract syntax~\cite{pfenning89pldi} is a powerful and
elegant approach to programming with names and binding that is
well-supported by higher-order logic programming languages such as
{\lprolog~\cite{nadathur98higher,DBLP:conf/cade/NadathurM99}} or
Twelf \cite{pfenning91logic,pfenning99system}.  In higher-order logic
programming, we consider logic programs to be formulas of a
higher-order logic such as Church's simple type
theory~\citeN{church40jsl} or the logical framework
LF~\cite{harper91jacm}.  
  Higher-order logic programming provides logically well-founded
  techniques for modularity and
  abstraction~\cite{miller89jlp,DBLP:conf/elp/Miller93} and provides
  advanced capabilities for programming with abstract syntax involving
  bound names and capture-avoiding substitution.

These capabilities are ideal for programming a wide variety of type
systems, program transformations, and theorem provers~\cite{hannan88iclp,pfenning91logic,felty93jar,nadathur98higher}.
Thus higher-order logic programming is an excellent tool for
prototyping and designing type systems and program transformations.

While this approach is elegant and powerful, it has some disadvantages
as well.  These disadvantages seem {tied} to higher-order abstract
syntax's main advantage: the use of constants of higher-order type to
describe object language binding syntax, meta-language variables to
encode object variables, and meta-language hypotheses and contexts to
encode object-language assumptions and contexts.
  In particular, the fact that object-language names ``disappear''
  into meta-level variables means that computations that involve
  comparing names (such as alpha-inequality) or generating fresh names
  (as in the semantics of references) seem to require quite different
  handling in a higher-order abstract syntax setting (using e.g.
  linearity~\cite{cervesato02ic}) than is typically done on paper.

  Another drawback of the higher-order approach is that ``elegant''
  encodings work well only when the inherent properties of the
  meta-language concepts are shared by the object language.  In
  particular, if the metalanguage's context is used for the context(s)
  of the object language, then the latter inherits the properties of
  the former, such as weakening and contraction.  This is, of course,
  no problem for the many programmming or logical calculi that have
  straightforward binding and context structure which fits the
  higher-order setting perfectly.

  However, many interesting systems have unusual contexts or binding
  behavior, especially substructural type
  systems~\cite{girard88,ohearn99bsl} and Floyd-Hoare-style logics of
  imperative programs~\cite{mason87lfcs,harel00,reynolds02separation}.
  These languages seem disproportionately difficult to program and
  reason about in pure higher-order logic (or LF).  Of course, such
  programs can still be written as higher-order logic programs, in the
  worst case by foregoing the use of higher-order abstract syntax.
  This can result in nondeclarative (and nonintuitive) programs which
  are not as convenient for experimentation or reasoning as one might
  like.

  One remedy is to extend the meta-language with new features that
  make it possible to encode larger classes of object languages
  elegantly.  Examples include linearity (Linear
  LF~\cite{cervesato02ic}) and monadic encapsulation of effects
  (Concurrent LF~\cite{watkins03types}).  In contrast, in
  \aprolog, many examples of substructural and concurrency calculi can
  be implemented without recourse to logical features beyond nominal
  logic, but also of course without the level of elegance and
  convenience offered by Linear or Concurrent LF.  However, the convenience of
  such extensions must be balanced against the effort needed to adapt
  the metatheory and implementations to support them.  

\subsubsection{Logic programming with higher-order patterns}\labelSec{comparison-ltas}

$L_\lambda$ is a restricted form of higher-order logic programming
introduced by \citeN{miller91jlc}.  In $L_\lambda$, occurrences of
meta-variables in unification problems are required to obey the
\emph{higher-order pattern constraint}: namely, each such
meta-variable may only occur as the head of an application to a
sequence of distinct \emph{bound} variables.  For example, $\lambda x.
F~x$ is a pattern but $\lambda x.F~x~x$ and $\lambda x.  x~(F X)$ are
not.  The higher-order pattern restriction guarantees that most
general unifiers exist, and that unification is decidable.

However, built-in capture-avoiding substitution for arbitrary terms is
not available in $L_\lambda$.  In full \lprolog, the beta-reduction
predicate can be encoded as
\begin{verbatim}
beta (app (lam (x\M x)) N) (M N).
\end{verbatim}
but this is not a higher-order pattern because of the subterm 
\verb|M N|.  Instead, substitution must be programmed explicitly in
$L_\lambda$, though this is not difficult:
\begin{verbatim}
beta (app (lam (x\E x)) E') E'' :- subst (x\E x) E' E''.
subst (x\x) E E.
subst (x\app (E1 x) (E2 x)) E (app E1' E2') 
  :- subst E1 E E1', subst E2 E E2'.
subst (x\lam y\E1 x y) E (lam y\E1' y) 
  :- pi y\ (subst (x\y) E y -> subst (x\E1 x y) E (E1' y)).
\end{verbatim}
This definition involves only higher-order patterns.  In $L_\lambda$,
the only substitutions permitted are {those of the form Miller calls}
$\beta_0$:
\[(\lambda x.M)~y = M[y/x]\]
that is, in which a bound variable is replaced with another bound
variable.

There are several interesting parallels between $L_\lambda$ and
\aprolog (and nominal unification and $L_\lambda$
unification~\cite{urban04tcs}).  The name-restricted fragment of
nominal logic programming which underlies the current \aprolog
implementation seems closely related to $L_\lambda$.  It seems
possible to translate many programs directly from one formalism to the
other, for example, by replacing local hypotheses with an explicit
context.  The proof-theoretic semantics in this paper may be useful
for further investigating this relationship.

  Miller and Tiu have investigated logics called \FOLDNabla and
  $LG^\omega$ which include a novel quantifier $\nabla$ that
  quantifies over ``generic'' objects~\cite{miller05tocl,tiu06lfmtp}.
  Miller and Tiu argue that $\nabla$ provides the right logical
  behavior to encode ``fresh name'' constraints such as arise in
  encoding (bi)similarity in the $\pi$-calculus.  As first observed by
  \citeN{miller05tocl}, the $\nabla$-quantifier has some, but not all
  properties in common with $\new$; this relationship has been
  explored further by several
  authors~\cite{gabbay04lics,cheney05fossacs,schoepp06lfmtp,tiu06lfmtp}.
However, \FOLDNabla has
primarily been employed as a foundation for encoding and reasoning
about languages, not as the basis of a logic programming language per
se.

\subsubsection{Qu-Prolog}

Qu-Prolog~\cite{staples89meta,cheng91higher,nickolas96tcs} is a
logic programming language with built-in support for object languages
with variables, binding, and capture-avoiding substitution.  It
extends Prolog's (untyped) term language with constant symbols
denoting object-level variables and a built-in simultaneous
capture-avoiding substitution operation $t\{t_1/x_1,\ldots,t_n/x_n\}$.
Also, a binary predicate $x~not\_free\_in~t$ is used to assert that an
object-variable $x$ does not appear in a term $t$.  Certain
identifiers can be declared as binders or \emph{quantifiers}; for
example, $lambda$ could be so declared, in which case the term
$lambda~x~t$ is interpreted as binding $x$ in $t$.  Unlike in
higher-order abstract syntax, quantifier symbols are not necessarily
$\lambda$-abstractions, so Qu-Prolog is not simply a limited form of
higher-order logic programming.  Qu-Prolog does not provide direct
support for name-generation; instead name-generation is dealt with by
the implementation during execution as in higher-order abstract
syntax.

Qu-Prolog is based on a classical theory of names and binding
described in terms of substitution.  Like higher-order unification,
Qu-Prolog's unification problem is undecidable, but in practice a
semidecision procedure based on delaying ``hard'' subproblems seems to
work well~\cite{nickolas96tcs}.

Qu-Prolog enjoys a mature implementation including a compiler for
Qu-Prolog written in Qu-Prolog.  Many interesting programs have be
written in Qu-Prolog, including interactive theorem provers,
client/server and database applications~\cite{clark01multithreading}.
Relations such as $\lambda$-term typability can be programmed
essentially the same as in \aprolog.  As with higher-order abstract
syntax, Qu-Prolog's built-in substitution operation is extremely
convenient.

Formal investigations of Qu-Prolog have been limited to the
operational semantics and unification algorithm.  There is no
denotational or proof-theoretic semantics explaining the behavior of
names and binding in Qu-Prolog.  Qu-Prolog is untyped and there is no
distinction between names and ordinary Prolog constants.  There is no
analogue of the $\new$-quantifier or the equivariance or freshness
principles.  It may be possible to define a clearer denotational
semantics for Qu-Prolog programs in terms of nominal logic.  This
could be useful for relating the expressiveness of \aprolog and
Qu-Prolog.  Conversely, it may be interesting to add a Qu-Prolog-like
built-in substitution operation (and associated unification
techniques) to \aprolog.

\subsubsection{Logic programming with binding algebras}

\citeN{fiore99lics} and \citeN{hofmann99lics} introduced \emph{binding
  algebras} and techniques for reasoning about abstract syntax with
binding using functor categories.  \citeN{hamana01tacs} developed a
unification algorithm and logic programming language for programming
with binding algebra terms involving name-abstraction $[a]t$,
name-ap\-pli\-ca\-tion $t@a$, name occurrences $var(a)$, injective
renamings {$\xi = [x_1:=y_1,x_2:=y_2,\ldots]$}, and first-order
function symbols and constants.

Hamana's unification algorithm unifies up to $\beta_0$-equivalence of
bound names with respect to name-application.  Hamana employs a type
system that assigns each term a type and a set of names that may
appear free in the term.  Hamana's unification algorithm appears to
generalize higher-order pattern unification; since names in
application sequences do not have to be distinct, however, most
general unifiers do not exist; for example $[x] F\conc x \conc
x \eq^? [y] G\conc y$ has two unifiers, $F = [x][y] y$ and $F = [x][y] x$.  

Many of the example programs of \refSec{examples} can also be
programmed using Hamana's programming language.  For example,
capture-avoiding substitution is given as an example by
\citeN{hamana01tacs}.  However, because binding algebras are based on
arbitrary renamings, rather than injective renamings, it may be
difficult to write programs such as {$aneq$ or $step$} that
rely on distinguishing {or generating} names.  In addition, since the
names free in a term must appear in the term's type, some programs may
require more involved type annotations or may be ruled out by the type
system.

\subsection{Functional programming with names and binding}

\subsubsection{$ML_\lambda$}

\citeN{miller90lf} also proposed a functional language extending
Standard ML to include an \emph{intensional function type} $\tau \To
\tau'$ populated by ``functions that can be analyzed at run-time'',
that is, higher-order patterns.  This language is called $ML_\lambda$
and supports functional programming with $\lambda$-term abstract
syntax using the intensional function type.  Since higher-order
pattern unification and matching are decidable, programs in
$ML_\lambda$ can examine the structure of intensional function values,
in contrast to ordinary function values which cannot be examined, only
applied to data.  \citeN{miller90lf}'s original proposal left many
issues open for future consideration; \citeN{pasalic00dali} developed
an operational semantics and prototype implementation of a language
called DALI, which was inspired by $ML_\lambda$.

\subsubsection{FreshML}

FreshML~\cite{pitts00mpc,shinwell03icfp,shinwell05tcs,pitts07popl,pottier07lics}
is a variant of ML (or Objective Caml) that provides built-in
primitives for names and binding based on nominal abstract syntax.
FreshML was an important source of inspiration for \aprolog.  At
present FreshML and \aprolog provide similar facilities for dealing
with nominal abstract syntax.  Arguably, because of the similarities
between higher-order patterns and nominal
terms~\cite{urban04tcs,cheney05unif}, FreshML can be viewed as an
alternative realization of $ML_\lambda$.

The main differences are 
\begin{itemize}
\item FreshML's treatment of name-generation uses side-effects, whereas
\aprolog uses nondeterminism.
\item There are no ground names in FreshML programs;
instead, names are always manipulated via variables.
\item FreshML currently provides more advanced forms of name-binding (such as
  binding a list of names simultaneously).
\item FreshML provides richer higher-order programming features.
\end{itemize}
Conversely, there are many programs that can be written cleanly in
\aprolog's logical paradigm but not so cleanly in FreshML's functional
paradigm, such as typechecking relations and nondeterministic
transition systems.

\subsubsection{Delphin}

Another language which draws upon $ML_\lambda$ is Delphin.  Delphin is
a functional language for programming with higher-order abstract
syntax and dependent types~\cite{schuermann05tlca}.  Because ordinary
recursion principles do not work for {many} higher-order
encodings~\cite{hofmann99lics}, Delphin provides novel features for
writing such programs (based on earlier work in the context of
Twelf~\cite{DBLP:conf/lpar/Schurmann01,DBLP:conf/csl/Schurmann01}).
This approach seems very powerful, but also potentially more complex
than nominal techniques.  For example, Delphin programs may be
nondeterministic and produce non-ground answers, because the
underlying higher-order matching problems needed for pattern matching
may lack most general unifiers.  At present a prototype called Elphin
{that supports the simply-typed case} has been implemented.

\subsubsection{\alphacaml}

\citeN{pottier05ml} has developed a tool for OCaml called \alphacaml.
\alphacaml\ translates high-level, OCaml-like specifications of the
binding structure of a language to ordinary OCaml type declarations
and code for performing pattern matching and fold-like traversals of
syntax trees.  \alphacaml\ uses a swapping-based nominal abstract
syntax technique internally, but these details typically do not need
to be visible to the library user.  Like FreshML, \alphacaml\ provides
forms of binding beyond binding a single variable; for example, its
binding specifications can describe pattern-matching and
$\mathbf{letrec}$ constructs.  

\subsubsection{FreshLib}

\citeN{cheney05icfp} developed FreshLib, a library for Haskell that
employs advanced generic programming techniques to provide nominal
abstract syntax for Haskell programs.  FreshLib provides common
operations such as capture-avoiding substitution and free-variables
functions as generic operations.  FreshLib also provides a richer
family of binding structures, as well as a type class-based interface
which permits users to define their own binding structures (such as
pattern matching binders).  Since Haskell is purely functional,
FreshLib code that performs fresh name generation has to be
encapsulated in a monad.



%% file: conclusions.tex
\section{Conclusions}\labelSec{concl}

Declarative programming derives much of its power from the fact that
programs have a clear mathematical meaning.  Name-binding and
name-generation are one of many phenomena which seem to motivate
abandoning declarativity in favor of expediency in practical Prolog
programming.  {On the other hand,
  although high-level programming with names and binding based on
  higher-order abstract syntax is compelling for many applications,
  sometimes its high level of abstraction is an obstacle to directly
  formalizing an informal system.}  As a result both first-order and
higher-order logic programs {sometimes} depart from the
declarative ideal when we wish to program with names and binding.

This paper investigates logic programming based on \emph{nominal
  logic}.  Nominal logic programs can be used to define a wide variety
of computations involving names, binding, and name generation
declaratively.  It provides many of the benefits of higher-order
abstract syntax, particularly built-in handling of renaming and
$\alpha$-equivalence, while still providing lower-level access to
names as ordinary data that can be generated and compared.  As a
result, nominal logic programs are frequently direct transcriptions of
what one would write ``on paper''.  {On the other hand,
  although nominal abstract syntax possesses 
  advantages not shared by any other technique, it does not currently
  provide all of the advantages of all previously explored
  techniques---the most notable example being the support for
  capture-avoiding substitution provided by higher-order abstract
  syntax.}

In this paper we have presented a variety of examples of nominal logic
programs, thoroughly investigated the semantics of nominal logic
programming, and presented some applications of the semantics.  This
work provides a foundation for future investigations, such as
developing practical techniques for nominal constraint solving,
{investigating extensions
  such as negation,} adding nominal abstract syntax as ``just another
constraint domain'' to existing, mature CLP implementations, and
analyzing or proving metatheoretic properties of core languages or
logics defined using nominal logic programs.



%% file: appendix.tex
\section{Proofs from \refSec{model-theoretic}}\labelApp{model-theoretic-proofs}

\begin{restateThm}{herbrand}
  A collection of program clauses is satisfiable in
 nominal logic if and only if it has {a Herbrand} model.
\end{restateThm}
\begin{proof}
  We note without proof that we can prenex-normalize all $\exists$ and
  $\new$ quantifiers in goals in $D$-formulas out to the top level as
  $\forall$ and $\new$ quantifiers respectively.  Then a collection of
  normalized $D$-formulas is a $\new\forall$-theory in the sense
  of~\cite[Theorem 6.17]{cheney06jsl}, so has a model iff it has {a Herbrand} model.
\end{proof}

\begin{restateLem}{herbrand-intersection}
  Let $\Delta$ be a program and $\MM$ a nonempty set of
  Herbrand models of $\Delta$.  Then $\HH = \Isect\MM$ is also {a Herbrand} model of $\Delta$.
\end{restateLem}
\begin{proof}
  We first note that the intersection of a collection of equivariant
  sets is still equivariant, so $\HH$ is {a Herbrand} model.  To prove
  it models $\Delta$, we show by mutual induction that 
  \begin{enumerate}
  \item For any program clause $D$, if $\forall M \in \MM. M \models D$
    then $\HH \models D$; and
  \item For any goal formula $G$, if $\HH \models G$ then $\forall M \in
    \MM.M \models G$.
  \end{enumerate}
  All the cases are standard except for $\new \Aa.G$ and $\new \Aa.D$.
  If $\forall M \in \MM. M \models \new \Aa.D$ then for each $M$, $M
  \models \swap{\Ab}{\Aa}D$ for all $\Ab$ not in $\supp(\new \Aa.D)$.
  Choose a $\Ab \not\in \supp(\new \Aa.D)$ such that $\forall M \in
  \MM. M \models \swap{\Ab}{\Aa}D$.  Appealing to the induction
  hypothesis, we obtain $\HH \models \swap{\Ab}{\Aa}D$.  By
  \refLem{some-any-new}, it follows that $\HH \models \new
  \Aa.D$.  The case for $\new \Aa.G$ is similar (but simpler).
\end{proof}

\begin{restateThm}{herbrand-models-atoms}
  Let $\Delta$ be a program.  Then $\HH_\Delta = \{A \in
  B_\LL \mid \Delta \models A\}$.
\end{restateThm}
\begin{proof}
  If $A \in \HH_\Delta$, then $A$ is valid in every Herbrand model of
  $\Delta$, so by \refThm{herbrand}, $A$ is valid in every model
  of $\Delta$.  Conversely, if $\Delta \models A$ then since $\HH_\Delta
  \models \Delta$ we have $\HH_\Delta \models A$; thus $A \in \HH_\Delta$.
\end{proof}

\begin{restateThm}{nominal-fix}
  Suppose $T:\powerset{B_\LL} \to \powerset{B_\LL}$ is equivariant and monotone.  Then 
  $\lfp(T) = \bigcap\{S \in \powerset{B_\LL}\mid T(S) \subseteq S\}$
  is the least fixed point of $T$ and is equivariant.  If, in
  addition, $T$ is continuous, then
  $\lfp(T) = T^\omega = \bigcup^\omega_{i = 0}T^i(\emptyset)$.
\end{restateThm}
\begin{proof}
  By the Knaster-Tarski fixed-point theorem, $\lfp(T)$ is the least
  fixed point of $T$.  To show that $\lfp(T)$ is equivariant, it
  suffices to show that $A \in \lfp(T) \implies \swap{\Aa}{\Ab}{A} \in
  \lfp(T)$.  Let $a,b$ be given and assume $A \in \lfp(T)$.
  Then for any pre-fixed point $S$ of $T$ (satisfying $T(S)
  \subseteq S$), we have $A \in S$.  Let such an $S$ be given.  Note
  that $T(\swap{\Aa}{\Ab}{S}) = \swap{\Aa}{\Ab}{T(S)} \subseteq
  \swap{\Aa}{\Ab}{S}$, so $\swap{\Aa}{\Ab}{S}$ is also a pre-fixed point of
  $T$.  Hence $A \in \swap{\Aa}{b}{S}$ so $\swap{\Aa}{b}{A} \in
  \swap{\Aa}{b}\swap{\Aa}{b}{S} = S$.  Since $S$ was an arbitrary
  pre-fixed point, it follows that $\swap{\Aa}{\Ab}{A} \in \lfp(T)$, as
  desired.

  The second part follows immediately from Kleene's fixed point
  theorem.
\end{proof}

\begin{restateLem}{one-step-mono-cont}
  For any program $\Delta$, $T_\Delta$ is monotone and continuous.
\end{restateLem}
\begin{proof}
  We prove by induction on the structure of $D$ that $T_D$ has the
  above properties.  Monotonicity is straightforward.  For continuity,
  let $S_0,S_1,\ldots,$ be an $\omega$-chain of subsets of $B_\LL$.
  The cases for $\true, \andd, \impp, \forall,\impp$, and atomic formulas
  follow standard arguments.  For $\new$, 
\begin{itemize}
\item Suppose $D = G \impp D'$.  Suppose that $A \in T_{G \impp
    D'}(\Union_i S_i)$.  If $\Union_i S_i \models G$ then $A \in
  T_{D'}(\Union_i S_i)$, and by induction $A \in \Union_i T_{D'}(S_i) =
  \Union_i T_{G \impp D'}(S_i)$.  Otherwise, $A \in \Union_i S_i =
  \Union_i T_{G \impp D'}(S_i)$.  This shows that $T_{G \impp
    D'}(\Union_i S_i) \subseteq \Union_i T_{G \impp D'}(S_i)$.  For the
  reverse direction, suppose $A \in \Union_i T_{G \impp D'}(S_i)$.  Then
  for some $i$, $A \in T_{G \impp D'}(S_i)$.  There are two cases.  If
  $S_i \models G$, then $A \in T_{D'}(S_i) = T_{G \impp D'} (S_i)
  \subseteq T_{G \impp D'}(\Union_i(S_i))$.  Otherwise, $A \in S_i =
  T_{G \impp D'}(S_i) \subseteq T_{G \impp D'}(\Union_i(S_i))$.

\item Suppose $D = \new \Aa.D'$.  Then we have
\[\begin{array}{rcll}
T_{\smallnew \Aa.D'}(\Union_i S_i) &=&
\Union_{\Ab{:}\nu\not\in \supp(\smallnew \Aa.D')}T_{\swap{\Aa}{\Ab}{D'}}(\Union_i S_i) & \text{Definition}\\
&=& \Union_{\Ab{:}\nu\not\in \supp(\smallnew \Aa.D')}\Union_i T_{\swap{\Aa}{\Ab}{D'}}(S_i) & \text{Induction hyp.}\\
&=& \Union_i \Union_{\Ab{:}\nu\not\in \supp(\smallnew \Aa.D')}T_{\swap{\Aa}{\Ab}{D'}}(S_i) & \text{Unions commute}\\
&=& \Union_i T_{\smallnew \Aa.D'}(S_i) & \text{Definition}
\end{array}\]
\end{itemize}
This completes the proof.
\end{proof}

\begin{restateLem}{one-step-equivariant}
  For any $\Aa,\Ab\in \BA$, $\swap{\Aa}{\Ab}T_D(S) =
  T_{\swap{\Aa}{\Ab} D}(\tran{\Aa}{\Ab} \act S)$.  In particular, if
  $\Delta$ is a closed program with $FV(\Delta) = \supp(\Delta) =
  \emptyset$, then $T_\Delta$ is equivariant.
\end{restateLem}
\begin{proof}
  The proof is by induction on the structure of $D$.  The cases for
  $\true,A,\andd$ are straightforward; for $\impp$ we need the easy
  observation that $S \models G \iff \tran{\Aa}{\Ab} \act S \models \tran{\Aa}{\Ab} \act G$.
  For $\forall X{:}\sigma.D$ formulas, observe that 
\[\begin{array}{rcll}
  \tran{\Aa}{\Ab} \act T_{\forall X.D}(S)
&=& \tran{\Aa}{\Ab} \act \bigcup_{t:\sigma} T_{D[t/X]}(S)  & \text{Definition}\\
  &=&
  \bigcup_{t:\sigma} \swap{\Aa}{\Ab}{T_{D[t/X]}( S)} & \text{Swapping commutes with union}\\
  &=&
  \bigcup_{t:\sigma} T_{(\tran{\Aa}{\Ab} \act D)[\tran{\Aa}{\Ab} \act t/X]}(\tran{\Aa}{\Ab} \act S) &\text{Induction hyp.}\\
  &=&
  \bigcup_{u:\sigma} T_{(\tran{\Aa}{\Ab} \act D)[u/X]}(\tran{\Aa}{\Ab} \act S) &\text{Change of variables ($u = \swap{\Aa}{\Ab}{t}$)}\\
  &=& T_{\tran{\Aa}{\Ab}
    \act \forall X.D}(\tran{\Aa}{\Ab} \act S) & \text{Definition.}
\end{array} \]
For $\new$, the argument is similar.
\end{proof}

\begin{restateLem}{fixed-points-are-models}
  If  $\MM$ is a fixed point of $T_\Delta$, then $\MM
  \models \Delta$.
\end{restateLem}
\begin{proof}
  We first prove by induction on the structure of $D$ that
  if $T_D(\MM) = \MM$ then $\MM \models D$.  
\begin{itemize}
\item If $D = \true$,
  trivially $\MM \models \true$.  

\item
If $D = A$, then clearly $\MM \cup
  \{A\} = T_A(\MM) = \MM$ implies $A \in \MM$ so $\MM \models A$.

\item If $D = D_1 \andd D_2$, then $T_{D_1 \andd D_2}(\MM) =
  T_{D_1}(\MM) \cup T_{D_2}(\MM) = \MM$ implies $T_{D_1}(\MM) =
  T_{D_2}(\MM) = \MM$ since $T_{D_1},T_{D_2}$ are monotone.  Then
  using the induction hypothesis $\MM \models D_1$ and $\MM \models
  D_2$, so $\MM \models D_1 \andd D_2$.

\item If $D = G \impp D'$, suppose that $\MM \models G$.  Then $T_{G
    \impp D'}(\MM) = T_{D'}(\MM) = \MM$ so by induction $\MM \models
  D'$.  Hence $\MM \models G \impp D'$.

\item For $D = \forall X{:}\sigma.D'$, note that $\MM = T_{\forall
    X.D'}(\MM) = \bigcup_{t:\sigma}T_{D'[t/X]}(\MM)$ implies
  $T_{D'[t/X]}(\MM) = \MM$ for every $t:\sigma$.  Hence by the
  induction hypothesis $\MM \models D'[t/X]$ for every $t:\sigma$;
  consequently $M \models \forall X.D'$.
\item
For $D = \new \Aa{:}\nu.D'$, note that $\MM =
  T_{\smallnew \Aa.D'}(\MM) = \bigcup_{\Ab:\nu \not\in
    \supp(\smallnew \Aa.D')}T_{\swap{\Aa}{\Ab}D'}(\MM)$ implies
  $T_{\swap{\Aa}{\Ab}D'}(\MM) = \MM$ for every fresh $\Ab$.  Hence
  by the induction hypothesis $\MM \models \swap{\Aa}{\Ab}D'$ for
  every fresh $\Ab$; consequently $M \models \new \Aa.D'$.
\end{itemize}

  Since any program $\Delta = \{D_1,\ldots,D_n\}$ is equivalent to a
  $D$-formula conjunction $D = D_1 \andd \cdots \andd D_n$, the
  desired result follows immediately.
\end{proof}

\begin{restateLem}{models-are-fixed-points}
  If $\MM \models \Delta$ then $\MM$ is a fixed point of $T_\Delta$.
\end{restateLem}
\begin{proof}
  Since $T_\Delta$ is monotone it suffices to show that $\MM$ is a
  pre-fixed point.  We first prove that for any $D$, if $\MM \models
  D$ then $T_D(\MM) \subseteq \MM$, by induction on the structure of
  $D$.
\begin{itemize}
\item If $D = \true$, clearly $T_\true(\MM) = \MM$.
\item If $D = A$ then since $\MM \models A$, we must have $A \in \MM$,
  so $T_A(\MM) = \MM \cup \{A\} = \MM$.
\item
If $D = D_1 \andd D_2$, then $T_{D_1\andd
    D_2}(\MM) = T_{D_1}(\MM) \cup
  T_{D_2}(\MM) \subseteq \MM$ since
  $T_{D_i}(\MM) \subseteq \MM$ by induction for $i =
  1,2$.  
\item For $D = G \impp D'$, since by assumption $\MM \models G \impp
  D$, there are two cases.  If $\MM \models G$, then $\MM \models D$,
  and by induction $T_{G \impp D}(\MM) = T_{ D}(\MM) \subseteq \MM$.
  On the other hand, if $\MM \not\models G$, then $T_{G \impp D}(\MM)
  = \MM$.
\item For $D = \forall X{:}\sigma.D'$, by assumption $\MM \models
  \forall X{:}\sigma.D'$ so we must have $\MM \models D[t/X]$ for all
  $t{:}\sigma$.  By induction $T_{D'[t/X]}(\MM) \subseteq \MM$ for any
  $t:\sigma$ so $\bigcup_{t:\sigma}T_{D'[t/X]}(\MM) \subseteq \MM$.
\item If $D = \new \Aa{:}\nu.D'$, by assumption $\MM \models \new
  \Aa{:}\nu.D'$ so $\MM \models \swap{\Aa}{\Ab}D'$ for any $\Ab
  \not\in \supp(\new \Aa.D')$.  By induction
  $T_{\swap{\Aa}{\Ab}D'}(\MM) \subseteq \MM$ for any $\Ab\not\in
  \supp(\new \Aa.D')$ so $\bigcup_{\Ab:\nu\not\in \supp(\smallnew
    \Aa.D')}T_{\swap{\Aa}{\Ab}D'}(\MM) \subseteq \MM$.
\end{itemize}

  To prove the lemma, take $\Delta = \{D_1,\ldots,D_n\}$ and $D = D_1
  \andd \cdots \andd D_n$. If $\MM \models \Delta$, then $\MM \models
  D$, so $T_D(\MM) \subseteq \MM$, whence
  $T_\Delta(\MM) \subseteq \MM$.
\end{proof}

\section{Proofs from \refSec{proof-theoretic}}\labelApp{proof-theoretic-proofs}

\begin{restateThm}{soundness}
~
\begin{enumerate}
\item
  If $\upfsdn{G}$ is derivable then $\sat{\Sigma}{\Delta,\nabla}{G}$.  
\item
If $\apfsdn{D}{G}$ is derivable then 
  $\sat{\Sigma}{\Delta,D,\nabla}{G}$.
\end{enumerate}
\end{restateThm}

\begin{proof}
  Induction on derivations.  The only novel cases involve $\new$.
\begin{itemize}
\item Suppose we have derivation
  \[  \infer[con]{\upfsdn{ C}}
  {\satsn{C}}
  \]
  Then $\satsn{C}$ implies $\sat{\Sigma}{\Delta,\nabla}{ C}$ as
  desired.

\item Suppose we have derivation
  \[
  \infer[\andR]{\upfsdn{G_1 \andd G_2}}
  {\upfsdn{G_1} & \upfsdn{G_2}}
\]
By induction, $\sat{\Sigma}{\Delta,\nabla}{ G_1}$ and $\sat{\Sigma}{\Delta,\nabla}{ G_2}$, so clearly $\sat{\Sigma}{\Delta,\nabla}{ G_1 \andd G_2}$.

\item Suppose we have derivation
  \[
  \infer[\orR_i]{\upfsdn{G_1 \orr G_2}}
  {\upfsdn{G_i}}
\]
By induction, $\sat{\Sigma}{\Delta,\nabla}{ G_i}$  so $\sat{\Sigma}{\Delta,\nabla}{ G_1 \orr G_2}$.

\item Suppose we have derivation
  \[
  \infer[\exR]{\upfsdn{\exists X{:}\sigma.G}}
  { \sat{\Sigma}{\nabla}{\exists X.C} & \upf{\Sigma,X}{\Delta}{\nabla,C}{G} }
\]
By induction, $\sat{\Sigma,X}{\Delta,\nabla,C}{ G}$.  Appealing to
\refLem{exists-sat}, we have $\sat{\Sigma}{\Delta,\nabla}{ \exists
X.C}$.

\item Suppose we have derivation
  \[\infer{\upfsdn{\new \Aa.G}}{\satsn{\new \Aa.C} & \upf{\Sigma\#\Aa}{\Delta}{\nabla,C}{G}}\]
  By induction we have that ${\Sigma\#\Aa}{\Delta,\nabla,C}{ G}$.
  Appealing to \refLem{new-sat}, we conclude $\sat{\Sigma}{\Delta,\nabla
  }{ \new \Aa.G}$.
\item Suppose we have derivation
  \[  \infer[sel]{\upfsdn{A}}{\apfsdn{D}{A} & (D \in \Delta)}\]
  Then by induction hypothesis (2), we have that
  $\sat{\Sigma}{\Delta,D,\nabla }{ A}$.  Since $D \in \Delta$, clearly
  $\sat{\Sigma}{\Delta }{ D}$ so we can deduce
  $\sat{\Sigma}{\Delta,\nabla}{A}$.
\end{itemize}

For the second part, proof is by induction on the derivation of $\apfsdn{D}{G}$.  
\begin{itemize}
\item Suppose we have derivation 
\[
  \infer[hyp]{\apfsdn{A'}{A}}{\satsn{A' \ev A}}
\]
We need to show $\sat{\Sigma}{\Delta,A',\nabla }{ A}$.  To see this,
suppose $\theta$ satisfies $\nabla$ and $\HH$ is {a Herbrand} model of
$\Delta,\theta(A')$.  Since $\satsn{A' \ev A}$, there must be a
permutation $\pi$ such that $\pi \act \theta(A') = \theta(A)$.
Moreover, since $\HH \models \theta(A')$, by the equivariance of $\HH$
we also have $\HH \models \pi \act \theta(A') $ so $\HH \models
\theta(A)$.  Since $\theta$ and $\HH$ were arbitrary, we conclude that
$\sat{\Sigma}{\Delta,A',\nabla }{ A}$.

\item Suppose we have derivation 
\[
  \infer[\andL_i]{\apfsdn{D_1 \andd D_2}{A}}{\apfsdn{D_i}{A}}
\]
By induction, we know that $\sat{\Sigma}{\Delta,D_i,\nabla }{ A}$, so
can conclude $\sat{\Sigma}{\Delta,D_1 \andd D_2,\nabla }{ A}$ by
\refLem{and-left}.
\item Suppose we have derivation 
\[
  \infer[\impL]{\apfsdn{G \impp D}{ A}}
  {\apfsdn{D}{A} & \upfsdn{G}}
\]
Then by induction, we have that $\sat{\Sigma}{\Delta,D,\nabla }{ A}$ and
$\sat{\Sigma}{\Delta,\nabla}{ G}$.  Then we can conclude
$\sat{\Sigma}{\Delta,G \impp D,\nabla }{ A}$ using \refLem{imp-left}.
\item Suppose we have derivation
\[
  \infer[\allL]{\apfsdn{\forall X{:}\sigma. D}{A}}
  {\sat{\Sigma}{\nabla}{\exists X.C} & \apf{\Sigma,X}{\Delta}{\nabla,C}{D}{A} }
\]
Then by induction, we have that $\sat{\Sigma,X}{\Delta,D,\nabla,C }{
A}$.  {We want} to conclude that $\sat{\Sigma}{\Delta,\forall X.D,\nabla }{A}$.  Suppose $\sat{\Sigma}{\theta }{ \nabla}$. Since $\sat{\Sigma}{\nabla}{ \exists X.C}$, we have that $\sat{\Sigma}{\theta }{ \exists X.C}$.  Thus, there exists a $t$ such that $\sat{\Sigma,X}{\theta[X \mapsto t] }{ C}$.  Therefore, $\sat{\Sigma,X}{\Delta,D,\theta[X \mapsto t]}{
 A}$.  Since $X$ appears only in $D$, by \refLem{all-left}, we
have that $\sat{\Sigma}{\Delta,\forall X.D,\theta }{ A}$.  Since
$\theta$ was an arbitrary valuation satisfying $\nabla$, it follows
that $\sat{\Sigma}{\Delta,\forall X.D,\nabla }{ A}$.

\item Suppose we have derivation 
  \[\infer{\apfsdn{\smallnew \Aa.D}{A}}{\satsn{\new \Aa.C} & \apf{\Sigma\#\Aa}{\Delta}{\nabla,C}{D}{A}}\]
  By induction, we know that $\sat{\Sigma\#\Aa}{\Delta,D,\nabla,C }{ A}$.
  Since $\satsn{\new \Aa.C}$ it follows that
  $\sat{\Sigma\#\Aa}{\Delta,D,\nabla,C }{ A}$, so by \refLem{deduction} we
  have $\sat{\Sigma\#\Aa}{\Delta,D,\nabla }{ A}$.  Moreover, by
  \refLem{new-left}, we can conclude $\sat{\Sigma}{\Delta,\new \Aa.D, \nabla  }{ A}$.
\end{itemize}
This completes the proof.
\end{proof}

\begin{restateProp}{completeness-step}
  For any $\Sigma,\Delta,G,D,i \geq 0$:
\begin{enumerate}
\item If $\sat{\Sigma}{T_\Delta^i,\theta }{ G}$ then there exists
  $\nabla$ such that $\sat{\Sigma}{\theta }{ \nabla}$ and $\upfsdn{G}$
  is derivable.
\item If $\sat{\Sigma}{T_{\theta(D)}(T_\Delta^i),\theta }{ A}$ but
  $\notsat{\Sigma}{T_\Delta^i,\theta }{ A}$ then there exists $\nabla$
  such that $\sat{\Sigma}{\theta }{ \nabla}$ and $\apfsdn{D}{A}$.
\end{enumerate}
\end{restateProp}
\begin{proof}
  For the first part, proof is by induction on $i$ and $G$; most cases
  are straightforward.  
\begin{itemize}
\item If $G = \true$ then trivially $\upf{\Sigma}{\Delta}{\cdot}{\true}$.

\item If $G = C$, a constraint, then $\sat{\Sigma}{T_\Delta^i,\theta }{
  C}$.  By definition, this means that $\models \theta(C)$ holds;
  equivalently, $\sat{\Sigma}{\theta }{ C}$.  Thus, taking $\nabla = C$,
  we obviously have
  \[\infer[con]{\upf{\Sigma}{\Delta}{C}{C}}{}\;.\]

\item If $G = A$ and $i = 0$, this case is vacuous since no atomic
  formulas are satisfied in the empty model $T^0_\Delta$.

\item If $G = A$ and $i > 0$, then there are two further cases.  If
  $\sat{\Sigma}{T_\Delta^{i-1},\theta }{ A}$ then we use part (1) of the
  induction hypothesis with $i-1$ to conclude $\upfsdn{A}$.  Otherwise
  $\notsat{\Sigma}{T_\Delta^{i-1},\theta }{ A}$.  This implies that
  $\theta(A) \in T_\Delta(T_\Delta^{i-1}) = \bigcup_{D \in
    \Delta}T_D(T_\Delta^{i-1})$, so we must have $\theta(A) \in
  T_D(T_\Delta^{i-1})$ for some $D \in\Delta$.  Since $D \in \Delta$ is closed, 
  we have $D = \theta(D)$, so  
  $\sat{\Sigma}{T_{\theta(D)}(T_\Delta^{i-1}),\theta }{ A}$ but
  $\notsat{\Sigma}{T_\Delta^{i-1} }{ A}$.  Induction hypothesis (2)
  applies and we can obtain a derivation of $\apfsdn{D}{A}$.  The
  following derivation completes this case:
\[\infer[sel]{\upfsdn{A}}{\apfsdn{D}{A} & (D \in\Delta)}\;.\]

\item If $G = G_1 \andd G_2$, then $\sat{\Sigma}{T_\Delta^i,\theta }{
  G_1 \andd G_2}$ implies $\sat{\Sigma}{T_\Delta^i,\theta }{ G_1}$ and
  $\sat{\Sigma}{T_\Delta^i,\theta }{ G_2}$, so by induction
  for some $\nabla_1,\nabla_2$, we have $\upf{\Sigma}{\Delta}{\nabla_1}{G_1}$, $\sat{\Sigma}{\theta}{\nabla_1}$, $\upfsdn{G_2}$, and $\sat{\Sigma}{\theta}{\nabla_2}$.  We can therefore conclude 
  \[\infer{\upf{\Sigma}{\Delta}{\nabla_1 \andd \nabla_2}{G_1 \andd G_2}}
  {\upf{\Sigma}{\Delta}{\nabla_1 \andd \nabla_2}{G_1} & \upf{\Sigma}{\Delta}{\nabla_1 \andd \nabla_2}{G_2}}\;.
  \]
since clearly $\sat{\Sigma}{\theta}{\nabla_1 \andd \nabla_2}$.

\item If $G = G_1 \orr G_2$, then $\sat{\Sigma}{T_\Delta^i,\theta }{
  G_1 \orr G_2}$ implies $\sat{\Sigma}{T_\Delta^i,\theta }{ G_i}$ for $i
  \in \{1,2\}$.  In either case, by induction $\upfsdn{G_i}$ and
  $\sat{\Sigma}{\theta}{\nabla}$ hold for some $\nabla$, so we deduce
  \[
  \infer{\upfsdn{G_1 \orr G_2}}{\upfsdn{G_i}}\;.
  \]

\item If $G = \exists X{:}\sigma.G'$, then $\sat{\Sigma}{T_\Delta^i,\theta  }{ \exists X{:}\sigma.G'}$ implies
  $\sat{\Sigma,X{:}\sigma}{T_\Delta^i,\theta[X\mapsto t] }{ G'}$ for
  some $t:\sigma$.  By induction, then, there exists $\nabla$ such
  that $\upf{\Sigma,X{:}\sigma}{\Delta}{\nabla}{G'}$ is derivable and
  $\sat{\Sigma,X}{\theta[X \mapsto t]}{\nabla}$.  We can therefore derive
  \[
  \infer{\upf{\Sigma}{\Delta}{\exists X.\nabla}{\exists X{:}\sigma.G'}}
  {\sat{\Sigma}{\exists X.\nabla}{\exists X.\nabla} 
    &
    \upf{\Sigma,X{:}\sigma}{\Delta}{\exists X.\nabla,\nabla}{G'}}
  \]
  using weakening to obtain the second subderivation.  Clearly $\sat{\Sigma,X}{\theta[X \mapsto t]}{\nabla}$ implies $\sat{\Sigma}{\theta}{\exists X.\nabla}$.
  
\item If $G = \new \Aa{:}\nu.G'$, assume without loss of generality
  $\Aa \not\in \Sigma$.  Then $\sat{\Sigma}{T_\Delta^i,\theta }{ \new
  \Aa.G'}$ implies $\sat{\Sigma\#\Aa}{T_\Delta^i,\theta }{ G'}$.  By
  induction, there exists $\nabla$ such that
  $\upf{\Sigma\#\Aa}{\Delta}{\nabla}{G'}$ is derivable and
  $\sat{\Sigma\#\Aa}{\theta}{\nabla}$.  We can therefore derive
  \[
  \infer{\upf{\Sigma}{\Delta}{\new \Aa.\nabla}{\new \Aa.G'}}{\sat{\Sigma}{\new \Aa.\nabla}{\new \Aa.\nabla} & 
    \upf{\Sigma,X{:}\sigma}{\Delta}{\new \Aa.\nabla,\nabla}{G'}}
  \]
  using weakening to obtain the second subderivation.  Clearly,
  $\sat{\Sigma\#\Aa}{\theta}{\nabla}$ implies
  $\sat{\Sigma}{\theta}{\new \Aa.\nabla}$
\end{itemize}

Similarly, the second part follows by induction on $D$, unwinding the
definition of $T_D$ in each case.
\begin{itemize}
\item If $D = \true$, then $\theta(\true) = \true$ and $T_\true(S) =
  S$; we cannot have both $\sat{\Sigma}{T_\Delta^i,\theta }{ A}$
  and $\notsat{\Sigma}{T_\Delta^i,\theta }{ A}$ so this case is vacuous.

\item If $D = A'$, then $T_{\theta(A')}(S) = S \cup \{\theta(A')\}$.
  Thus, if $\sat{\Sigma}{T_\Delta^i \cup \{\theta(A')\},\theta }{ A}$
  but $\notsat{\Sigma}{T_\Delta^i,\theta }{ A}$, then we must have
  $\theta(A) = \theta(A')$.  This clearly implies $\sat{\Sigma}{\theta  }{ A \eq A'}$, so taking $\nabla = A \eq A'$, clearly $\satsn{A
    \ev A'}$ and we can derive
  \[
  \infer{\apfsd{A \eq A'}{A'}{A}}{\sat{\Sigma}{A \eq A'}{A \ev A'}}\;.
  \]

\item If $D = D_1 \andd D_2$, then $\theta(D) = \theta(D_1) \andd
  \theta(D_2)$, and $T_{\theta(D_1) \andd \theta(D_2)}(S) =
  T_{\theta(D_1)}(S) \union T_{\theta(D_2)}(S)$, and
  $\sat{\Sigma}{T_{\theta(D_1)}(T_\Delta^i)\cup T_{\theta(D_2)}(T_\Delta^i),\theta }{
  A}$.  Then we must have $\sat{\Sigma}{T_{\theta(D_j)}(T_\Delta^i),\theta }{
  A}$ for $j \in \{1,2\}$.  In either case, by induction there exists
  $\nabla$ such that $\sat{\Sigma}{\theta }{ \nabla}$ and
  $\apfsdn{D_j}{A}$, so we can conclude
  \[
  \infer{\apfsdn{D_1 \andd
      D_2}{A}}{\apfsdn{D_j}{A}}
  \]

\item If $D = G \impp D'$, then $\theta(D) = \theta(G) \impp
  \theta(D')$.  There are two cases.  If $\sat{\Sigma}{T^i_\Delta,\theta  }{ G}$, then $T_{\theta(G) \impp \theta(D')}(T^i_\Delta) =
  T_{\theta(D')}(T^i_\Delta)$ so
  $\sat{\Sigma}{T_{\theta(D')}(T_\Delta^i),\theta }{ A}$.  By induction
  hypothesis (1), it follows that there exists a $\nabla$ such that
  $\upfsdn{G}$ and $\sat{\Sigma}{\theta}{\nabla}$; by induction
  hypothesis (2) there also exists a $\nabla'$ such that
  $\apf{\Sigma}{\Delta}{\nabla'}{D'}{A}$ and
  $\sat{\Sigma}{\theta}{\nabla'}$.  Using weakening and the $\impL$
  rule, we conclude
  \[\infer{\apf{\Sigma}{\Delta}{\nabla,\nabla'}{G \impp D'}{A}}
  {\upf{\Sigma}{\Delta}{\nabla,\nabla'}{G} & 
    \apf{\Sigma}{\Delta}{\nabla,\nabla'}{D'}{A}}\]
  which suffices since $\sat{\Sigma}{\theta}{\nabla,\nabla'}$.

  Otherwise, if $\notsat{\Sigma}{T^i_\Delta,\theta }{ G}$, then $T_{\theta(G)
    \impp \theta(D')}(T^i_\Delta) = T^i_\Delta$.  Then this case is
  vacuous since we cannot have both $\sat{\Sigma}{T^i_\Delta,\theta }{
  A}$ and $\notsat{\Sigma}{T^i_\Delta,\theta }{ A}$.

\item If $D = \forall X{:}\sigma.D'$, assume without loss of
  generality that $X \not\in Dom(\Sigma) \cup Dom(\theta)$.  Observe
  that $\theta(D) = \forall X{:}\sigma.\theta(D')$.  Since $T_{\forall
    X{:}\sigma.\theta(D')}(S) =
  \bigcup_{t{:}\sigma}T_{\theta(D')[t/X]}(S)$, we must have
  $\sat{\Sigma}{\bigcup_{t{}{}\sigma}T_{\theta(D')[t/X]}(T_\Delta^i),\theta  }{ A}$.  Hence, there must be a $t : \sigma$ such that
  $\theta(A) \in T_{\theta(D')[t/X]}(T_\Delta^i)$; choose a particular
  $t:\sigma$.  Consequently,
  $\sat{\Sigma}{T_{\theta(D')[t/X]}(T^i_\Delta),\theta }{ A}$.
  Moreover, since $X$ is not present in $\Sigma,A,\theta$, this is
  equivalent to $\sat{\Sigma,X}{T_{\theta[X \mapsto    t](D')}(T^i_\Delta),\theta[X \mapsto t] }{ A}$.  By induction,
  there must exist a $\nabla$ such that $\sat{\Sigma,X}{\theta,[X
    \mapsto t]}{\nabla}$ and $\apf{\Sigma,X}{\Delta}{\nabla}{D'}{A}$
  holds.  Hence, $\sat{\Sigma}{\theta}{\exists X.\nabla}$ so we can
  conclude by deriving
  \[\infer{\apf{\Sigma}{\Delta}{\exists X.\nabla}{\forall X{:}\sigma.D'}{A}}
  {\sat{\Sigma}{\exists X.\nabla}{\exists X.\nabla} & 
    \apf{\Sigma,X{:}\sigma}{\Delta}{\exists X.\nabla,\nabla}{D'}{A}}\]

\item If $D = \new \Aa{:}\nu.D'$, assume without loss of generality
  that $\Aa \not\in \Sigma,\theta,A$.  Then $\theta(D)=\new
  \Aa.\theta(D')$ and since $T_{\smallnew \Aa.\theta(D')}(S) =
  \bigcup_{\Ab \not\in \supp(\smallnew
    \Aa.\theta(D'))}T_{\swap{\Aa}{\Ab}{\theta(D')}}(S)$, so we must
  have $\sat{\Sigma}{\bigcup_{\Ab \not\in \supp(\smallnew    \Aa.\theta(D'))}T_{\swap{\Aa}{\Ab}\theta(D')}(T_\Delta^i),\theta   }{ A}$.  By definition, this means that $\theta(A) \in
  \bigcup_{\Ab \not\in \supp(\smallnew
    \Aa.\theta(D'))}T_{\swap{\Aa}{\Ab}\theta(D')}(T_\Delta^i)$.  Since
  by assumption $\Aa \not\in \Sigma,\theta,A$ and $\Aa \not\in
  \supp(\new \Aa.D')$, we must have $\theta(A) \in
  T_{\swap{\Aa}{\Aa}\theta(D')}(T_\Delta^i)$. Note that
  $\swap{\Aa}{\Aa}{\theta(D')} = \theta(D')$, and $\theta :
  \Sigma\#\Aa$, hence $\sat{\Sigma\#\Aa }{  T_{\theta(D')}(T_\Delta^i),\theta }{ A}$.  Consequently, by
  induction, there exists a $\nabla$ such that
  $\sat{\Sigma}{\theta}{\nabla}$ and
  $\apf{\Sigma\#\Aa}{\Delta}{\nabla}{D'}{A}$.  Therefore, we have
  \[\infer{\apf{\Sigma}{\Delta}{\new \Aa.\nabla}{\smallnew \Aa.D'}{A}}
  {\sat{\Sigma}{\new \Aa.\nabla}{\new \Aa.\nabla} & 
    \apf{\Sigma\#\Aa}{\Delta}{\new \Aa.\nabla,\nabla}{D'}{A}}\]
  Moreover, clearly $\sat{\Sigma\#\Aa}{\theta}{\nabla}$ implies
  $\sat{\Sigma}{\theta}{\new \Aa.\nabla}$.
\end{itemize}
This exhausts all cases and completes the proof.
\end{proof}

\begin{restateThm}{residuated-soundness}
~
\begin{enumerate}
  \item If $\rupf{\Sigma}{\Delta}{G}{C}$ then
    $\upf{\Sigma}{\Delta}{C}{G}$.  
  \item If $\upf{\Sigma}{\Delta}{\nabla}{G}$
    and $\rapf{\Sigma}{\Delta}{D}{A}{G}$ then
    $\apf{\Sigma}{\Delta}{\nabla}{D}{A}$.
\end{enumerate}
\end{restateThm}

\begin{proof}
Both parts are by structural induction on derivations.  
\begin{itemize}
\item If the derivation is of the form
\[
\infer[con]{\rupfsd{C}{C}}{}
\]
then deriving  $\upf{\Sigma}{\Delta}{C}{C}$ is immediate.
\item For derivation
\[
  \infer[\andR]{\rupfsd{G_1 \andd G_2}{C_1 \andd C_2}}
  {\rupfsd{G_1}{C_1} & \rupfsd{G_2}{C_2}}
\]
by induction we have $\upfsd{C_1}{G_1}$ and $\upfsd{C_2}{G_2}$.  
Weakening both sides, we have $\upfsd{C_1 \andd C_2}{G_1}$ and $\upfsd{C_1 \andd C_2}{G_2}$, so can derive 
\[\infer{\upfsd{C_1 \andd C_2}{G_1 \andd G_2}}{\upfsd{C_1 \andd C_2}{G_1} & \upfsd{C_1 \andd C_2}{G_2}}\]

\item
For derivation
\[
  \infer[\orR_i]{\rupfsd{G_1 \orr G_2}{C}}
  {\rupfsd{G_i}{C}}
\]
by induction we have $\upfsd{C}{G_i}$, so can derive 
\[
\infer{\upfsd{C}{G_1\orr G_2}}{\upfsd{C}{G_i}}
\]
\item For derivation 
\[ \infer[\exR]{\rupfsd{\exists
    X{:}\sigma.G}{\exists X.C}} { \rupf{\Sigma,X}{\Delta}{G}{C} }
\]
by induction, we have $\upf{\Sigma,X}{\Delta}{C}{G}$.  Weakening this derivation, we obtain
\[
\infer{\upfsd{\exists X.C}{\exists X.G}}{\sat{\Sigma}{\exists X.C}{\exists X.C} & \upf{\Sigma,X}{\Delta}{\exists X.C, C}{G}}
\]
\item For derivation
\[
\infer[\newR]{\rupfsd{\new\Aa.G}{\new \Aa.C}} {
  \rupf{\Sigma\#\Aa}{\Delta}{G}{C} }\]
by induction, we have $\upf{\Sigma\#\Aa}{\Delta}{C}{G}$.  Weakening this derivation, we obtain
\[
\infer{\upfsd{\new \Aa.C}{\new \Aa.G}}{\sat{\Sigma}{\new \Aa.C}{\new \Aa.C} & \upf{\Sigma\#\Aa}{\Delta}{\new \Aa.C, C}{G}}
\]\item For derivation

\[ \infer[back]{\rupfsd{A}{C}}
{\rapfsd{D}{A}{G}  & \rupfsd{G}{C}& (D \in \Delta)}
\]
by induction on the second derivation, we know that $\upfsd{C}{G}$
holds.  By induction hypothesis (2) on the first subderivation, it
follows that $\apfsd{C}{D}{A}$ holds.  Hence, since $D \in \Delta$, we can conclude
\[
\infer{\upfsd{C}{A}}{\apfsd{C}{D}{A} & (D \in \Delta)}
\]
\end{itemize}

For part (2), we reason simultaneously by induction on the structure of the two derivations.
\begin{itemize}
\item For derivations
\[
\infer[hyp]{\rapfsd{A'}{A}{ A \ev A'}}{} \qquad \infer{\upfsdn{A \ev A'}}{\deduce{\satsn{A \ev A'}}{\E}}
\]
it follows that 
\[\infer{\apfsdn{A'}{A}}{\satsn{A \ev A'}}\]
\item For derivations
\[
\infer[\andL_i]{\rapfsd{D_1 \andd D_2}{A}{G}}{\deduce{\rapfsd{D_i}{A}{G}}{\D}}
\qquad \deduce{\upfsdn{G}}{\E}
\]
by induction using $\D,\E$ we have $\apfsdn{D_i}{A}$ so we can conclude
\[
\infer[\andL_i]{\apfsdn{D_1 \andd D_2}{A}}{\apfsdn{D_i}{A}}
\]
\item For derivations
\[
\infer[\impL]{\rapfsd{G_1 \impp D}{ A}{G_1 \andd G_2}}{\deduce{\rapfsd{D}{A}{G_2}}{\D}}
\qquad \infer{\upfsdn{G_1 \andd G_2}}{\deduce{\upfsdn{G_1}}{\E_1} & \deduce{\upfsdn{G_2}}{\E_2}}
\]
By the induction hypothesis applied to $\D$ and $\E_2$, we have
$\apfsdn{D}{A}$.  Then we can conclude
\[
\infer{\apfsdn{G_1 \impp D}{A}}{\deduce{\upfsdn{G_1}}{\E_1}        & \apfsdn{D}{A}}
\]
\item For derivations
\[
\infer[\allL]{\rapfsd{\forall X{:}\sigma. D}{A}{\exists X.G' }}
{\deduce{\rapf{\Sigma,X}{\Delta}{D}{A}{G' }}{\D}} 
\qquad
\infer{\upfsdn{\exists X.G'}}{\satsn{\exists X.C} & \deduce{\upf{\Sigma,X}{\Delta}{\nabla,C}{G'}}{\E}}
\]
we can apply the induction hypothesis applied to subderivations
$\D,\E$ to obtain $\apf{\Sigma,X}{\Delta}{\nabla,C}{D}{A}$; hence, we
can conclude
\[
\infer{\apfsdn{\forall X.D}{A}}{\satsn{\exists X.C} & \apf{\Sigma,X}{\Delta}{\nabla,C}{D}{A}}
\]
\item For derivations
\[
\infer[\newL]{\rapfsd{\smallnew \Aa. D}{A}{\new \Aa.G}} {
  \deduce{\rapf{\Sigma\#\Aa}{\Delta}{D}{A}{G}}{\D} }
\qquad
\infer{\upfsdn{\new \Aa.G}}{\satsn{\new \Aa.C} & \deduce{\upf{\Sigma\#\Aa}{\Delta}{\nabla,C}{G}}{\E}}
\]
by induction on $\D,\E$ we can derive $\apf{\Sigma\#\Aa}{\Delta}{\nabla,C}{D}{A}$; hence we can conclude
\[
\infer{\apfsdn{\smallnew \Aa.D}{A}}{ \satsn{\new \Aa.C} & \apf{\Sigma\#\Aa}{\Delta}{\nabla,C}{D}{A}}
\]
\end{itemize}
This exhausts all possible cases, so the proof is complete.
\end{proof}

\begin{restateThm}{residuated-completeness}
~
\begin{enumerate}
  \item   If $\upf{\Sigma}{\Delta}{\nabla}{G}$ then there exists a constraint $C$ such that
    $\rupf{\Sigma}{\Delta}{G}{C}$ and $\satsn{C}$.
  \item If $\apf{\Sigma}{\Delta}{\nabla}{D}{A}$ then there exists goal $G$
    and constraint $C$ such that $\rapf{\Sigma}{\Delta}{D}{A}{G}$ and
    $\rupf{\Sigma}{\Delta}{G}{C}$ and $\satsn{C}$.
\end{enumerate}
\end{restateThm}
\begin{proof}
  Again, the proof is by structural induction on derivations.  The
  main subtlety is the construction of $C$ in each case.
\begin{itemize}
\item Case $con$
\[
  \infer[con]{\upfsdn{ C}}
  {\satsn{C}}
\]
Then clearly, we immediately derive
\[
\infer{\rupfsd{C}{C}}{}
\]
since $\satsn{C}$.
\item Case $\andR$ 
\[
  \infer[\andR]{\upfsdn{G_1 \andd G_2}}
  {\upfsdn{G_1} & \upfsdn{G_2}}
\]
By induction, we have $C_1$ such that $\rupfsd{G_1}{C_1}$ and $\satsn{C_1}$;
and $C_2$ such that $\rupfsd{G_2}{C_2}$ and $\satsn{C_2}$.  We can conclude that
\[
\infer{\rupfsd{G_1 \andd G_2}{C_1 \andd C_2}}{\rupfsd{G_1}{C_1} & \rupfsd{G_2}{C_2}}
\]
observing that $\satsn{C_1 \andd C_2}$ follows from $\satsn{C_1}$ and
$\satsn{C_2}$.
\item Case $\orR_i$
\[
  \infer[\orR_i]{\upfsdn{G_1 \orr G_2}}
  {\upfsdn{G_i}}
\]
By induction, we have $C$ such that $\rupfsd{G_i}{C}$ and $\satsn{C}$;
we can conclude by deriving 
\[\infer{\rupfsd{G_1 \orr G_2}{C}}
  {\rupfsd{G_i}{C}}
\]
\item Case $\exR$
\[
  \infer[\exR]{\upfsdn{\exists X{:}\sigma.G}}
  { \sat{\Sigma}{\nabla}{\exists X.C} & \upf{\Sigma,X}{\Delta}{\nabla,C}{G} }
\]
By induction, we know that $\rupf{\Sigma,X}{\Delta}{G}{C'}$ holds for
some $C'$ satisfying $\sat{\Sigma,X}{\nabla,C}{C'}$.  We may derive
\[
\infer{\rupfsd{\exists X.G}{\exists X.C'}}{\rupf{\Sigma,X}{\Delta}{G}{C'}}
\]
To complete this case, we need to show that $\satsn{\exists X.C'}$.
This follows by \refLem{exists-sat}.

\item Case $\newR$
\[
  \infer[\newR]{\upfsdn{\new\Aa.G}}
  {\sat{\Sigma}{\nabla}{\new \Aa.C} & \upf{\Sigma\#\Aa}{\Delta}{\nabla,C}{G} }
\]
By induction, we have $\rupf{\Sigma\#\Aa}{\Delta}{G}{C'}$
holds for some $C'$ such that $\sat{\Sigma\#\Aa}{\nabla,C}{C'}$.
We may derive
\[
\infer{\rupfsd{\new \Aa.G}{\new \Aa.C'}}{\rupf{\Sigma\#\Aa}{\Delta}{G}{C'}}
\]
Finally, to show that $\satsn{\new \Aa.C'}$, we appeal to \refLem{new-sat}.
\item Case $sel$
\[
  \infer[sel]{\upfsdn{A}}{\apfsdn{D}{A} & (D \in \Delta)}
\]

By induction hypothesis (2), there exists $C$ and $G$ such that
$\rapfsd{D}{A}{G}$, $\rupfsd{G}{C}$ and $\satsn{C}$.  Therefore, we can conclude by deriving
\[\infer{\rupfsd{A}{C}}{\rapfsd{D}{A}{G} & \rupfsd{G}{C} & (D \in \Delta)}
\]

\end{itemize}

Now we consider the cases arising from part (2).
\begin{itemize}
\item Case $hyp$
\[
  \infer[hyp]{\apfsdn{A'}{A}}{\satsn{A \ev A'}}
\]
Then we take $G = A \ev A' = C$ and derive
\[
\infer{\rapfsd{A'}{A}{A \ev A'}}{} \qquad \infer{\rupfsd{A \ev A'}{A \ev A'}}{}
\]
which suffices since $\satsn{A \ev A'}$.
\item Case $\andL_i$
\[
  \infer[\andL_i]{\apfsdn{D_1 \andd D_2}{A}}{\apfsdn{D_i}{A}}
\]
Then, by induction, we have $C$ and $G$ such that
$\rapfsd{D_i}{A}{G}$, $\rupfsd{G}{C}$, and $\satsn{C}$.  It suffices to
replace the first derivation with
\[
\infer{\rapfsd{D_1 \andd D_2}{A}{G}}{\rapfsd{D_i}{A}{G}}
\]
\item Case $\impL$
\[
  \infer[\impL]{\apfsdn{G \impp D}{ A}}
  {\apfsdn{D}{A} & \upfsdn{G}}
\]
Then, by induction on the first subderivation, we have $C'$ and $G'$
such that $\rapfsd{D}{A}{G'}$, $\rupfsd{G'}{C'}$, and $\satsn{C'}$.  By
induction on the second subderivation, we have $\rupfsd{G}{C}$ and
$\satsn{C}$ for some $C$.  To conclude, we derive
\[
\infer{\rapfsd{G \impp D}{A}{G \andd G'}}{\rapfsd{D}{A}{G'}}
\qquad
\infer{\rupfsd{G \andd G'}{C \andd C'}}{\rupfsd{G}{C} & \rupfsd{G'}{C'}}
\]
since $\satsn{C \andd C'}$ follows from $\satsn{C}$ and
$\satsn{C'}$.
\item Case $\allL$
\[
  \infer[\allL]{\apfsdn{\forall X{:}\sigma. D}{A}}
  {\sat{\Sigma}{\nabla}{\exists X.C} & \apf{\Sigma,X}{\Delta}{\nabla,C}{D}{A} }
\]
By induction hypothesis (2) applied to the second subderivation, there
exist $C'$ and $G'$ such that $\rapf{\Sigma,X}{\Delta}{D}{A}{G'}$ and
$\rupf{\Sigma,X}{\Delta}{G'}{C'}$ and $\sat{\Sigma,X}{\nabla,C}{C'}$.
We may therefore derive
\[
\infer{\rapfsd{\forall X.D}{A}{\exists X.G'}}{\rapf{\Sigma,X}{\Delta}{D}{A}{G'}}
\qquad
\infer{\rupfsd{\exists X.G'}{\exists X.C'}}{\rupf{\Sigma,X}{\Delta}{G'}{C'}}
\]
and conclude by observing that $\satsn{\exists X.C'}$ follows from
  existing assumptions by \refLem{exists-sat}.

\item Case $\newL$
\[
  \infer[\newL]{\apfsdn{\smallnew \Aa{:}\nu. D}{A}}
  {\sat{\Sigma}{\nabla}{\new \Aa.C} & \apf{\Sigma\#\Aa}{\Delta}{\nabla,C}{D}{A} }
\]
By induction, we can obtain a goal $G$ and constraint $C'$ such that
$\rapf{\Sigma\#\Aa}{\Delta}{D}{A}{G'}$ and
$\rupf{\Sigma\#\Aa}{\Delta}{G'}{C'}$ and
$\sat{\Sigma\#\Aa}{\nabla,C}{C'}$.  Clearly, we may now derive
\[
\infer{\rapfsd{\smallnew\Aa.D}{A}{\new \Aa.G'}}{\rapf{\Sigma\#\Aa}{\Delta}{D}{A}{G'}}\qquad
\infer{\rupfsd{\new \Aa.G'}{\new \Aa.C'}}{\rupf{\Sigma\#\Aa}{\Delta}{G'}{C'}}
\]
To conclude, we need to verify that $\sat{\Sigma}{\nabla}{ \new
  \Aa.C'}$.  This follows by \refLem{new-sat}.
\end{itemize}
This exhausts all cases and completes the proof.
\end{proof}

\section{Proofs from \refSec{operational}}\labelApp{operational-proofs}

\begin{restateProp}{trans-sound}
  If $\st{\Sigma}{\Gamma}{\nabla} \too \st{\Sigma'}{\Gamma'}{\nabla'}$
  and $\rupf{\Sigma'}{\Delta}{\vec{G'}}{\vec{C'}}$ then there exist
  $\vec{C}$ such that 
\begin{enumerate}
\item $\rupf{\Sigma}{\Delta}{\vec{G}}{\vec{C}}$ and
\item  $\sat{\Sigma'}{\nabla',\vec{C'} }{ \nabla,\vec{C}}$.
\end{enumerate}
\end{restateProp}
\begin{proof}
  Assume $\rupf{\Sigma'}{\Delta}{\vec{G'}}{\vec{C'}}$ is derivable.
  Proof is by case decomposition on the possible transition steps.
  
\begin{itemize}
\item Case $(B)$: If the backchaining rule is used,
\[
\st{\Sigma}{A,\vec{G_0}}{ \nabla}\too\st{\Sigma}{G', \vec{G_0} }{ \nabla}
\]
where $\rapfsd{D}{A}{G'}$ for some $D \in \Delta$, then we have
$\Sigma' = \Sigma$; $\vec{G} = A,\vec{G_0}$; $\vec{G'} =
G',\vec{G_0}$; $\vec{C'} = \vec{C} = C',\vec{C_0}$; and $\nabla' =
\nabla$.  We can extract a subderivation of $\rupfsd{G'}{C'}$ so for
(1) we derive $\rupfsd{A,\vec{G_0}}{C',\vec{C_0}}$ using the $back$
rule.  Part (2) is trivial.

\item Case $(C)$: If the constraint rule is used, we have
\[
 \st{\Sigma}{C, \vec{G'}}{ \nabla} \too \st{\Sigma}{\vec{G'} }{ \nabla, C}  
\]
where $\nabla,C$ is satisfiable.
Then $\Sigma' = \Sigma; \nabla' = \nabla,C; $ $\vec{G} = C,\vec{G'}$;
and $\vec{C} = C,\vec{C'}$.  For (1), we can derive using rule $con$
$\upfsd{C,\vec{G'}}{C,\vec{C'}}$; part (2) is trivial.

\item Case  $(\true)$: If the operational rule for $\true$ is used, we have
\[
 \st{\Sigma}{ \true,\vec{G'}}{ \nabla} \too \st{\Sigma}{\vec{G'} }{ \nabla}
\]
Then $\Sigma' = \Sigma; \nabla' = \nabla; \vec{C} = \true,\vec{C'}$;
for (1), $\rupfsd{\true,\vec{G'}}{\true,\vec{C'}}$ can be derived
using $\trueR$, while part (2) is trivial.

\item Case $(\andd)$:
\[
\st{\Sigma}{ G_1\andd G_2,\vec{G_0}}{ \nabla} \too \st{\Sigma}{G_1,G_2,\vec{G_0} }{ \nabla}  
\]
Then $\Sigma = \Sigma'$; $\nabla' = \nabla$; $\vec{G} = G_1\andd
G_2,\vec{G_0}$; $\vec{G'} = G_1,G_2,\vec{G_0} $; and $\vec{C'} =
C_1,C_2,\vec{C_0}$.  Set $\vec{C} = C_1 \andd C_2,\vec{C_0}$.  For
(1), $\rupfsd{G_1\andd G_2,\vec{G_0}}{C_1\andd C_2,\vec{C_0}}$ is
derivable using $\andR$; moreover, for (2), observe that
$\sat{\Sigma}{\nabla,C_1,C_2 }{ \nabla,C_1 \andd C_2}$.

\item Case  $(\orr_i)$:
\[
 \st{\Sigma}{G_1\orr G_2,\vec{G_0}}{ \nabla} \too \st{\Sigma}{G_i,\vec{G_0} }{ \nabla}  
\]
Then $\Sigma = \Sigma'$; $\nabla' = \nabla$; $\vec{G} = G_1\orr
G_2,\vec{G_0}$; $\vec{G'} = G_i,\vec{G_0} $; and $\vec{C'} = C,\vec{C_0}$;
so set $\vec{C} = \vec{C'}$.  For (1), $\rupfsd{G_1 \orr
  G_2,\vec{G_0}}{C,\vec{C_0}}$ follows using $\orR$, while (2) is
trivial.

\item Case  $(\exists)$:
\[
\st{\Sigma}{ \exists X{:}\sigma. G,\vec{G_0}}{ \nabla} \too 
\st{\Sigma,X{:}\sigma}{G,\vec{G_0} }{ \nabla}   
\]
Then $\Sigma' = \Sigma,X$; $\nabla' = \nabla$; $\vec{G} = \exists
X.G,\vec{G_0}$; $\vec{G'} = G,\vec{G_0}$; $\vec{C'} = C,\vec{C_0}$, so
set $\vec{C} = \exists X.C, \vec{C_0}$.  We can therefore derive
$\rupfsd{\exists X.G,\vec{G_0}}{\exists X.C,\vec{C_0}}$ for part (1).
For part (2), we observe that $\sat{\Sigma,X}{\nabla,C,\vec{C_0} }{
\exists X.C,\vec{C_0}}$.
\item Case  $(\new)$: Similar to the case for $(\exists)$.
\[
 \st{\Sigma}{ \new \Aa{:}\nu. G,\vec{G_0}}{ \nabla} \too \st{\Sigma\#\Aa{:}\nu}{G,\vec{G_0} }{ \nabla} 
\]
Then $\Sigma' = \Sigma\#\Aa$; $\nabla' = \nabla$, $\vec{G} = \new
\Aa.G,\vec{G_0}$; $\vec{G'} = G,\vec{G_0}$; $\vec{C'} = C,\vec{C_0}$,
so set $\vec{C} = \new \Aa.C,\vec{C_0}$.  For part (1), derive
$\rupfsd{\new \Aa.G,\vec{G_0}}{\new \Aa.C,\vec{C_0}}$ using $\newR$.
For part (2), observe that $\sat{\Sigma\#\Aa}{\nabla,C,\vec{C_0} }{
\new \Aa.C,\vec{C_0}}$.
\end{itemize}
This completes the proof.
\end{proof}

\begin{restateThm}{op-sound}
  if  $\st{\Sigma}{\vec{G}}{\nabla} \too^*
  \st{\Sigma'}{\emptyset}{\nabla'}$ then there exists $\vec{C}$
  such that $\sat{\Sigma'}{\nabla' }{ \nabla,\vec{C}}$ and
  $\rupf{\Sigma}{\Delta}{\vec{G}}{\vec{C}}$.  
\end{restateThm}
\begin{proof}
  Proof is by induction on the number of transition steps.  If no
  steps are taken, then $\vec{G}$ is empty and $\nabla' = \nabla$, so
  taking $\vec{C}$ to be empty, the conclusion is trivial.  Otherwise
  we have a step
  \[
  \st{\Sigma}{\vec{G}}{\nabla} \too 
  \st{\Sigma_0}{\vec{G_0}}{\nabla_0} \too^*
  \st{\Sigma'}{\emptyset}{\nabla'}\;.
  \]
  By induction, there exists $\vec{C_0}$ such that $\sat{\Sigma'}{\nabla'
  }{ \nabla_0,\vec{C_0}}$ and
  $\rupf{\Sigma_0}{\Delta}{\vec{G_0}}{\vec{C_0}}$.  Using
  \refProp{trans-sound}, we can construct $\vec{C}$ such that
  $\rupf{\Sigma}{\Delta}{\vec{G}}{\vec{C}}$ and
  $\sat{\Sigma_0}{\nabla_0,\vec{C_0}}{\nabla,\vec{C}}$.  Moreover, using
  weakening and deduction, we can conclude that $\sat{\Sigma'}{\nabla' }{
  \nabla,\vec{C}}$.
\end{proof}

\begin{restateProp}{trans-complete}
  For any nonempty $\vec{G}$ and satisfiable $\nabla$, $\vec{C}$, if we
  have derivations $\vec{\D} ::
  \rupf{\Sigma}{\Delta}{\vec{G}}{\vec{C}}$ then for some $\Sigma'$,
  $\nabla'$, and $\vec{C'}$ we have
\begin{enumerate}
\item $\st{\Sigma}{\vec{G}}{\nabla} \too \st{\Sigma'}{\vec{G'}}{\nabla'}$,
\item $\D' :: \rupf{\Sigma'}{\Delta}{\vec{G'}}{\vec{C'}}$, where $\vec{\D'} <^* \vec{\D}$
\item $\exists \Sigma[\nabla] \models \exists \Sigma'[\nabla']$
\end{enumerate}
\end{restateProp}
\begin{proof}
  Let $\vec{G}, \vec{C}, \nabla$ be given as above.  Since $\vec{G}$ is nonempty, we must have $\vec{G} = G,\vec{G_0}$ and
  $\vec{C} = C,\vec{C_0}$.  Proof is by case decomposition of the
  derivation of  $\rupf{\Sigma}{\Delta}{G}{C}$.
\begin{itemize}
  \item Suppose the derivation is of the form
\[
  \infer[con]{\rupfsd{C}{C}}{}
\]
thus, $\vec{G} = C,\vec{G'}$ and $\vec{C} = C,\vec{C'}$.  Then set
$\Sigma' = \Sigma$; $\nabla' = \nabla,C$.  We can take the step
$\st{\Sigma}{C,\vec{G_0}}{\nabla} \too
\st{\Sigma}{\vec{G_0}}{\nabla,C}$.  For (2), we already have smaller
derivations $\rupf{\Sigma}{\Delta}{\vec{G_0}}{\vec{C_0}}$ and for (3),
observe that $\exists\Sigma[\nabla,(C,\vec{C_0})] \models \exists
\Sigma[(\nabla,C),\vec{C_0}]$.

\item Case $\trueR$:  
If the derivation is of the form
\[
  \infer[\trueR]{\rupfsd{\true}{\true}}{}
\]
then $\vec{G} = \true,\vec{G'}$ and $\vec{C} = \true,\vec{C'}$.
Setting $\Sigma' = \Sigma, \nabla' = \nabla$, clearly
$\st{\Sigma}{\true,\vec{G'}}{\nabla}\too
\st{\Sigma}{\vec{G'}}{\nabla}$.  For (2), we already have smaller
derivations $\rupfsd{\vec{G'}}{\vec{C'}}$ and for (3),
$\exists\Sigma[\nabla,\true,\vec{C_0}] \models
\exists\Sigma[\nabla,\vec{C_0}]$.

\item Case $\andR$: If the derivation is of the form
\[
  \infer[\andR]{\rupfsd{G_1 \andd G_2}{C_1 \andd C_2}}
  {\rupfsd{G_1}{C_1} & \rupfsd{G_2}{C_2}}
\]
Thus, $\vec{G} = G_1 \andd G_2,\vec{G_0}$ and $\vec{C} = C_1 \andd
C_2,\vec{C_0}$.  Setting $\sigma = \Sigma$; $\nabla' = \nabla$;
$\vec{G'} = G_1,G_2,\vec{G_0}$; and $\vec{C} = C_1,C_2,\vec{C_0}$, we
can take the operational step $\st{\Sigma}{G_1\andd
  G_2,\vec{G_0}}{\nabla} \too \st{\Sigma}{G_1,G_2,\vec{G_0}}{\nabla}$.
In addition, for (2) we have subderivations
$\rupfsd{G_1,G_2,\vec{G_0}}{C_1,C_2,\vec{C_0}}$ and for (3),
$\exists\Sigma[\nabla,C_1\andd C_2,\vec{C_0}] \models
\exists\Sigma[\nabla,C_1, C_2,\vec{C_0}]$, as desired.

\item Case $\orR_i$: If the derivation is of the form
\[
  \infer[\orR_i]{\rupfsd{G_1 \orr G_2}{C}}
  {\rupfsd{G_i}{C}}
\]
then $\vec{G} = G_1 \orr G_2$ and $\vec{C} = C,\vec{C_0}$.  Setting
$\Sigma' = \Sigma;\nabla' = \nabla; \vec{G'} = G_i,\vec{G_0}$; and $\vec{C'}
= C,\vec{C_0}$; we can take the operational step $\st{\Sigma}{G_1 \orr
  G_2}{\nabla} \too \st{\Sigma}{G_i}{\nabla}$.  Moreover, we have for
part (2) immediate subderivations
$\rupfsd{G_i,\vec{G_0}}{C,\vec{C_0}}$ and part (3) is trivial.
\item Case $\exR$: For a derivation of the form
\[
  \infer[\exR]{\rupfsd{\exists X{:}\sigma.G}{\exists X.C}}
  {  \rupf{\Sigma,X}{\Delta}{G}{C} }
\]
we have $\vec{G} = \exists X.G,\vec{G_0}$ and $\vec{C} = \exists
X.C,\vec{C_0}$.  Setting $\Sigma' = \Sigma,X; \nabla' = \nabla;
\vec{G'} = G,\vec{G_0}; \vec{C'} = C,\vec{C_0}$; we can take the
operational step $\st{\Sigma}{\exists X.G,\vec{G_0}}{\nabla} \too
\st{\Sigma,X}{G,\vec{G_0}}{\nabla}$.  Moreover, for part (2), from the
given derivations we can obtain subderivations
$\rupf{\Sigma,X}{\Delta}{G,\vec{G_0}}{C,\vec{C_0}}$ .  For part (3),
observe that $\exists\Sigma[\nabla,\exists X.C,\vec{C_0}]
\models\exists\Sigma,X[ \nabla,C,\vec{C_0}]$ since $X$ is not free in
$\nabla,\vec{C_0}$.
\item Case $\newR$: In this case, the derivation is of the form
\[
  \infer[\newR]{\rupfsd{\new\Aa.G}{\new \Aa.C}}
  { \rupf{\Sigma\#\Aa}{\Delta}{G}{C} }
\]
$\vec{G} = \new \Aa.G,\vec{G_0}$ and $\vec{C} = \new \Aa.C,\vec{C_0}$.
Setting $\Sigma' = \Sigma\#\Aa; \nabla' = \nabla; \vec{G'} =
G,\vec{G_0}; \vec{C'} = C,\vec{C_0}$; we can take the operational step
$\st{\Sigma}{\new \Aa.G,\vec{G_0}}{\nabla} \too
\st{\Sigma\#\Aa}{G,\vec{G_0}}{\nabla}$.  In addition, for (2) we can obtain
smaller subderivations of
$\rupf{\Sigma\#\Aa}{\Delta}{G,\vec{G_0}}{C,\vec{C_0}}$ from the given
derivations, and for (3) observe that $\exists\Sigma[\nabla,\new \Aa.C,\vec{C_0}]
\models\exists\Sigma\#\Aa[ \nabla,C,\vec{C_0}]$ since $\Aa$ is not free in
$\nabla,\vec{C_0}$.

\item Case $back$: For a derivation of the form
\[
  \infer[back]{\rupfsd{A}{C}}
  {\rapfsd{D}{A}{G'}  & \rupfsd{G'}{C}& (D \in \Delta)}
\]
we have $\vec{G} = A,\vec{G_0}$ and $\vec{C} = C,\vec{C_0}$.  Set
$\Sigma = \Sigma'$; $\vec{G'} = G',\vec{G_0}$; $\vec{C'} =
C,\vec{C_0}$; $\nabla' = \nabla$.  Using the first subderivation, we
can take a backchaining step $\st{\Sigma}{A,\vec{G_0}}{\nabla} \too
\st{\Sigma}{G',\vec{G_0}}{\nabla}$.  Moreover, for part (2), using the
second subderivation we obtain a smaller derivation
$\rupfsd{G',\vec{G_0}}{C,\vec{C_0}}$, and part (3) is trivial.
\end{itemize}
This completes the proof.
\end{proof}

\begin{restateThm}{op-complete}
  If $\rupf{\Sigma}{\Delta}{\vec{G}}{\vec{C}}$ and $\nabla,\vec{C}$ is
  satisfiable then for some $\Sigma'$ and $\nabla'$, we have
  $\st{\Sigma}{\vec{G}}{\nabla} \too^*
  \st{\Sigma'}{\emptyset}{\nabla'}$ and $\exists \Sigma[\nabla,\vec{C}] \models\exists\Sigma'[\nabla']$.
\end{restateThm}
\begin{proof}
The proof is by induction on the length of $\vec{G}$ and the sizes of 
the derivations $\vec{\D}$ of $\rupf{\Sigma}{\Delta}{\vec{G}}{\vec{C}}$.  If $\vec{G}$ 
is empty, then we are done.  Otherwise, using \refProp{trans-complete},
there exist $\Sigma_0$, $\vec{G_0}$, $\vec{C_0}$, and $\nabla_0$, such that 
\[
\st{\Sigma}{\vec{G}}{\nabla} \too \st{\Sigma_0}{\vec{G_0}}{\nabla_0}
\quad
\deduce{\rupf{\Sigma_0}{\Delta}{\vec{G_0}}{\vec{C_0}}}{\vec{\D'}}
\quad
\exists \Sigma[\nabla,\vec{C}] \models \exists\Sigma_0[\nabla_0,\vec{C_0}]
\]
The derivations $\vec{\D'}$ are smaller than $\vec{\D}$, and the
satisfiability of $\nabla,\vec{C}$ implies that $\nabla_0,\vec{C_0}$
is also satisfiable, so the induction hypothesis applies.
Accordingly, construct $\Sigma', \nabla'$ such that
\[\st{\Sigma_0}{\vec{G_0}}{\nabla_0} \too^* \st{\Sigma'}{\emptyset}{\nabla'}
\quad
\exists \Sigma[\nabla_0,\vec{C_0}] \models\exists\Sigma'[\nabla']
\]
Chaining the transitions and entailments, we conclude
\[
\st{\Sigma}{\vec{G}}{\nabla} \too \st{\Sigma_0}{\vec{G_0}}{\nabla_0} \too^*\st{\Sigma'}{\emptyset}{\nabla'}
\quad
\exists \Sigma[\nabla,\vec{C}] \models \exists\Sigma_0[\nabla_0,\vec{C_0}]\models\exists\Sigma'[\nabla']
\]
as desired.
\end{proof}

\section{Proofs from \refSec{applications-eq-resolution}}\labelApp{eq-resolution-proofs}

\begin{restateLem}{eq-swapping}
  Let $\Delta$ be a \newgoal program and $\pi$ be a type-preserving
  permutation of names in $\Sigma$.
\begin{enumerate}  
\item If $\equpfsdn{G}$ then $\equpfsdn{\pi \act G}$.  
\item If $\eqapfsdn{D}{A}$ then $\eqapfsdn{\pi \act D}{\pi \act A}$.
\end{enumerate}  
\end{restateLem}
\begin{proof}
  By induction on derivations.  
\begin{itemize}

\item For case $con$, we transform derivations as follows:
\[
  \infer[con]{\equpfsdn{ C}}
  {\satsn{C}}
\longmapsto
  \infer[con]{\equpfsdn{ \pi \act C}}
  {\satsn{\pi \act C}}
\]
since $\satsn{C}$ implies $\satsn{\pi \act C}$.
\item For case $\trueR$, we transform
\[
  \infer[\trueR]{\equpfsdn{\true}}{}
\longmapsto
  \infer[\trueR]{\equpfsdn{\pi \act \true}}{}
\]
since $\pi \act \true = \true$.
\item For case $\andR$, note that $\pi \act(G_1 \andd G_2) = \pi \act
  G_1 \andd \pi \act G_2$, so we transform
\[
\infer[\andR]{\equpfsdn{G_1 \andd G_2}}
  {\deduce{\equpfsdn{G_1}}{\D_1} & \deduce{\equpfsdn{G_2}}{\D_2}}
\longmapsto
\infer[\andR]{\equpfsdn{\pi \act(G_1 \andd G_2)}}
  {\deduce{\equpfsdn{\pi \act G_1}}{\D_1'} & \deduce{\equpfsdn{\pi \act G_2}}{\D_2'}}
\]
where by induction $\D_i :: \equpfsdn{G_i} \longmapsto \D_i' ::\equpfsdn{\pi \act G_i}$
for $i \in \{1,2\}$.
\item For case $\orR_i$ ($i \in \{1,2\}$), note that $\pi \act (G_1
  \orr G_2) = \pi \act G_1 \orr \pi \act G_2$, so we have
\[
  \infer[\orR_i]{\equpfsdn{G_1 \orr G_2}}
  {\deduce{\equpfsdn{G_i}}{\D}}
\longmapsto
  \infer[\orR_i]{\equpfsdn{\pi \act(G_1 \orr G_2)}}
  {\deduce{\equpfsdn{\pi \act G_i}}{\D'}}
\]
where by induction 
$\D :: \equpfsdn{G_i} \longmapsto \D' ::\equpfsdn{\pi \act G_i}$
\item For case $\exR$, we have 
\[
\infer{\equpfsdn{\exists X.G}}
{\sat{\Sigma}{\nabla}{\exists X.C[X]} & 
\deduce{\equpf{\Sigma,X}{\Delta}{\nabla,C[X]}{G}}{\D}}
\]
Note that $\pi \act \exists X.G[X] = \exists X. \pi \act G[\pi^{-1} \act X]$.  By
induction,
\[\deduce{\equpf{\Sigma,X}{\Delta}{\nabla,C[X]}{G}}{\D}
\longmapsto \deduce{\equpf{\Sigma,X}{\Delta}{\nabla,C[X]}{\pi \act G[X]}}{\D'}\;.\]
Since $\pi$ is invertible, we can substitute $Y = \pi
\act X$ to obtain $\D''::\equpf{\Sigma,Y}{\Delta}{\nabla,C[\pi^{-1}  \act
  Y]}{\pi \act G[\pi^{-1} \act Y]}$; moreover, clearly,
$\sat{\Sigma}{\nabla}{\exists Y.C[\pi^{-1}\act Y]}$, so we can
conclude 
\[
\infer{\equpfsdn{\pi \act \exists X.G}}
{\satsn{\exists Y.C[\pi^{-1}\act Y]}
&
\deduce{\equpf{\Sigma,Y}{\Delta}{\nabla,C[\pi ^{-1} \act
  Y]}{\pi \act G[\pi^{-1} \act Y]}}{\D''}}\;.
\]
\item For case $\newR$, we have derivation
\[
\infer[\newR]{\equpfsdn{\new\Aa{:}\nu.G}}
{\sat{\Sigma}{\nabla}{\new \Aa.C} & 
  \deduce{\equpf{\Sigma\#\Aa}{\Delta}{\nabla,C}{G}}{\D} }
\longmapsto
\infer[\newR]{\equpfsdn{\pi \act(\new\Aa{:}\nu.G)}}
{\sat{\Sigma}{\nabla}{\new \Aa.C} & 
  \deduce{\equpf{\Sigma\#\Aa}{\Delta}{\nabla,C}{\pi \act G}}{\D'} }
\]
since $\pi \act \new \Aa{:}\nu.G = \new \Aa{:}\nu.\pi \act G$, (since,
without loss, $\Aa\not\in FN(\Sigma) \cup \supp(\pi)$).  The
derivation $\D'::\equpf{\Sigma\#\Aa}{\Delta}{\nabla,C}{\pi \act G}$ is
obtained by induction.
\item For case $sel$, 
\[
  \infer[sel]{\equpfsdn{A}}{\deduce{\eqapfsdn{D}{A}}{\D} & (D \in \Delta)}
\longmapsto
  \infer[sel]{\equpfsdn{\pi \act A}}{\deduce{\eqapfsdn{D}{\pi \act A}}{\D'} & (D \in \Delta)}
\]
using induction hypothesis (2) to derive $\D'$ from $\D$, and the fact that $\pi \act D = D$
(because $D \in \Delta$ is closed).
\end{itemize}
For part (2), all cases are straightforward; cases $hyp$ and $\newL$ are
of interest.
\begin{itemize}
\item Case $hyp$
\[
  \infer[hyp]{\eqapfsdn{A'}{A}}{\satsn{A' \eq A}}
\longmapsto 
  \infer[hyp]{\eqapfsdn{\pi \act A'}{\pi \act A}}{\satsn{\pi \act A' \eq \pi \act A}}
\]
since $\sat{\Sigma}{A' \eq A}{\pi \act A' \eq \pi \act A}$.
\item Case $\andL_i$
\[
  \infer[\andL_i]{\eqapfsdn{D_1 \andd D_2}{A}}{\eqapfsdn{D_i}{A}}
\longmapsto 
  \infer[\andL_i]{\eqapfsdn{\pi \act(D_1 \andd D_2)}{\pi \act A}}{\eqapfsdn{\pi \act D_i}{\pi \act A}}
\]
since $\pi \act (D_1 \andd D_2) = \pi \act D_1 \andd \pi \act D_2$.
The subderivations are constructed by induction.
\item Case $\impL$
\[
  \infer[\impL]{\eqapfsdn{G \impp D}{ A}}
  {\eqapfsdn{D}{A} & \equpfsdn{G}}
\longmapsto
  \infer[\impL]{\eqapfsdn{\pi \act(G \impp D)}{ \pi \act A}}
  {\eqapfsdn{\pi \act D}{\pi \act A} & \equpfsdn{\pi \act G}}
\]
where the subderivations are obtained by induction; this suffices
because $\pi \act (G \impp D) = \pi \act G \impp \pi \act D$.
\item Case $\allL$: We have
\[
  \infer[\allL]{\eqapfsdn{\forall X{:}\sigma. D}{A}}
  {\sat{\Sigma}{\nabla}{\exists X.C[X]} & \eqapf{\Sigma,X}{\Delta}{\nabla,C[X]}{D[X]}{A} }
\]
The argument is  similar to that for $\exR$ for part (1).
By induction we have $\eqapf{\Sigma,X}{\Delta}{\nabla,C[X]}{\pi \act D[X]}{\pi \act A}$.  Substituting $Y = \pi \act X$, we have $\eqapf{\Sigma,Y}{\Delta}{\nabla,C[\pi^{-1} \act Y]}{\pi \act D[\pi ^{-1} \act Y]}{\pi \act A}$.
and $\sat{\Sigma}{\nabla}{\exists Y.C[\pi^{-1} \act Y]}$.  
It follows that
\[
  \infer[\allL]{\eqapfsdn{\pi \act \forall Y{:}\sigma. D}{\pi \act A}}
  { \sat{\Sigma}{\nabla}{\exists Y.C[\pi^{-1} \act Y]}
& 
 \eqapf{\Sigma,Y}{\Delta}{\nabla,C[\pi^{-1} \act Y]}{\pi \act D[\pi ^{-1} \act Y]}{\pi \act A}}
\]
since $\pi \act \forall Y{:}\sigma. D = \forall Y. \pi \act D[\pi^{-1}
\act Y]$.
\item The case for $\newL$ is vacuous because no formulas $\new
  \Aa.D$ can appear in a \newgoal program.
\end{itemize}
This completes the proof.
\end{proof}

\begin{restateThm}{nice-eq-equivariance}
  If $\Delta$ is \newgoal then
\begin{enumerate}
\item If $\upfsdn{G}$ is derivable, then $\equpfsdn{G}$ is derivable.
\item If $\apfsdn{D}{A}$ is derivable, there exists a $\pi$ such that
  $\eqapfsdn{\pi \act D}{A}$ is derivable.
\end{enumerate}
\end{restateThm}
\begin{proof}
  The proof is by induction on derivations.  
For part (1), the most interesting case is $sel$; the rest are
straightforward {and omitted}.
\begin{itemize}
\item
For $sel$, we have
\[\infer{\upfsdn{A}}{\apfsdn{D}{A}}\]
for some closed $D \in \Delta$.
  By induction hypothesis (2), for some $\pi$, $\eqapfsdn{\pi \act D}{A}$ holds.  However,
  since $D$ is closed, $\pi \act D = D \in \Delta$ so we may conclude
  \[\infer[sel]{\equpfsdn{A}}{\eqapfsdn{D}{A} & (D \in \Delta)}\]
%
\end{itemize}
For part (2), the interesting cases are $hyp$ and $\newL$; {the others are omitted.}
\begin{itemize}
\item For $hyp$, we have
\[\infer[hyp]{\apfsdn{
    A'}{A}}{\sat{\Sigma}{\nabla}{A' \ev A}}
\]
By definition $\sat{\Sigma}{\nabla}{A' \ev A}$ means there exists a
$\pi$ such that $\satsn{\pi \act A' \eq A}$, so 
\[\infer[hyp]{\eqapfsdn{\pi \act
  A'}{A}}
{\satsn{\pi \act A' \eq A}}
\]
\item Case $\newL$ is vacuous, since no instance of $\newL$ can occur in 
a derivation involving a \newgoal program.
\end{itemize}
This completes the proof.
\end{proof}
